%%%%%%%%%%%%%%%%%%%%%
\def\yyy{} %reminder
%%%%%%%%%%%%%%%%%%%%%%
% packages
%%%%%%%%%%%%%%%%%%%%%%
\input amssym
\input harvmac
\input texdraw
\input xymatrix
\input xyarrow
%\input epsf
%\draftmode
\noblackbox

%%%%%%%%%%%%%%%%%%%%%%
%  Figures
%%%%%%%%%%%%%%%%%%%%%%
%
\def\lfig#1{%  this is to call the figure in the text
\let\labelflag=#1%
\def\numb@rone{#1}%
\ifx\labelflag\UnDeFiNeD%
{\xdef#1{\the\figno}%
\writedef{#1\leftbracket{\the\figno}}%
\global\advance\figno by1%
}\fi{\hyperref{}{figure}{{\numb@rone}}{Fig.~{\numb@rone}}}}
\def\figboxinsert#1#2#3{%
\let\flag=#1
\ifx\flag\UnDeFiNeD
  {\xdef#1{\the\figno}
   \writedef{#1\leftbracket{\the\figno}}
   \global\advance\figno by1}
\fi
\vbox{\bigskip \centerline{#3} \smallskip
\leftskip 4pc \rightskip 4pc
\noindent\ninepoint\sl \baselineskip=11pt
{\bf{\hyperdef\hypernoname{figure}{{#1}}{Fig.~{#1}}}.~}#2
\smallskip}
\bigskip}

%%%%%%%%%%%%%%%%%%%%%%
%  Tables
%%%%%%%%%%%%%%%%%%%%%%
\newcount\tabno
\tabno=1
\def\ltab#1{%
\let\labelflag=#1%
\def\numb@rone{#1}%
\ifx\labelflag\UnDeFiNeD{%
  \xdef#1{\the\tabno}%
  \writedef{#1\leftbracket{\the\tabno}}%
  \global\advance\tabno by1%
}%
\fi%
{\hyperref{}{table}{{\numb@rone}}{Table~{\numb@rone}}}}
\def\tabinsert#1#2#3{%
\let\flag=#1
\ifx\flag\UnDeFiNeD
  {\xdef#1{\the\tabno}
   \writedef{#1\leftbracket{\the\tabno}}
   \global\advance\tabno by1 }
\fi
\vbox{\bigskip \centerline{#3} \smallskip
\leftskip 4pc \rightskip 4pc
\noindent\ninepoint\sl \baselineskip=11pt
{\bf{\hyperdef\hypernoname{table}{{#1}}{Table~{#1}}}.~}#2
\smallskip}
\bigskip}

%%%%%%%%%%%%%%%%%%%%%%
% hypertex
%%%%%%%%%%%%%%%%%%%%%%
\newif\ifhypertex
\ifx\hyperdef\UnDeFiNeD
\hypertexfalse
\message{[HYPERTEX MODE OFF]}
\def\hyperref#1#2#3#4{#4}
\def\hyperdef#1#2#3#4{#4}
\def\hypernoname{}
\def\e@tf@ur#1{}

\else
\hypertextrue
\message{[HYPERTEX MODE ON]}

\fi

%%%%%%%%%%%%%%%%%%%%%%
% Section commands
%%%%%%%%%%%%%%%%%%%%%%
%
\def\newsec#1{\global\advance\secno by1\message{(\the\secno. #1)}
\global\subsecno=0\eqnres@t\noindent{\bf\the\secno. #1}
\writetoca{{\bf\secsym} {\bf #1}}\par\nobreak\medskip\nobreak}
\def\eqnres@t{\xdef\secsym{\the\secno.}\global\meqno=1\bigbreak\bigskip}
\def\sequentialequations{\def\eqnres@t{\bigbreak}}\xdef\secsym{}
\def\subsec#1{\global\advance\subsecno by1\message{(\secsym\the\subsecno. #1)}
\ifnum\lastpenalty>9000\else\bigbreak\fi
\noindent{\it\secsym\the\subsecno. #1}\writetoca{\string\quad 
{\secsym\the\subsecno.} {#1}}\par\nobreak\medskip\nobreak}
\def\subsubsec#1{\message{(#1)}
\ifnum\lastpenalty>9000\else\bigbreak\fi
\noindent{\it #1}\writetoca{\string\quad\quad {#1}}\par\nobreak\medskip\nobreak}
\def\appendix#1#2{\global\meqno=1\global\subsecno=0\xdef\secsym{\hbox{#1.}}
\bigbreak\bigskip\noindent{\bf Appendix #1. #2}\message{(#1. #2)}
\writetoca{{\bf Appendix {#1.} {#2}}}\par\nobreak\medskip\nobreak}
\newif\iffileexists

%
%
%%
%%% Table of Contents %%%
%
\def\listtoc{\centerline{\bf Table of Contents}\nobreak\medskip{\baselineskip=12pt
\parskip=0pt\catcode`\@=11 
%\inputiffileexists{toc.tmp}
%%%%%%%%%%%%%%%%%%%%%%%%%%%%%%%%%%%%
\noindent {\fam \bffam \tenbf 1.} {\fam \bffam \tenbf Introduction} \leaderfill{2} \par 
\noindent {\fam \bffam \tenbf 2.} {\fam \bffam \tenbf Preliminaries: M-theory on Calabi--Yau fourfolds} \leaderfill{5} \par 
\noindent {\fam \bffam \tenbf 3.} {\fam \bffam \tenbf Conifold transitions in local Calabi--Yau fourfolds} \leaderfill{8} \par 
\noindent \quad{3.1.} {Small resolution phases $\Xlsharp$} \leaderfill{8} \par 
\noindent \quad{3.2.} {Deformed phase $\Xlflat$} \leaderfill{11} \par 
\noindent \quad{3.3.} {Classification of $G$-flux on the local geometries} \leaderfill{13} \par 
\noindent \quad{3.4.} {Non-dynamical flux constraints for the phase transitions} \leaderfill{17} \par 
\noindent \quad{3.5.} {M-theory three-form $C$-field and Cheeger-Simons cohomology} \leaderfill{20} \par 
\noindent \quad{3.6.} {Flat directions of the superpotential and Abel-Jacobi map} \leaderfill{22} \par 
\noindent \quad{3.7.} {Local transitions for special configurations} \leaderfill{28} \par 
\noindent {\fam \bffam \tenbf 4.} {\fam \bffam \tenbf Conifold transition in global Calabi--Yau fourfolds} \leaderfill{36} \par 
\noindent \quad{4.1.} {A simple example: Extremal transition for the sextic} \leaderfill{37} \par 
\noindent \quad{4.2.} {M-theory transitions via topological surgery} \leaderfill{42} \par 
\noindent \quad{4.3.} {M-theory transitions via the Clemens--Schmid exact sequence} \leaderfill{44} \par 
\noindent \quad{4.4.} {$G$-flux quantization condition} \leaderfill{45} \par 
\noindent \quad{4.5.} {Non-Abelian gauge groups and relation to F-theory} \leaderfill{47} \par 
\noindent {\fam \bffam \tenbf 5.} {\fam \bffam \tenbf M-theory phases in the effective $N=2$ three-dimensional field theory} \leaderfill{53} \par 
\noindent \quad{5.1.} {$N=2$ three-dimensional field theory} \leaderfill{54} \par 
\noindent \quad{5.2.} {Effective $N=2$ field theory for M-theory on smooth fourfolds with flux} \leaderfill{56} \par 
\noindent \quad{5.3.} {Singularities, charged matter, and the $5d\to 3d$ reduction} \leaderfill{57} \par 
\noindent \quad{5.4.} {The dynamics of 3d $U(1)$ gauge theories for general matter fields of charge $\pm 1$} \leaderfill{60} \par 
\noindent \quad{5.5.} {Comparing M-theory conifold transitions with 3d field theory phase structure} \leaderfill{62} \par 
\noindent {\fam \bffam \tenbf 6.} {\fam \bffam \tenbf Conclusions} \leaderfill{66} \par 
\noindent {\fam \bffam \tenbf Appendix {A.} {Calabi--Yau fourfolds and collection of (co)homology data}} \leaderfill{69} \par 
\noindent {\fam \bffam \tenbf Appendix {B.} {Conifold flop transitions in local Calabi--Yau fourfolds}} \leaderfill{74} \par 
\noindent {\fam \bffam \tenbf Appendix {C.} {The quarternionic Hopf fibration and the Milnor fibration}} \leaderfill{76} \par 
\noindent {\fam \bffam \tenbf Appendix {D.} {The Clemens-Schmid exact sequence}} \leaderfill{79} \par 
\noindent \quad{\hbox {D.}1.} {Triple-point-free Clemens-Schmid exact sequences} \leaderfill{79} \par 
\noindent \quad{\hbox {D.}2.} {Conifold transition in Calabi--Yau threefolds} \leaderfill{82} \par 
\noindent \quad{\hbox {D.}3.} {Conifold transition along a genus $g$ curve in global Calabi--Yau fourfolds} \leaderfill{84} \par 
%%%%%%%%%%%%%%%%%%%%%%%%%%%%%%%%%%%%
\catcode`\@=12\bigbreak\bigskip}}
%

%%%%%%%%%%%%%%%%%%%%%%
% Sizes
%%%%%%%%%%%%%%%%%%%%%%
\def\mn{\the\secno.\the\subsecno}
\baselineskip=16pt plus 2pt minus 1pt
\parskip=2pt plus 16pt minus 1pt

%%%%%%%%%%%%%%%%%%%%%%
% TeXdraw
%%%%%%%%%%%%%%%%%%%%%%
\def\DrawDiag#1{ 
{\btexdraw 
\drawdim pt \linewd 0.75 \lpatt ()
\arrowheadtype t:F \arrowheadsize l:8 w:4 
\textref h:C v:C
#1 
\etexdraw} }

%%%%%%%%%%%%%%%%%%%%%%
% macros
%%%%%%%%%%%%%%%%%%%%%%

% small font in math environment
\font\fivesm=cmr10  scaled 350 
\font\sevensm=cmr10  scaled 500
\font\tensm=cmr10 scaled 700      
\font\fivems=cmmi10 scaled 350 
\font\sevenms=cmmi10 scaled 500
\font\tenms=cmmi10 scaled 700           
\def\smmath{
  \textfont0=\tensm\scriptfont0=\sevensm\scriptscriptfont0=\fivesm  
  \textfont1=\tenms\scriptfont1=\sevenms\scriptscriptfont1=\fivems}
% 'double' letters
\def\IB{{\Bbb B}}
\def\IC{{\Bbb C}}
\def\IH{{\Bbb H}}

\def\IP{{\Bbb P}}
\def\IQ{{\Bbb Q}}
\def\IR{{\Bbb R}}

\def\IZ{{\Bbb Z}}
% calligraphic letters

\def\cB{{\cal B}}
\def\cC{{\cal C}}

\def\cE{{\cal E}}

\def\cI{{\cal I}}
\def\cJ{{\cal J}}

\def\cL{{\cal L}}

\def\cO{{\cal O}}
\def\cP{{\cal P}}

\def\cS{{\cal S}}
\def\cT{{\cal T}}

\def\cX{{\cal X}}

\def\al{\alpha}
\def\be{\beta}
\def\Ga{\Gamma}
\def\la{\lambda}

\def\eps{\epsilon}

\def\om{\omega}

%%%%
\def\frac#1#2{{#1 \over #2 }}
\def\fc#1#2{{#1 \over #2 }}
\def\slashedD{{\slash\!\!\!\! D}}
\def\subsubsecc#1{\noindent {\it #1} \hfil\break}

%%%%% AMS macros %%%%
\def\mathcal#1{{\cal #1}}
\def\mathbb#1{{\Bbb #1}}
\def\operatorname#1{{\rm #1}}
\def\text#1{{\rm #1}}
\def\overset#1#2{{\buildrel#1\over#2}}

%%%%% Note specific macros %%%%%%
\def\-{\phantom{-}}
\def\Xs{X^\sharp}
\def\Xf{X^\flat}
\def\Xc{X^c}

\def\Xlsharp{\widetilde\Xs}
\def\Xlsharpi#1{\widetilde {X^\sharp_#1}}
\def\Xlflat{\widetilde\Xf}
\def\Xlsing{{\widetilde X}^{\rm sing}}
\def\Cg{\cC}
\def\GS{{G^\sharp}}
\def\GSi#1{{G^\sharp_#1}}
\def\GF{{G^\flat}}
\def\CCg#1{\cC_{#1}}
\def\Ssharp{S^\sharp}
\def\Svform{d{\rm vol}(\Ssharp)}
\def\SZ{{S_G}}
\def\mod{{\, \rm mod \,}}
\def\Wtot{{\cal W}}
\def\bq#1{b^\flat_{#1}}
\def\sentry#1{\hfil$\smmath{\vphantom{Xg}#1}$\hfil}
\def\qt{q_-}\def\qp{q_+}
\def\JC{\cJ} % supercurrent
\def\LM{J}   % bosonic component of linear multiplets

%%%%%%%%%%%%%%%%%%%%%%
% references
%%%%%%%%%%%%%%%%%%%%%%
\lref\Chow{
W.-L. Chow, 
``On compact complex analytic varieties,''
Amer. J. Math. {\bf 71} (1949) 893–-914. 
}
\lref\AlimZA{
M.~Alim, M.~Hecht, H.~Jockers, P.~Mayr, A.~Mertens and M.~Soroush,
``Type II/F-theory Superpotentials with Several Deformations and N=1 Mirror Symmetry,''
JHEP {\bf 1106}, 103 (2011).
[arXiv:1010.0977 [hep-th]].
}
\lref\AtiyahAA{
  M.~F.~Atiyah, ``Riemann surfaces and spin structures,'' 
  Ann.\ Sci. {\'E}cole Norm. Sup. (4), 47 (1971).}
\lref\BaerAA{
  C.~B\"ar and P.~Schmutz, ``Harmonic spinors on Riemann surfaces,''
  Ann.\ Global\ Anal.\ Geom.\ {\bf 10}, 263, (1992).}
\lref\GuilleminAA{
V.~Guillemin and S.~Sternberg, ``Birational equivalence in the symplectic
  category,'' Invent.\ Math.\ {\bf 97}, 485 (1989).}
\lref\HitchinAA{N.~Hitchin, ``Harmonic spinors,'' 
  Advances\ in\ Math.\ {\bf 14}, 1 (1974).}
\lref\MartensAA{H.~Martens, ``Varietes of special divisors on a curve II,''
  J.\ Reine\ Angew.\ Math.\ {\bf 233}, 89 (1968).}  
%%% Clemens-Schmid appendix Refs %%%
\lref\SchmidAA{
W.~Schmid, ``Variation of {H}odge structure: the singularities of the period
  mapping,'' Invent. Math. {\bf 22} 211-319 (1973).}
\lref\ClemensAA{
C.~H. Clemens, ``Degeneration of {K}\"ahler manifolds,'' Duke Math. J. {\bf
  44}, 215-290 (1977).}
\lref\MorrisonAA{
D.~R. Morrison, ``The {C}lemens--{S}chmid exact sequence and applications,''
  Topics in Transcendental Algebraic Geometry (P.~Griffiths, ed.), Annals of
  Math. Studies {\bf 106}, Princeton University Press, Princeton, 101-119 (1984).}
\lref\GriffithsAA{
P.~Griffiths and W.~Schmid, ``Recent developments in {H}odge theory: a
  discussion of techniques and results,'' 
  Discrete subgroups of Lie groups and applications to moduli (Internat. Colloq., Bombay, 1973), Oxford Univ. Press, 31-127 (1975).}
\lref\CornalbaAA{  
  M.~Cornalba and P.~A. Griffiths, ``Some transcendental aspects of algebraic
  geometry,'' Algebraic geometry (Humboldt State Univ., Arcata, Calif., 1974),
  Proc. Sympos. Pure Math. {\bf 29}, Amer. Math. Soc., 3-110 (1975).}
\lref\PerssonAA{  
  U.~Persson, ``On degenerations of algebraic surfaces,'' Memoirs Amer.
  Math. Soc., {\bf 189} (1977).}
\lref\CrauderAA{  
B.~Crauder and D.~R. Morrison, ``Triple-point-free degenerations of surfaces
  with {K}odaira number zero,'' The Birational Geometry of Degenerations
  (R.~Friedman and D.~R. Morrison, eds.), Progress in Math. {\bf 29},
  Birkh\"auser, Boston, Basel, Stuttgart, 353-386 (1983).}
\lref\CrauderAB{
B.~Crauder and D.~R. Morrison, ``Minimal models and degenerations of surfaces with {K}odaira
  number zero,'' Trans. Amer. Math. Soc. {\bf 343}, 525-558 (1994).}
\lref\ClemensAB{
  C.~H. Clemens, ``Double solids,'' Adv. in Math. {\bf 47}, 107-230 (1983).}
\lref\ClemensAC{
C.~H. Clemens, ``Homological equivalence, modulo algebraic equivalence, is not
  finitely generated,'' Publ. Math. IHES {\bf 58}, 19-38 (1983).}
\lref\FriedmanAA{  
R.~Friedman, ``Simultaneous resolution of threefold double points,'' Math.
  Ann. {\bf 274} 671-689 (1986).}
\lref\CollinucciGZ{
  A.~Collinucci and R.~Savelli,
  ``On Flux Quantization in F-Theory,''
JHEP {\bf 1202}, 015 (2012).
[arXiv:1011.6388 [hep-th]].}
\lref\CollinucciAS{
  A.~Collinucci and R.~Savelli,
  ``On Flux Quantization in F-Theory II: Unitary and Symplectic Gauge Groups,''
[arXiv:1203.4542 [hep-th]].}
\lref\MorrisonAB{
D.~R. Morrison, ``Through the looking glass,'' Mirror Symmetry {III} (D.~H.
  Phong, L.~Vinet, and S.-T. Yau, eds.), AMS/IP Stud. Adv. Math. {\bf 10},
  International Press, Cambridge, 263-277 (1999) [arXiv:alg-geom/9705028].}
\lref\KempfAA{  
G.~Kempf, F.~F. Knudsen, D.~Mumford, and B.~Saint-Donat, ``Toroidal
  embeddings. {I},'' Lecture Notes in Math. {\bf 339}, Springer-Verlag, Berlin (1973).}
\lref\Conway{J. H. Conway, N. J. A. Sloane, ``Sphere Packings, Lattices and Groups''. Springer 1988.}
\lref\OoguriCK{
  H.~Ooguri, Y.~Oz, Z.~Yin,
  ``D-branes on Calabi--Yau spaces and their mirrors,''
Nucl.\ Phys.\  {\bf B477}, 407-430 (1996).
[arXiv:hep-th/9606112].}
\lref\KlemmTS{
  A.~Klemm, B.~Lian, S.~S.~Roan and S.~T.~Yau,
  ``Calabi--Yau fourfolds for M- and F-theory compactifications,''
  Nucl.\ Phys.\  B {\bf 518}, 515 (1998)
  [arXiv:hep-th/9701023].}
\lref\MayrSH{
  P.~Mayr,
  ``Mirror symmetry, N = 1 superpotentials and tensionless strings on
  Calabi--Yau four-folds,''
  Nucl.\ Phys.\  B {\bf 494}, 489 (1997)
  [arXiv:hep-th/9610162].}
\lref\MumfordAA{
D.~Mumford, ``Theta-characteristics on algebraic curves,'' Ann. Ecole Norm. Sup. (4) {\bf 4} 181-192 (1971).}
\lref\MorrisonBM{
  D.~R.~Morrison and J.~Walcher,
  ``D-branes and normal Functions,''
[arXiv:0709.4028 [hep-th]].}
\lref\WalcherCalculations{
J.~Walcher,
  ``Calculations for mirror symmetry with D-branes,''
JHEP {\bf 0909}, 129 (2009)
[arXiv:0904.4095 [hep-th]].}
\lref\JMW{
H.~Jockers, P.~Mayr and J.~Walcher,
  ``On N=1 4d effective couplings for F-theory and heterotic vacua,''
Adv.\ Theor.\ Math.\ Phys.\  {\bf 14}, 1433 (2010)
[arXiv:0912.3265 [hep-th]].}
%%%%%%%% Inspires %%%%%%%%%%
%
%\AharonyBX
\lref\AharonyBX{
  O.~Aharony, A.~Hanany, K.~A.~Intriligator, N.~Seiberg and M.~J.~Strassler,
  ``Aspects of N = 2 supersymmetric gauge theories in three dimensions,''
  Nucl.\ Phys.\  B {\bf 499}, 67 (1997)
  [arXiv:hep-th/9703110].
  %%CITATION = NUPHA,B499,67;%%
}
%\BeasleyDC
\lref\BeasleyDC{
  C.~Beasley, J.~J.~Heckman and C.~Vafa,
  ``GUTs and exceptional branes in F-theory - I,''
  JHEP {\bf 0901}, 058 (2009)
  [arXiv:0802.3391 [hep-th]].
  %%CITATION = JHEPA,0901,058;%%
}
%\BeasleyKW
\lref\BeasleyKW{
  C.~Beasley, J.~J.~Heckman and C.~Vafa,
  ``GUTs and exceptional branes in F-theory - II: Experimental predictions,''
  JHEP {\bf 0901}, 059 (2009)
  [arXiv:0806.0102 [hep-th]].
  %%CITATION = JHEPA,0901,059;%%
}
%\BeckerAY
\lref\BeckerAY{
  K.~Becker, M.~Becker, D.~R.~Morrison, H.~Ooguri, Y.~Oz, Z.~Yin,
  ``Supersymmetric cycles in exceptional holonomy manifolds and Calabi--Yau 4 folds,''
  Nucl.\ Phys.\  {\bf B480}, 225-238 (1996).
  [arXiv:hep-th/9608116].
  %%CITATION = hep-th/9608116%%
}
%\BeckerGJ
\lref\BeckerGJ{
  K.~Becker and M.~Becker,
  ``M-theory on eight-manifolds,''
  Nucl.\ Phys.\  B {\bf 477}, 155 (1996)
  [arXiv:hep-th/9605053].
  %%CITATION = NUPHA,B477,155;%%
}
%\BerglundUY
\lref\BerglundUY{
  P.~Berglund, S.~H.~Katz, A.~Klemm and P.~Mayr,
  ``New Higgs transitions between dual N = 2 string models,''
  Nucl.\ Phys.\  B {\bf 483}, 209 (1997)
  [arXiv:hep-th/9605154].
  %%CITATION = NUPHA,B483,209;%%
}
%\BraunZM
\lref\BraunZM{
  A.~P.~Braun, A.~Collinucci and R.~Valandro,
  ``G-flux in F-theory and algebraic cycles,''
  Nucl.\ Phys.\ B {\bf 856}, 129 (2012).
  [arXiv:1107.5337 [hep-th]].
%%CITATION = arXiv:1107.5337%%
}
%\CurioBVA
\lref\CurioBVA{
  G.~Curio and R.~Y.~Donagi,
  ``Moduli in N=1 heterotic / F theory duality,''
Nucl.\ Phys.\ B {\bf 518}, 603 (1998).
[hep-th/9801057].
%%CITATION = hep-th/9801057%%
}
%\KrauseYH
\lref\KrauseYH{
  S.~Krause, C.~Mayrhofer and T.~Weigand,
  ``Gauge Fluxes in F-theory and Type IIB Orientifolds,''
[arXiv:1202.3138 [hep-th]].
%%CITATION = arXiv:1202.3138%%
}
%\GreeneVM
\lref\GreeneVM{
  B.~R.~Greene, D.~R.~Morrison, M.~R.~Plesser,
  ``Mirror manifolds in higher dimension,''
  Commun.\ Math.\ Phys.\  {\bf 173}, 559-598 (1995).
  [arXiv:hep-th/9402119 [hep-th]].
%%CITATION = CLNS-93-1253%%
}
%\GukovES
\lref\GukovES{
 S.~Gukov and D.~Tong,
 ``D-brane probes of special holonomy manifolds, and dynamics of N = 1 three-dimensional gauge theories,''
JHEP {\bf 0204}, 050 (2002).
[arXiv:hep-th/0202126].
%%CITATION = hep-th/0202126%%
}
%\GukovZG
\lref\GukovZG{
  S.~Gukov, J.~Sparks, D.~Tong,
  ``Conifold transitions and five-brane condensation in M theory on spin(7) manifolds,''
Class.\ Quant.\ Grav.\  {\bf 20}, 665-706 (2003).
[hep-th/0207244].
%%CITATION = hep-th/0207244%%
}
%\MarsanoIX
\lref\MarsanoIX{
  J.~Marsano, N.~Saulina and S.~Sch\"afer-Nameki,
  ``A Note on G-Fluxes for F-theory Model Building,''
JHEP {\bf 1011}, 088 (2010).
[arXiv:1006.0483 [hep-th]].
%%CITATION = arXiv:1006.0483%%
}
%\MarsanoHV
\lref\MarsanoHV{
J.~Marsano and S.~Sch\"afer-Nameki,
``Yukawas, G-flux, and Spectral Covers from Resolved Calabi-Yau's,''
JHEP {\bf 1111}, 098 (2011).
[arXiv:1108.1794 [hep-th]].
%%CITATION = arXiv:1108.1794%%
}
%\PoppitzHR
\lref\PoppitzHR{
  E.~Poppitz and M.~Unsal,
  ``Index theorem for topological excitations on $R^3\times S^1$ and Chern-Simons theory,''
JHEP {\bf 0903}, 027 (2009).
[arXiv:0812.2085 [hep-th]].
%%CITATION = arXiv:0812.2085%%
}
%\DonagiCA
\lref\DonagiCA{
  R.~Donagi, M.~Wijnholt,
  ``Model building with F-theory,''
[arXiv:0802.2969 [hep-th]].
%%CITATION = arXiv:0802.2969%%
}
%\DiaconescuJV
\lref\DiaconescuJV{
  D.~E.~Diaconescu, R.~Donagi and T.~Pantev,
  ``Geometric transitions and mixed Hodge structures,''
Adv.\ Theor.\ Math.\ Phys.\  {\bf 11}, 65--89 (2007)
  [arXiv:hep-th/0506195].
  %%CITATION = HEP-TH/0506195;%%
}
%\BrunnerBF
\lref\BrunnerBF{
  I.~Brunner, M.~Lynker and R.~Schimmrigk,
  ``Dualities and phase transitions for Calabi--Yau threefolds and  fourfolds,''
  Nucl.\ Phys.\ Proc.\ Suppl.\  {\bf 56B}, 120 (1997)
  [arXiv:hep-th/9703182].
  %%CITATION = NUPHZ,56B,120;%%
}
%\BrunnerBU
\lref\BrunnerBU{
  I.~Brunner, M.~Lynker and R.~Schimmrigk,
  ``Unification of M- and F-theory Calabi--Yau fourfold vacua,''
  Nucl.\ Phys.\  B {\bf 498}, 156 (1997)
  [arXiv:hep-th/9610195].
  %%CITATION = NUPHA,B498,156;%%
}
%\CandelasUG
\lref\CandelasUG{
  P.~Candelas, P.~S.~Green and T.~H\"ubsch,
  ``Rolling among Calabi--Yau vacua,''
  Nucl.\ Phys.\  B {\bf 330}, 49 (1990).
  %%CITATION = NUPHA,B330,49;%%
}
%\DoreyRB
\lref\DoreyRB{
  N.~Dorey, D.~Tong,
  ``Mirror symmetry and toric geometry in three-dimensional gauge theories,''
JHEP {\bf 0005}, 018 (2000).
[hep-th/9911094].
%%CITATION = hep-th/9911094%%
}
%\DiaconescuCN
\lref\DiaconescuCN{
  D.~-E.~Diaconescu, R.~Entin,
 ``Calabi--Yau spaces and five-dimensional field theories with exceptional gauge symmetry,''
Nucl.\ Phys.\  {\bf B538}, 451-484 (1999).
[hep-th/9807170].
%%CITATION = hep-th/9807170%%
}
%\HaackJZ
\lref\HaackJZ{
  M.~Haack and J.~Louis,
  ``M-theory compactified on Calabi--Yau fourfolds with background flux,''
  Phys.\ Lett.\  B {\bf 507}, 296 (2001)
  [arXiv:hep-th/0103068].
  %%CITATION = PHLTA,B507,296;%%
}
%\KatzTH
\lref\KatzTH{
  S.~H.~Katz and C.~Vafa,
  ``Geometric engineering of N=1 quantum field theories,''
  Nucl.\ Phys.\  B {\bf 497}, 196 (1997)
  [arXiv:hep-th/9611090].
  %%CITATION = NUPHA,B497,196;%%
}
%\AffleckAS
\lref\AffleckAS{
  I.~Affleck, J.~A.~Harvey and E.~Witten,
  ``Instantons and (Super)Symmetry Breaking in (2+1)-Dimensions,''
Nucl.\ Phys.\ B {\bf 206}, 413 (1982)..
%%CITATION = PRINT-82-0478 (PRINCETON)%%
}
%\BatyrevHM
\lref\BatyrevHM{
  V.~V.~Batyrev,
  ``Dual polyhedra and mirror symmetry for Calabi--Yau hypersurfaces in toric varieties,''
  J.\ Alg.\ Geom.\  {\bf 3}, 493 (1994)
[arXiv:alg-geom/9310003].
}
%\BelovZE
\lref\BelovZE{
  D.~Belov, G.~W.~Moore,
  ``Classification of Abelian spin Chern-Simons theories,''
[arXiv:hep-th/0505235].
%%CITATION = hep-th/0505235%%
}
%\AspinwallNU
\lref\AspinwallNU{
  P.~S.~Aspinwall, B.~R.~Greene and D.~R.~Morrison,
  ``Calabi-Yau moduli space, mirror manifolds and space-time topology change in string theory,''
Nucl.\ Phys.\ B {\bf 416}, 414 (1994).
[arXiv:hep-th/9309097].
%%CITATION = hep-th/9309097%%
}
%\GreeneHU
\lref\GreeneHU{
  B.~R.~Greene, D.~R.~Morrison and A.~Strominger,
  ``Black hole condensation and the unification of string vacua,''
  Nucl.\ Phys.\  B {\bf 451}, 109 (1995)
  [arXiv:hep-th/9504145].
  %%CITATION = NUPHA,B451,109;%%
}
%\KapustinHK
\lref\KapustinHK{
  A.~Kapustin, N.~Saulina,
  ``Topological boundary conditions in abelian Chern-Simons theory,''
Nucl.\ Phys.\  {\bf B845}, 393-435 (2011).
[arXiv:1008.0654 [hep-th]].
%%CITATION = arXiv:1008.0654%%
}
%\KlemmBJ
\lref\KlemmBJ{
  A.~Klemm, W.~Lerche, P.~Mayr, C.~Vafa and N.~P.~Warner,
  ``Self-Dual strings and N=2 supersymmetric field theory,''
  Nucl.\ Phys.\  B {\bf 477}, 746 (1996)
  [arXiv:hep-th/9604034].
  %%CITATION = NUPHA,B477,746;%%
}
%\IntriligatorEX
\lref\IntriligatorEX{
  K.~A.~Intriligator and N.~Seiberg,
  ``Mirror symmetry in three-dimensional gauge theories,''
Phys.\ Lett.\ B {\bf 387}, 513 (1996).
[hep-th/9607207].
%%CITATION = hep-th/9607207%%
}
%\GreeneTX
\lref\GreeneTX{
  B.~R.~Greene and Y.~Kanter,
  ``Small volumes in compactified string theory,''
  Nucl.\ Phys.\  B {\bf 497}, 127 (1997)
  [arXiv:hep-th/9612181].
  %%CITATION = NUPHA,B497,127;%%
}
%\deBoerKR
\lref\deBoerKR{
  J.~de Boer, K.~Hori, Y.~Oz,
  ``Dynamics of N=2 supersymmetric gauge theories in three-dimensions,''
Nucl.\ Phys.\  {\bf B500}, 163-191 (1997).
[arXiv:hep-th/9703100].
%%CITATION = hep-th/9703100%%
}
%\DunneQY
\lref\Dunne{
  G.~V.~Dunne,
  ``Aspects of Chern-Simons theory,''
[hep-th/9902115].
%%CITATION = hep-th/9902115%%
}
%\DiaconescuUA
\lref\DiaconescuUA{
  D.~-E.~Diaconescu and S.~Gukov,
  ``Three-dimensional N=2 gauge theories and degenerations of Calabi-Yau four folds,''
Nucl.\ Phys.\ B {\bf 535}, 171 (1998).
[hep-th/9804059].
%%CITATION = hep-th/9804059%%
}
%\CachazoPR
\lref\CachazoPR{
  F.~Cachazo and C.~Vafa,
  ``N=1 and N=2 geometry from fluxes,''
[arXiv:hep-th/0206017].
%%CITATION = hep-th/0206017%%
}
%\EguchiFM
\lref\EguchiFM{
  T.~Eguchi, N.~P.~Warner, S.~-K.~Yang,
  ``ADE singularities and coset models,''
  Nucl.\ Phys.\  {\bf B607}, 3-37 (2001).
  [arXiv:hep-th/0105194].
%%CITATION = hep-th/0105194%%
}
%\GukovYA
\lref\GukovYA{
  S.~Gukov, C.~Vafa and E.~Witten,
  ``CFT's from Calabi--Yau four-folds,''
  Nucl.\ Phys.\  B {\bf 584}, 69 (2000)
  [Erratum-ibid.\  B {\bf 608}, 477 (2001)]
  [arXiv:hep-th/9906070].
  %%CITATION = NUPHA,B584,69;%%
}
%\GreeneDH
\lref\GreeneDH{
  B.~R.~Greene, D.~R.~Morrison and C.~Vafa,
  ``A geometric realization of confinement,''
  Nucl.\ Phys.\  B {\bf 481}, 513 (1996)
  [arXiv:hep-th/9608039].
  %%CITATION = NUPHA,B481,513;%%
}
%\GrimmFX
\lref\GrimmFX{
  T.~W.~Grimm and H.~Hayashi,
  ``F-theory fluxes, Chirality and Chern-Simons theories,''
  JHEP {\bf 1203}, 027 (2012).
  [arXiv:1111.1232 [hep-th]].
%%CITATION = arXiv:1111.1232%%
}
%\IntriligatorPQ
\lref\IntriligatorPQ{
  K.~A.~Intriligator, D.~R.~Morrison and N.~Seiberg,
  ``Five-dimensional supersymmetric gauge theories and degenerations of
  Calabi--Yau spaces,''
  Nucl.\ Phys.\  B {\bf 497}, 56 (1997)
  [arXiv:hep-th/9702198].
  %%CITATION = NUPHA,B497,56;%%
}
%\KatzHT
\lref\KatzHT{
  S.~H.~Katz, D.~R.~Morrison and M.~Ronen Plesser,
  ``Enhanced gauge symmetry in type II string theory,''
  Nucl.\ Phys.\  B {\bf 477}, 105 (1996)
  [arXiv:hep-th/9601108].
  %%CITATION = NUPHA,B477,105;%%
}
%\KatzXE
\lref\KatzXE{
  S.~H.~Katz and C.~Vafa,
  ``Matter from geometry,''
Nucl.\ Phys.\ B {\bf 497}, 146 (1997).
[arXiv:hep-th/9606086].
%%CITATION = hep-th/9606086%%
}
%\KatzFH
\lref\KatzFH{
  S.~H.~Katz, A.~Klemm and C.~Vafa,
  ``Geometric engineering of quantum field theories,''
Nucl.\ Phys.\ B {\bf 497}, 173 (1997).
[arXiv:hep-th/9609239].
%%CITATION = hep-th/9609239%%
}
%\KatzEQ
\lref\KatzEQ{
  S.~Katz, P.~Mayr and C.~Vafa,
  ``Mirror symmetry and exact solution of 4-D N=2 gauge theories: 1.,''
Adv.\ Theor.\ Math.\ Phys.\  {\bf 1}, 53 (1998).
[arXiv:hep-th/9706110].
%%CITATION = hep-th/9706110%%
}
%\KlemmKV
\lref\KlemmKV{
  A.~Klemm and P.~Mayr,
  ``Strong coupling singularities and non-abelian gauge symmetries in $N=2$
  String Theory,''
  Nucl.\ Phys.\  B {\bf 469}, 37 (1996)
  [arXiv:hep-th/9601014].
  %%CITATION = NUPHA,B469,37;%%
}
%\AtiyahQF
\lref\AtiyahQF{
  M.~Atiyah, E.~Witten,
  ``M theory dynamics on a manifold of G(2) holonomy,''
Adv.\ Theor.\ Math.\ Phys.\  {\bf 6}, 1-106 (2003).
[hep-th/0107177].
%%CITATION = hep-th/0107177%%
}
%\DiaconescuCN
\lref\DiaconescuCN{
  D.~-E.~Diaconescu and R.~Entin,
  ``Calabi--Yau spaces and five-dimensional field theories with exceptional gauge symmetry,''
  Nucl.\ Phys.\ B {\bf 538}, 451 (1999)
  [arXiv:hep-th/9807170].
  %%CITATION = HEP-TH/9807170;%%
}
%\MorrisonXF
\lref\MorrisonXF{
  D.~R.~Morrison and N.~Seiberg,
  ``Extremal transitions and five-dimensional supersymmetric field  theories,''
  Nucl.\ Phys.\  B {\bf 483}, 229 (1997)
  [arXiv:hep-th/9609070].
  %%CITATION = NUPHA,B483,229;%%
}
%\RedlichDV
\lref\RedlichDV{
  A.~N.~Redlich,
  ``Parity Violation And Gauge Noninvariance Of The Effective Gauge Field
  Action In Three-Dimensions,''
  Phys.\ Rev.\  D {\bf 29}, 2366 (1984).
  %%CITATION = PHRVA,D29,2366;%%
}
%\SeibergBD
\lref\SeibergBD{
  N.~Seiberg,
  ``Five dimensional SUSY field theories, non-trivial fixed points and  string
  dynamics,''
  Phys.\ Lett.\  B {\bf 388}, 753 (1996)
  [arXiv:hep-th/9608111].
  %%CITATION = PHLTA,B388,753;%%
}
%\SeibergRS
\lref\SeibergRS{
  N.~Seiberg and E.~Witten,
  ``Electric - magnetic duality, monopole condensation, and confinement in N=2 supersymmetric Yang-Mills theory,''
Nucl.\ Phys.\ B {\bf 426}, 19 (1994), [Erratum-ibid.\ B {\bf 430}, 485 (1994)].
[arXiv:hep-th/9407087].
%%CITATION = hep-th/9407087%%
}
%\SethiES
\lref\SethiES{
  S.~Sethi, C.~Vafa and E.~Witten,
  ``Constraints on low-dimensional string compactifications,''
  Nucl.\ Phys.\  B {\bf 480}, 213 (1996)
  [arXiv:hep-th/9606122].
  %%CITATION = NUPHA,B480,213;%%
}
%\StromingerCZ
\lref\StromingerCZ{
  A.~Strominger,
  ``Massless black holes and conifolds in string theory,''
  Nucl.\ Phys.\  B {\bf 451}, 96 (1995)
  [arXiv:hep-th/9504090].
  %%CITATION = NUPHA,B451,96;%%
}
%\TongKY
\lref\TongKY{
  D.~Tong,
  ``Dynamics of N=2 supersymmetric Chern-Simons theories,''
JHEP {\bf 0007}, 019 (2000).
[hep-th/0005186].
%%CITATION = hep-th/0005186%%
}
%\WittenMD
\lref\WittenMD{
  E.~Witten,
  ``On flux quantization in M-theory and the effective action,''
  J.\ Geom.\ Phys.\  {\bf 22}, 1 (1997)
  [arXiv:hep-th/9609122].
  %%CITATION = JGPHE,22,1;%%
}
%\WittenQB
\lref\WittenQB{
  E.~Witten,
  ``Phase transitions In M-theory and F-theory,''
  Nucl.\ Phys.\  B {\bf 471}, 195 (1996)
  [arXiv:hep-th/9603150].
  %%CITATION = NUPHA,B471,195;%%
}
%\WittenYC
\lref\WittenYC{
  E.~Witten,
  ``Phases of N = 2 theories in two dimensions,''
  Nucl.\ Phys.\  B {\bf 403}, 159 (1993)
  [arXiv:hep-th/9301042].

  %%CITATION = NUPHA,B403,159;%%
}
%\WittenZE
\lref\WittenZE{
  E.~Witten,
  ``Topological Quantum Field Theory,''
  Commun.\ Math.\ Phys.\  {\bf 117}, 353 (1988).
  %%CITATION = CMPHA,117,353;%%
}
\lref\CheegerSimons{
  J.~Cheeger and J.~Simons,
  ``Differential characters and geometric invariants,''
  Lecture Notes in Math. 
%(Geometry and Topology) 
{\bf 1167}, 50-80 (1985).
}
\lref\HopkinsSinger{
  M.~J.~Hopkins and I.~M.~Singer,
  ``Quadratic functions in geometry, topology, and M-theory,''
  J.\ Diff.\ Geom.\ {\bf 70}, 329 (2005)
  [arXiv:math/0211216 [math.AT]].
}
%\MooreJV
\lref\MooreJV{
G.~W.~Moore,
``Anomalies, Gauss laws, and Page charges in M-theory,''
Comptes Rendus Physique\ {\bf 6}, 251  (2005).
[arXiv:hep-th/0409158].
%%CITATION = hep-th/0409158%%
}
%\VafaWI
\lref\VafaWI{
  C.~Vafa,
  ``Superstrings and topological strings at large N,''
J.\ Math.\ Phys.\  {\bf 42}, 2798 (2001).
[arXiv:hep-th/0008142].
%%CITATION = hep-th/0008142%%
}
%\CachazoJY
\lref\CachazoJY{
  F.~Cachazo, K.~A.~Intriligator and C.~Vafa,
  ``A Large N duality via a geometric transition,''
Nucl.\ Phys.\ B {\bf 603}, 3 (2001).
[arXiv:hep-th/0103067].
%%CITATION = hep-th/0103067%%
}
%\HoriZJ
\lref\HoriZJ{
  K.~Hori, H.~Ooguri and C.~Vafa,
  ``NonAbelian conifold transitions and N=4 dualities in three-dimensions,''
Nucl.\ Phys.\ B {\bf 504}, 147 (1997).
[arXiv:hep-th/9705220].
%%CITATION = hep-th/9705220%%
}%
%%%%%%%% Books %%%%%%%%%%
%
\lref\CoxVI{
D.~A.~Cox and S.~Katz,
``Mirror Symmetry and Algebraic Geometry,''
Mathematical Surveys and Monographs, Vol. 68,
American Mathematical Society, Providence, (2000).
}
\lref\GreenMN{
  M.~B.~Green, J.~H.~Schwarz and E.~Witten,
  ``Superstring Theory. Vol. 2: Loop Amplitudes, Anomalies And Phenomenology,''
  Cambridge Monographs On Mathematical Physics,
  Cambridge University Press, (1987).
}
%\MilnorAA
\lref\MilnorAA{
J.~Milnor, ``Singular Points of Complex Hypersurfaces,'' 
Princeton University Press, (1968).
}
%\GriffithsPAG
\lref\GriffithsPAG{
  P.~Griffiths and J.~Harris,
  ``Principles of Algebraic Geometry,''
  John Wiley \& Sons, (1978).
}
%\WessCP
\lref\WessCP{
  J.~Wess and J.~Bagger,
  ``Supersymmetry and supergravity,''
  Princeton University Press, (1992).
}
%%%%

%%%%%%%%%%%%%%%%%%%%%%
% end references
%%%%%%%%%%%%%%%%%%%%%%

%%%%%%%%%%%%%%% Title Page %%%%%%%%%%%%%
\vskip-1in
\Title{\vbox{
\hbox{\tt BONN-TH-2012-06}
\hbox{\tt LMU-ASC 20/12}
\hbox{\tt UCSB Math 2012-15}
\hbox{\tt UCSD-PTH-11-16}
\vskip-0.5in 
}}
{\vbox{
\centerline{Conifold Transitions in M-theory on Calabi--Yau Fourfolds}\vskip0.2cm
\centerline{with Background Fluxes}
\vskip -0.3in
}}
\centerline{\bf {Kenneth Intriligator${}^{1}$, Hans Jockers${}^{2}$,  Peter Mayr${}^{3}$, David R. Morrison${}^{4}$, M. Ronen Plesser${}^{5}$}}
\bigskip
\bigskip
\centerline{${}^1${\it Department of Physics, University of California, San Diego, La Jolla, CA 92093, USA}}
\vskip0.5ex
\centerline{${}^2${\it Bethe Center for Theoretical Physics, Physikalisches Institut Universit\"at Bonn, 53115 Bonn, Germany}}
\vskip0.5ex
\centerline{${}^3${\it Arnold Sommerfeld Center for Theoretical Physics, Ludwig-Maximilians-Universit\"at, 80333 Munich, Germany}}
\vskip0.5ex
\centerline{${}^4${\it Departments of Mathematics and Physics, University of California, Santa Barbara, CA 93106, USA}}
\vskip0.5ex
\centerline{${}^5${\it Center for Geometry and Theoretical Physics, Box 90318, Duke University, Durham, NC 27708, USA}}
\bigskip
\vskip 0.25in
\centerline{\bf Abstract}
We consider topology changing transitions for M-theory compactifications on Calabi--Yau fourfolds with background $G$-flux. The local geometry of the transition is generically a genus $g$ curve of conifold singularities, which engineers a 3d gauge theory with four supercharges, near the intersection of Coulomb and Higgs branches. We identify a set of canonical, minimal flux quanta which solve the local quantization condition on $G$ for a given geometry, including new solutions in which the flux is neither of horizontal nor vertical type. A local analysis of the flux superpotential shows that the potential has flat directions for a subset of these fluxes and the topologically different phases can be dynamically connected. For special geometries and background configurations, the local transitions extend to extremal transitions between global fourfold compactifications with flux. By a circle decompactification the M-theory analysis identifies consistent flux configurations in four-dimensional F-theory compactifications and flat directions in the deformation space of branes with bundles.
\bigskip
\Date{\sl {March 2012}}
\vfill\eject

%%%%%%%%%%%%%%%%
%%% Table of Contents %%%
\listtoc
%\writetoc
\vfill\eject 
%%%%%%%%%%%%%%%%
%%%%%%%%%%%%%%%%

%%%%%%%%%%%%%%%%%%%%%%%%%%%%%%%%%%
\newsec{Introduction}\seclab\secIntro
%%%%%%%%%%%%%%%%%%%%%%%%%%%%%%%%%% 
As seen in many examples over the years, there is an intriguing interplay between the geometry of 
 string theory compactifications 
(in the presence of branes, or near singularities)
and the dynamics of supersymmetric quantum field theories in various dimensions.  The geometry of Calabi--Yau  manifolds and their moduli spaces can determine the {\it non-perturbative} vacuum manifold and the spectrum of BPS particles of a field theory. Extremal transitions in the geometry can map to transitions between different branches of low-energy supersymmetric gauge theories. Examples with 8 supercharges include transition between Higgs and Coulomb branches of abelian \refs{\StromingerCZ,\GreeneHU,\GreeneDH} and non-abelian \refs{\KatzHT,\KlemmKV,\BerglundUY} four-dimensional $N=2$ gauge theories, three-dimensional $N=4$ theories \refs{\HoriZJ} and five-dimensional $N=1$ theories \refs{\WittenQB,\MorrisonXF\IntriligatorPQ,\DiaconescuCN}.  For theories with four supercharges, i.e., $N=1$ in four or $N=2$ in three dimensions, an essential new ingredient is needed to match between deformations of a Calabi--Yau geometry and field theory dynamics: a choice of background fluxes or background branes in needed on  top of the geometry, which induces an $N=1$ superpotential in the field theory. An  example is the fluxed conifold transition in a Calabi--Yau threefold, leading to confining glueball superpotentials in the associated non-abelian gauge theories \refs{\VafaWI,\CachazoJY}.

In this paper, we study extremal transitions in M- and F-theory compactifications on Calabi--Yau fourfolds, whose low energy theories are described by a certain class of supersymmetric theories with four supercharges. The three-dimensional $N=2$ theories have again Coulomb and Higgs branches meeting at singular points of the moduli space, as discussed e.g. in ref.~\AharonyBX. This turns out to have a nice parallel description in the fourfold geometry, where a set of four-cycles shrinks and another set of four-cycles grows, e.g.:
\eqn\dict{\eqalign{\hbox{flop transition:} \quad S^\sharp _1\to S^\sharp _2\qquad &\simeq\qquad \hbox{Coulomb}_1\to \hbox{Coulomb}_2\cr 
\hbox{conifold transition:} \quad S^\sharp \to S^\flat \qquad &\simeq\qquad \hbox{Coulomb}\ \to \hbox{Higgs} }}
Here $S^\sharp$ denotes an algebraic four-cycle, whereas $S^\flat$ has a generically non-zero volume with respect to the holomorphic $(4,0)$-form $\Omega$. The spectrum of additional light BPS particles near the transition locus arises on the $S^\sharp$ side from M2 branes wrapping small 2-cycles in the local geometry. It can eventually be computed in the topologically twisted theory associated with M-theory compactification on the local geometry by counting the number of holomorphic sections of certain line bundles, similarly as in ref.~\KatzHT.

As mentioned above, a key element for four supercharges is the dependence of the spectrum and the superpotential obstruction on background flux. In an M-theory compactification on the fourfold one has to specify on top of the fourfold geometry $X$ the background four-form flux $G$, which induces the geometric GVW superpotential \GukovYA. Interestingly enough, a quantization condition for $G$ enforces a non-zero flux on a four-cycle $S\in H_4(X,\IZ)$ -- and a non-zero superpotential -- if the second Chern class $c_2(X)$ evaluated on a $S$ is not even \WittenMD. The {\it local} aspects of this quantization condition have not been studied systematically so far. We identify a set of canonical, minimal flux quanta which solve the quantization condition for a given four-cycle geometry.

As a consequence of the flux superpotential one expects that topology changing transitions between Calabi--Yau fourfolds will  generally
be obstructed, except 
when a judicious choice of flux quanta has been made that solves 
the quantization condition. If a four-cycle $S$ with minimal flux is affected by a transition with a 3d field theory interpretation, these flux quanta should correspond to choices in the field theory and the superpotential obstruction should be matched by the 3d spectrum and vacuum structure. Both expectations turn out to be true via a beautiful correspondence between sections of certain line bundles in the geometry and meson operators in the field theory.

If $X$ is elliptically fibered and a four-dimensional F-theory limit exists, the $G$-flux is replaced by, amongst others, gauge flux on 7-branes wrapping algebraic four-cycles $D$ in $X$. In this case one can resort to the detailed results on the field theory spectra and potentials obtained from a local computation in refs.~\refs{\BeasleyDC,\BeasleyKW,\DonagiCA}. As expected, after a circle compactification these results match with those obtained from M2 branes in M-theory as discussed in this note. The perhaps new point on F-theory obtained here from the M-theory analysis are local solutions to the flux quantization condition and the interplay of spectra, flux quanta and superpotential obstructions near the transition point. These are needed to determine the dynamics on Higgs-branches and recombination of 7-branes in a local model. Our solutions have a number of parallels to some recent discussions 
in the literature of flux quanta in F-theory \refs{\MarsanoHV,\BraunZM,\KrauseYH}.

The low-energy gauge theory gauge fields that we consider all arise from the eleven dimensional M-theory 3-form $C$ field on two-cycles, with electrically charged matter from M2 branes wrapped on two-cycles.  Many early works have already explored aspects of connecting M-theory on fourfolds with three-dimensional $N=2$ gauge theory dynamics with M-theory on fourfolds.  To cite one example, ref. \DiaconescuUA\ explored connecting the three-dimensional $N=2$ gauge theory non-perturbative results, including instanton-generated superpotentials \AffleckAS\ and the higher $N_f$ SQED and SQCD results of \AharonyBX, with M-theory realizations from euclidean M5 branes wrapping six-cycles.   Other early works explored aspects of the connection between M-theory background G-flux (which spontaneously breaks parity) and three-dimensional Chern-Simons terms, as well as type II or M-theory realizations of 3d mirror symmetry \IntriligatorEX, and gauge theory moduli space or phase transitions via geometric transitions, see, e.g., refs.~\refs{\GukovYA, \GukovES, \GukovZG}.

In this paper we mostly focus on singularities which, when approached through the moduli space of K\"ahler structures, arise in codimension one.  The light M2-brane states arise from two-cycles in one homology class and their charge singles out a $U(1)$ gauge group.  To decouple the dynamics of these degrees of freedom we work in the limit in which all other cycles are large, including the curve $\Cg$ to which $S^\sharp$ shrinks at the transition.

 Much of the literature on string or M-theory realizations of three-dimensional gauge theories uses spacetime filling D2 or M2 branes at special points in the geometry.  That is not our focus here.  Including $M$ spacetime filling M2 branes introduces additional fields and dynamics in the low-energy three dimensional quantum field theory.  In particular, there are moduli fields for varying the points where the M2 branes are located in the fourfold.  We here focus on small $M$ and, when $M$ is non-zero, on the regions of the moduli space where the M2 branes are not near the fourfold singularity,  and hence they do not participate in or affect the conifold transitions.  In terms of the three dimensional field theory, the degrees of freedom coming from the M2 branes are, in this region of moduli space, decoupled from those that we study, associated with the geometric singularity.     Interesting new degrees of freedom will become light when the M2 branes are near the singularity, and can potentially participate in the conifold transitions; 
we will not discuss that here in detail, leaving it for future work.  We will also not discuss here the interesting large $M$ limit, where the backreaction of the M2 branes on the geometry leads to  M-theory on $AdS_4\times H_7$.  We will see here that, even without including M2 branes in the conifold transition dynamics, there is already  a lot of rich structure.

The organization of this paper is as follows:
In Section 2 we set the stage for the subsequent analysis and collect various aspects of $G$-flux of M-theory on Calabi--Yau fourfold geometries. In Section 3 we set up local Calabi--Yau fourfold geometries so as to model the extremal fourfold transitions of interest. We analyze consistency conditions for $G$-fluxes imposed by the quantization and tadpole cancellation conditions. We examine the structure and the flat directions of the flux-induced potentials. In Section 4 we embed the local fourfold geometries into global Calabi--Yau fourfolds, for which we again examine the behavior of (consistently quantized) $G$-fluxes together with their flux-induced scalar potentials, as we go through extremal M-theory transitions. We discuss our findings both for generic Calabi--Yau fourfolds with a genus $g$ curve of conifold singularities and for explicit Calabi--Yau fourfold examples. In particular we find particular flux configurations with flat directions through the extremal transition. Finally, we comment on the relationship of our M-theory configurations to similar F-theory compactifications. 
In Section 5 we discuss the associated three-dimensional low energy theory, which reproduces the M-theory phase structure obtained by geometric means in the previous sections.

 %%%%%%%%%%%%%%%%%%%%%%%%%%%%%%%%%%
\newsec{Preliminaries: M-theory on Calabi--Yau fourfolds}%
\seclab\secConiTrans
%%%%%%%%%%%%%%%%%%%%%%%%%%%%%%%%%%
%
In a compactification of M-theory on a Calabi--Yau fourfold $X$ to three dimensions, one has to specify the background flux for the four-form $G$. We first collect a few basic facts on $G$ and the superpotential induced by it. 
Most importantly, $G$ is not exactly integral, but satisfies the shifted quantization condition \WittenMD
\eqn\Gflux{ G_{\IZ}={G \over 2\pi} - {c_2(X) \over 2} \in H^4(X,\IZ) \ . } 
This condition ensures {\it locally} that the integral of $G_{\IZ}$ over an arbitrary cycle $S\in H_4(X,\IZ)$ is an integer.
Globally, the shift by $c_2(X)/2$ is related to the integrality of the M2 brane tadpole
\eqn\Acon{ M = {\chi(X) \over 24} - {1\over 2} \int_X {G\over 2\pi} \wedge {G \over 2\pi}  \ . }
Only if $c_2(X)/2$ is an integral class, $\chi(X) \over 24$ is an integer and $G$ can be consistently set to zero by including $M$ space-filling (anti-)M2 branes \refs{\WittenMD,\SethiES}.\foot{To avoid supersymmetry breaking by anti-M2 branes, one needs positive M2 brane charge
$
M= -60-{1\over 2}\int_X G_{\IZ}(G_{\IZ}+c_2(X)) \geq 0\ .
$ 
Self-dual flux components $G$ with $\int G\wedge G >0$ reduce
the number of M2 branes $M$. If $M$ gets negative, or if there are anti-self-dual flux components, supersymmetry is generically broken.}

The conditions for unbroken supersymmetry have been derived in ref.~\BeckerGJ\ and phrased in terms of a superpotential\foot{As we will review and discuss in Section~5\yyy, $W(\Omega)$ is a true superpotential, for Higgs branch fields in $\Omega$, whereas $\widetilde W(J)$ is not a superpotential but rather gives Chern-Simons terms and masses for Coulomb branch moduli in $J$.  Given the different nature of $W(\Omega)$ and $W(J)$, adding then together in $\Wtot$ is perhaps curious; $\Wtot$, though,  can contribute to the central charge of $1/4$-BPS objects.  Moreover, as in \GukovYA, it is tempting to introduce an additional massive field, $\Phi$, with ${\cal W}(\Omega, J, \Phi)$ having various $\vev{\Phi}$ minima that give ${\cal W}(\Omega, J)$ for the various allowed $G$ fluxes.} in ref.~\GukovYA
\eqn\Wpots{
\Wtot=\int \fc{G}{2\pi} \wedge (\Omega+\fc{1}{2}J^2)=W(\Omega)+\widetilde W(J)\ .
}
Here $W(\Omega)$ and $\widetilde W(J)$ are the parts of the integral depending on the holomorphic (4,0)-form $\Omega$ and the K\"ahler form $J$, respectively. Minimization with respect to the complex structure deformations and K\"ahler deformations requires $G$ to be of Hodge type (2,2) and primitive, i.e., $J\wedge G=0$, which in turn implies that $G$ is self-dual, $*G=G$ \GukovYA. 

The superpotential $W(\Omega)$ vanishes if $G$ is dual to an algebraic cycle
and the twisted superpotential $\widetilde W(J)$ vanishes if $G$ is 
Poinar\'e dual to a special Lagrangian cycle.  There is a decomposition
of $H_4(X,\IZ)$ into `vertical' and `horizontal' sublattices with the property
that $W(\Omega)$ always vanishes if $G$ is in the vertical sublattice
and $\widetilde W(J)$ always vanishes if $G$ is in the horizontal sublattice.
(See Appendix~A for a thorough discussion of these sublattices.)
It is therefore tempting to decompose $G$ into these pieces
\eqn\nosplit{{G\over 2\pi}\  = \ {G_V\over 2\pi}+{G_H \over 2\pi}\ , }
and treat the pieces separately.
Unfortunately, this decomposition does not work within the lattice
$H^4(X,\IZ)$ -- rational coefficients must be introduced (except perhaps in exceptional cases).\foot{It has been suggested in ref.~\BeckerAY\ that any missing cycle classes might be provided by Cayley submanifolds calibrated by the form $\Omega_0+\fc{1}{2}J^2$ (see \Wpots), where $\Omega_0$ a fixed representative for the class of $\Omega$. Note that if calibrated cycles do not provide a basis for $H_4(X,\IQ)$, the superpotential~\Wpots\ is only valid on the sublocus of deformation space, where $G$ is dual to a sum of calibrated cycles with rational coefficients, including the special case of a split flux $G=G_V+G_H$.} The generic solution to the quantization condition \Acon\ will therefore arise from a {\it mixed} $G$ flux, where the decomposition \nosplit\ of $G_\IZ$ is not defined over the integers.

As will be studied in some detail below, $G$ fluxes of mixed type also give a new important class of supersymmetric vacua. The standard solution to \Gflux\ is to shift $G$ by a class $\om$ dual to an algebraic cycle, with $\om-c_2(X)/2\in H^4(X,\IZ)$. This flux of a split type generates a twisted superpotential $\widetilde W(J)$ for the K\"ahler moduli and the condition $\widetilde W(J)=0=d\widetilde W(J)$ would exclude supersymmetric vacua for $h^{1,1}(X)=1$ and greatly reduce the number of vacua in general. On the other hand, a mixed flux allows to cancel the $c_2(X)/2$ anomaly {\it locally} on each cycle in $H_4(X,\IZ)$, as required, while at the same time the twisted superpotential can be identically zero and there is no restriction on the K\"ahler moduli at all.

In the following we will study topology changing transitions between two Calabi--Yau fourfolds $X^\sharp$ and $X^\flat$, where a number of algebraic four-cycles $S^\sharp_i$ shrink on the $X^\sharp$ side, and a number of new, generically non-algebraic cycles  $S^\flat_k$ appear on the $X^\flat$ side.\foot{Various aspects of transitions of this type have been previously studied in refs.~\refs{\SethiES,\BrunnerBU,\BraunZM,\KrauseYH}.} As indicated in Appendix~A, all of the Hodge numbers of $X^\sharp$ and $X^\flat$ are determined by three specific ones (which also determine the Euler characteristic $\chi$,
the signature $\sigma$, and the decomposition of $h^{2,2}$ into self-dual and anti-self-dual parts).  The changes are: 
\eqn\hnchange{
{\delta \chi \over 6} ={\delta \sigma \over 2} = \delta h^{1,1}
-\delta h^{1,2}+\delta h^{1,3}\ ,\quad 
\delta h^{2,2}_- = \delta h^{1,1}\ ,\quad 
\delta h^{2,2}_+ = 3\delta h^{1,1}-2\delta h^{1,2}+4\delta h^{1,3} \ .
}
If $\delta \chi$ is not a multiple of 24, integrality of the M2 brane charges $M(X^\flat)$ and $M(X^\sharp)$ requires a jump of flux quanta during the transition. 
 (In fact, if we keep locations of the space-filling M2-branes far away
from the transition, {\it any}\/ change in $\chi$ will require a jump
of flux quanta.) 
A priori it is not obvious whether this prohibits the transition or whether there is a physical effect that causes this jump. As argued below, transitions are possible if there are appropriate new massless states at the transition point that induce a jump of flux by a one-loop effect. 

Since the quantization condition \Acon\ implies a {\it local} constraint on each four-cycle, the non-zero jump of flux must appear on four-cycles that take part in the transition. A simple intersection argument shows that 
if $T^\sharp$ is any four-cycle which transversally intersects a vanishing four-cycle $S^\sharp$, the value of $c_2$ jumps as
\eqn\ctwochang{
\int_{T^\sharp} c_2(X^\sharp)=\int_{T^\flat}c_2(X^\flat)+\delta c_2\ ,\qquad
\delta c_2 = {(T^\sharp.S^\sharp)\over(S^\sharp.S^\sharp)}\int_{S^\sharp} c_2(X^\sharp)\ .
}
Here $T^\flat$ is a cycle which replaces the $T^\sharp$ after the local surgery that describes transition from $X^\sharp$ to $X^\flat$; detailed examples in local and global geometries will be considered in Sections~3\yyy\ and 4.4\yyy. If $\delta c_2$ is odd, quantization will require a non-zero flux on either $T^\sharp$ or on $T^\flat$.

%%%%%%%%
\newsec{Conifold transitions in local Calabi--Yau fourfolds}%
\seclab\secTransLocal
%%%%%%%%
%
After these preliminaries, we turn to a detailed study of the local model for the generic transition, which describes a higher-dimensional analogue of the familiar extremal transitions at isolated conifold points in Calabi--Yau threefolds \refs{\StromingerCZ,\GreeneHU}. The double point of a threefold is given by the equation
\eqn\ConSing{ x_1\,x_2 - x_3\,x_4 \,=\,  0 \ , }
in terms of the complex coordinates $x_\ell$ of $\IC^4$. The fourfold analog of a conifold point arises from fibering this conifold singularity over a genus $g$ curve $\Cg$. To accomplish that, the coordinates $x_\ell$ are taken to be sections of line bundles $\cL_\ell$ over the curve $\Cg$, and the singular local Calabi--Yau fourfold $\Xlsing$ is given by the hypersurface~\ConSing\ in the five-dimensional complex variety $\cL_1 \oplus \cL_2 \oplus \cL_3 \oplus \cL_4 \rightarrow \Cg$. The line bundles are required to obey the relation $\cL_1 \otimes \cL_2 \simeq \cL_3 \otimes \cL_4$ so that $x_1\,x_2 - x_3\,x_4$ in eq.~\ConSing\ is a well-defined section.

Analogously to a conifold singularity in a threefold, the singular Calabi--Yau fourfold  $\Xlsing$ can be smoothed by either a small resolution or by a deformation along the curve $\Cg$ of double points. We denote the resulting Calabi--Yau fourfolds by $\Xlsharpi a$, $a=1,2$, and $\Xlflat$, respectively. As in the case of Calabi--Yau threefolds, the two distinct small resolutions $\Xlsharpi a$ are related by a flop \AspinwallNU, whereas the small resolution $\Xlsharp$ and the deformation $\Xlflat$ are connected by an extremal transition.\foot{We often neglect the index of the small resolution $\Xlsharpi a$, if the distinction between the two resolved phases is not relevant. Also, we refer to local (non-compact) Calabi--Yau fourfolds with a tilde in order to distinguish them from global (compact) Calabi--Yau fourfolds written without a tilde, which will appear later on.}

%%%
\subsec{Small resolution phases $\Xlsharp$}
%%%
%
The  local fourfold $\Xlsharp$ is a fibration of the resolved conifold $\cO(-1)\oplus\cO(-1)\rightarrow \IP^1$ over the genus $g$ base curve $\Cg$, and it contains the compact complex surface $\Ssharp \subset \Xlsharp$, which is a $\IP^1$ fibration over the curve $\Cg$
\eqn\Zfib{\xymatrix{\IP^1  \ar[r] & \Ssharp \ar[d] \cr & \Cg } \ .}
The $\IP^1$-fiber arises from the small resolution of the conifold singularity \ConSing, i.e.,
\eqn\bMat{ \pmatrix{ x_1 & x_4 \cr x_3 & x_2 } \pmatrix{ u \cr v } \,=\, 0 \ . }
Here $[u:v]$ are the homogeneous coordinates of the $\IP^1$-fibers and they transform -- up to an overall tensoring with an arbitrary line bundle -- as sections of $\cL_4$ and $\cL_1$. Therefore, the affine coordinates $z={u\over v}$ and $w={v\over u}$ of the two coordinate patches of the $\IP^1$-fibers are sections of $\cL_4 \otimes \cL_1^{-1}$ and $\cL_1 \otimes \cL_4^{-1}$, respectively.  The line bundle $\cL_1 \otimes \cL_4^{-1}$ has degree $n$, 
\eqn\nDeg{ n \,=\, \deg \cL_1 - \deg \cL_4 \ , }
which is an integral parameter in the geometry that determines the intersection numbers.    

The Picard group of the surface $\Ssharp$ is generated by the divisor classes $F$ of the generic fiber and $C$ of the base curve, given by the zero section $z=0$. They have the intersection numbers\foot{Alternatively, we could have chosen the divisor $C'$ in terms of the zero section $w=0$, such that $F.C'=1$ and $C'.C'=-n'$ with $n'=-n=\deg \cL_4 - \deg \cL_1$. The two divisor classes $C$ and $C'$ are related by $C'=C+nF$ and do not intersect as $C.C'=0$.}

\eqn\DivInt{ F.F \,=\, 0 \ , \qquad F.C\,=\, 1 \ , \qquad C.C\,=\, - n \ . }
The Euler characteristic of the fibration~\Zfib\ is given by
\eqn\SEuler{ \chi(\Ssharp) \,=\, \chi(\IP^1) \, \chi(\Cg) \,=\, 4-4g \ , }
and, with the help of the adjunction formula, we find for the total Chern class of $\Ssharp$
\eqn\SCclass{ c(\Ssharp) \,=\, 1 + \left( 2 [C] - (2g-2-n) [F] \right) + (4-4g) \Svform \ . } 
Here $[C]$ and $[F]$ denotes the $(1,1)$-form in $H^{1,1}(\Ssharp)$ corresponding to the divisor classes $C$ and $F$, while $\Svform$ is the volume form generating $H^4(\Ssharp,\IZ)$. The self-intersection of the surface $S^\sharp$ is 
\eqn\intSelfi{S^\sharp.S^\sharp\,=\,2-2g \ , }
and its K\"ahler volume, measured in terms of the K\"ahler form $J(\Ssharp) = J^F ([C]+ n [F]) + J^C [F]$, is given by 
\eqn\VolS{ {\rm Vol}(\Ssharp) 
   \,=\, {1\over 2} \int_{\Ssharp} J(\Ssharp) \wedge J(\Ssharp)
   \,=\, {n\over 2} (J^F)^2 +J^FJ^C \ . }
The K\"ahler parameters $J^F$ and $J^C$ measure the volume of the $\IP^1$ fiber and the curve $\Cg$.

For the local fourfold geometry $\Xlsharp$ we need to take into account the non-compact normal bundle directions $N\Ssharp \simeq \Xlsharp$ over the surface $\Ssharp$. The normal bundle $N\Ssharp$ restricted to the generic fiber $F$ is the resolved conifold $\cO(-1)\oplus\cO(-1)\rightarrow \IP^1$, whereas the normal bundle restricted to zero section $z=0$ is the bundle $\cL_1\oplus\cL_3$. Hence, the first Chern class of the normal bundle is given by
\eqn\NBsec{ c_1(N\Ssharp) \,=\, - 2 [C'] + (\deg \cL_1 + \deg\cL_3) [F] \,=\, - 2 [C] + (\deg \cL_2 + \deg \cL_4) [F] \ . }
For the small resolution to yield a local smooth Calabi--Yau fourfold, i.e., $c_1(\Xlsharp)=0$, it is required that $c_1(N\Ssharp)=-c_1(\Ssharp)$. Thus with eq.~\nDeg\ we arrive at the Calabi--Yau condition for the small resolution $\Xlsharp$ {}\foot{Note that due to the relation~\ConSing\ we arrive at the same conclusion, if we derive the Calabi--Yau condition with the generators $F$ and $C'$ of the Picard group.}
\eqn\CYcon{ \deg \cL_1 + \deg \cL_2\,=\, \deg \cL_3 + \deg \cL_4 \,=\, \deg K_\Cg \,=\, 2g - 2 \ , }
and the second Chern class of the fourfold is determined to be
\eqn\XlCclass{ 
   c_2(\Xlsharp)\,=\,- (2-2g)\, \Svform \ . 
   }

Analogously to the analyzed small resolution~\bMat, we can carry out the other small resolution described by the blow-up
\eqn\bMatii{ \pmatrix{ x_1 & x_3 \cr x_4 & x_2 } \pmatrix{ s \cr t } \,=\, 0 \ , }
where now the homogeneous coordinates $[ s  : t  ]$ transform as sections of $\cL_3 \otimes \cL_1^{-1}$. Then the twisting of the $\IP^1$-bundle is captured by the integer $\deg \cL_1 - \deg \cL_3$.

The two distinct small resolutions $\Xlsharpi1$ and $\Xlsharpi2$ are related by a flop transition. We can explicitly model this flop transition by describing the conifold fibers of the genus $g$ curve $\Cg$ as a symplectic quotient $V /\!\!/ U(1)$ as in refs.~\refs{\GuilleminAA,\WittenYC}. We relegate the detailed analysis of the flop transition to Appendix~B\yyy. Here we record that the volume integral over the surface $S^\sharp_2$, measured in terms of the K\"ahler coordinates $J_1^F$ and $J_1^C$ of the phase $\Xlsharpi1$, reads
\eqn\VolSiii{
 {1\over2} \int_{S^\sharp_2} J(S^\sharp_1)\wedge J(S^\sharp_1)\,=\,
   -{1\over 2}(n-(2g-2))(J^F_1)^2 - J^F_1 J^C_1 \ . }
This is the negative of \VolS, except for the shift by $(g-1)$ in the first term. As we will see, this shift represents a quantum correction to the twisted superpotential and the Chern-Simons coefficient in the three-dimensional gauge theory, whereas $n$ determines the {\it classical} coefficient. The shift becomes important as we trace the flux-induced twisted superpotential through the flop transition and we return to this aspect in Section~5.

%%%
\subsec{Deformed phase $\Xlflat$}
%%%
We obtain the deformed conifold geometry by deforming the conifold singularity \ConSing\ by 
\eqn\DefConSing{ x\,y - u\,v = \epsilon \ , }
In the context of the local Calabi--Yau fourfold $\Xlflat$, the deformation parameter $\epsilon$ is again a section of the line bundle $\cL_1 \otimes \cL_2 \simeq \cL_3 \otimes \cL_4$, which, according to the Calabi--Yau condition is the canonical line bundle $K_\Cg$. 

The canonical line bundle has $g$ global holomorphic sections and as a consequence contributes $g$ directions to the deformation space ${\rm Def}(\Xlflat)$ of the  Calabi--Yau fourfold $\Xlflat$. Since $\epsilon$ transforms as a section of a line bundle of degree $(2g-2)$, a generic global holomorphic section has $(2g-2)$ isolated zeros along the curve $\Cg$. For a generic deformed conifold fiber -- that is to say for a fiber where the deformation section $\epsilon$ is non-zero -- the singular conifold fiber is replaced by a deformed conifold fiber $T^*S^3$. The $(2g-2)$ fibers, which are located at the zeros of the section $\epsilon$, remain singular conifold fibers. As (generically) these $(2g-2)$ fibers are isolated, the total space of the deformed Calabi--Yau fourfold $\Xlflat$ is smooth, even in the vicinity of singular conifold fibers. As a result, the Euler characteristic of the deformed Calabi--Yau fourfold $\Xlflat$ reads
\eqn\XlflatEuler{ \chi(\Xlflat) \,=\, \chi(S^3) \chi(\widetilde\Cg) + (2g-2) \,=\, 2g - 2 \ , }
where $\widetilde\Cg$ is obtained by removing the vicinities of the curve $\Cg$ where $\epsilon$ becomes zero. 

From the described local fourfold geometry $\Xlflat$, we identify homologically non-trivial four-cycles. We obtain $2g$ four-cycles $A^\flat_n$, $n=1,\ldots, 2g$ of topology $S^1\times S^3$ by transporting (generic) $S^3$-fibers along a non-trivial one-cycle (which avoids the $(2g-2)$ singular points) on the base $\Cg$. By transporting $S^3$-fibers along the path connecting two singular points $p_k$, $k=0,\ldots,(2g-3)$, we arrive at non-trivial four-cycle of topology $S^4$. There are $(2g-3)$ inequivalent such four-cycles $B^\flat_\ell$, which may be constructed by considering the paths $p_0$ -- $p_\ell$, $\ell=1, \ldots, (2g-3)$. Thus we arrive at
\eqn\fourcyclesXlf{ H_4(\Xlflat,\IZ) \,=\, 
  \langle\!\langle A^\flat_1,\ldots,A^\flat_{2g},B^\flat_1, \ldots B^\flat_{2g-3}  \rangle\!\rangle
  \,\simeq\, \IZ^{4g-3} \ .}
This basis of four-cycles is depicted in \lfig\Basisi.

%%%%
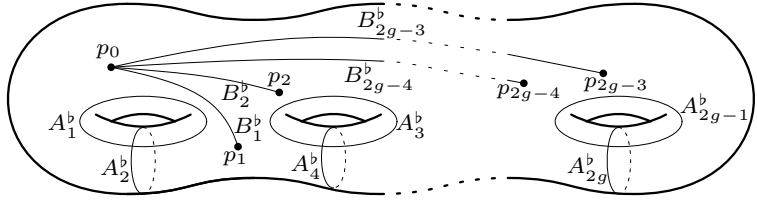
\figboxinsert\Basisi{Depicted are the zeros $p_\ell$ of the deformation section $\epsilon$ on the genus $g$ curve $\Cg$. The depicted paths, supplemented by $S^3$ in the fiber, give rise to the four-cycles $A^\flat_n$ and $B^\flat_\ell$, which furnish a basis of $H_4(\Xlflat,\IZ)$.}{
\DrawDiag{%
% Genus g curve
%%%%%%%%%
\move (40 0) \clvec (-10 0)(-10 60)(40 60)
\move (40 0) \clvec (60 0)(60 5)(80 5)
\move (40 0) \clvec (60 0)(60 5)(80 5)
\move (80 5) \clvec (100 5)(100 0)(120 0)
\move (40 60) \clvec (60 60)(60 55)(80 55)
\move (80 55) \clvec (100 55)(100 60)(120 60)
\lpatt(1 4) % dotted
\move (120 60) \clvec (140 60)(140 55)(160 55)
\move (120 0) \clvec (140 0)(140 5)(160 5)
\lpatt() 
\move (160 55) \clvec (180 55)(180 60)(200 60)
\move (160 5) \clvec (180 5)(180 0)(200 0)
\move (200 0) \clvec (250 0)(250 60)(200 60)
% holes
\move (45 20) \bsegment \move (-15 5) \clvec (-5 0)(5 0)(15 5) \move (-10 3) \clvec (-3 6)(3 6)(10 3) \esegment
\move (105 20) \bsegment \move (-15 5) \clvec (-5 0)(5 0)(15 5) \move (-10 3) \clvec (-3 6)(3 6)(10 3) \esegment
\move (195 20) \bsegment \move (-15 5) \clvec (-5 0)(5 0)(15 5) \move (-10 3) \clvec (-3 6)(3 6)(10 3) \esegment
% A cycles
%%%%%%%
\linewd 0.15
\move (45 23) \lellip rx:20 ry:8 \move (105 23) \lellip rx:20 ry:8 \move (195 23) \lellip rx:20 ry:8
\move (45 21) \clvec (40 21)(40 0)(45 0) \move (105 21) \clvec (100 21)(100 2)(105 2)  \move (195 21) \clvec (190 21)(190 0)(195 0)
\lpatt(1 2)
\move (45 21) \clvec (50 21)(50 0)(45 0) \move (105 21) \clvec (110 21)(110 2)(105 2) \move (195 21) \clvec (200 21)(200 0)(195 0)
\lpatt ()
% B cycles
%%%%%%%
\move (35 40) \clvec (60 35)(70 30)(75 15)
\move (35 40) \clvec (60 38)(70 37)(88 32)
\move (35 40) \clvec (80 50)(100 50)(120 49) \move (35 40) \clvec (80 43)(100 43)(120 42)
\lpatt (1 4) \move (120 49) \clvec (130 49)(140 47)(160 44) \move (120 42) \clvec (130 42)(140 38)(160 36) \lpatt ()
\move (160 44) \lvec (190 38) \move (160 36) \lvec(165 35)
\move (35 40) \fcir f:0 r:1.3
\move (75 15) \fcir f:0 r:1.3
\move (88 32) \fcir f:0 r:1.3
\move (190 38) \fcir f:0 r:1.3
\move (165 35) \fcir f:0 r:1.3
% captions
%%%%%%%
\move (35 45) \htext{${}_{p_0}$}
\move (75 10) \htext{${}^{p_1}$}
\move (89 35) \htext{${}^{p_2}$}
\move (167.5 31) \htext{${}^{p_{2g-4}}$}
\move (195 33.5) \htext{${}^{p_{2g-3}}$}
\move (20.5 22) \htext{${}_{A^\flat_1}$} \move (37 8) \htext{${}_{A^\flat_2}$}
\move (130 22) \htext{${}_{A^\flat_3}$} \move (97 9) \htext{${}_{A^\flat_4}$}
\move (226 27) \htext{${}_{A^\flat_{2g-1}}$} \move (186 8) \htext{${}_{A^\flat_{2g}}$}
\move (79 21) \htext{${}_{B^\flat_1}$} 
\move (75 31.5) \htext{${}_{B^\flat_2}$} 
\move (120 36.5) \htext{${}_{B^\flat_{2g-4}}$}
\move (124 54) \htext{${}_{B^\flat_{2g-3}}$}
}}
%%%%

For the analysis of background fluxes in the fourfold $\Xlflat$, later on we need to derive the intersection numbers for the basis elements \fourcyclesXlf. By construction the four-cycles $A^\flat_n$ do not intersect the four-cycles $B^\flat_\ell$, i.e., $A^\flat_n.B^\flat_\ell=0$. Furthermore, the four-cycles $A^\flat_n$ have vanishing\foot{While the $S^1$ cycles depicted in Fig. 1 clearly have non-zero intersections, the associated four-cycles can be deformed within the $S^3$ fiber so that they do not intersect.}  intersection numbers among themselves, i.e., $A^\flat_n.A^\flat_m=0$. The intersections among the $B^\flat_\ell$ cycles turn out to be 
\eqn\intnumbers{
   B^\flat_n.B^\flat_n\,=\, 2 \ , \qquad B^\flat_n.B^\flat_m\,=\,1 \ (n\ne m) \ , }
which yield the intersection matrix
\eqn\IPairing{
  \cI\,=\,\left( B^\flat_n.B^\flat_m \right) \,=\, \pmatrix{ 2 & 1 & \cdots & 1 & 1 \cr 1 & 2 & \cdots & 1 & 1  \cr
    \vdots & \vdots & \ddots & \vdots & \vdots \cr
    1 & 1 & \cdots & 2 & 1 \cr 1 & 1 & \cdots & 1 & 2 } \  .
   }
These intersections are derived in detail in Appendix~C\yyy{} by carefully examining the structure of the shrinking $S^3$-fibers in the vicinity of the points $p_\ell$. 

Instead of the cycles $B^\flat_\ell$ we can also work with the integral cycles $\hat B^\flat_\ell, \ell=1,\ldots, 2g-3$, which are constructed by considering the paths $p_0$ -- $p_1$, $p_1$ -- $p_2$, $\ldots$, $p_{2g-4}$ -- $p_{2g-3}$ depicted in \lfig\Basisii. Then starting from the intersection matrix~\IPairing\ with respect to the basis $B^\flat_\ell$, it is straightforward to determine the intersection matrix $\hat\cI$ of the basis $\hat B^\flat_\ell$:\foot{We have chosen our conventions such that the intersection matrix $\hat\cI$ gives rise to the Cartan matrix of $SU(N)$. This differs from the conventions used in ref.~\GukovYA, where the four-cycles $\hat B^\flat_\ell$ are oriented in such a way that the off-diagonal entries of $\hat\cI$ become positive.}
\eqn\IntCartan{ \hat\cI \,=\,\left( \hat B^\flat_n. \hat B^\flat_m \right) \,=\,
   \pmatrix{ 
     \-2 & -1 & \-0 & \cdots & \-0 & \-0 \cr 
     -1 & \-2 & -1 & \cdots & \-0 & \-0  \cr
     \-0 & -1 & \-2 & \cdots & \-0 & \-0 \cr
    \-\vdots & \-\vdots & \-\vdots & \ddots & \-\vdots & \-\vdots \cr
    \-0 & \-0 & \-0 & \cdots & \-2 & -1 \cr 
    \-0 & \-0 & \-0 & \cdots & -1 & \-2 } \  .
   }
Note that $\hat\cI$  is just the Cartan matrix of $H=SU(N)$, $N=2g-2$, with the cycles $\hat B^\flat_\ell$ corresponding to the roots of $H$, and this is precisely the homology lattice of the local $A$-type singularity studied in \refs{\GukovYA,\EguchiFM} in connection with 2d Kazama--Suzuki CFTs. The difference here is that the Landau-Ginzburg field lives on the Riemann surface $\Cg$ instead of the complex plane and the paths between the $p_\ell$ define points in the Jacobian of $\Cg$. In the fourfold geometry, periods of $\Omega$ on $\hat B^\flat_n$ are defined up to addition of $A^\flat_m$ periods.
 
%%%%
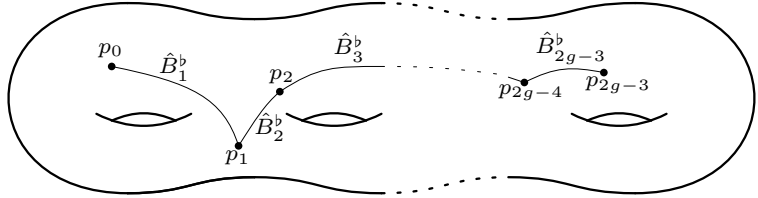
\figboxinsert\Basisii{Depicted are the paths of the group-theoretic basis of cycles $\hat B^\flat_\ell$.}{
\DrawDiag{%
% Genus g curve
%%%%%%%%%
\move (40 0) \clvec (-10 0)(-10 60)(40 60)
\move (40 0) \clvec (60 0)(60 5)(80 5)
\move (40 0) \clvec (60 0)(60 5)(80 5)
\move (80 5) \clvec (100 5)(100 0)(120 0)
\move (40 60) \clvec (60 60)(60 55)(80 55)
\move (80 55) \clvec (100 55)(100 60)(120 60)
\lpatt(1 4) % dotted
\move (120 60) \clvec (140 60)(140 55)(160 55)
\move (120 0) \clvec (140 0)(140 5)(160 5)
\lpatt() 
\move (160 55) \clvec (180 55)(180 60)(200 60)
\move (160 5) \clvec (180 5)(180 0)(200 0)
\move (200 0) \clvec (250 0)(250 60)(200 60)
% holes
\move (45 20) \bsegment \move (-15 5) \clvec (-5 0)(5 0)(15 5) \move (-10 3) \clvec (-3 6)(3 6)(10 3) \esegment
\move (105 20) \bsegment \move (-15 5) \clvec (-5 0)(5 0)(15 5) \move (-10 3) \clvec (-3 6)(3 6)(10 3) \esegment
\move (195 20) \bsegment \move (-15 5) \clvec (-5 0)(5 0)(15 5) \move (-10 3) \clvec (-3 6)(3 6)(10 3) \esegment
% B cycles
%%%%%%%
\linewd 0.15
\move (35 40) \clvec (60 35)(70 30)(75 15)
\clvec (80 23)(83 28)(88 32)
\clvec (100 40)(105 40)(120 40)
\lpatt (1 4) \move (120 40) \clvec (130 40)(140 40)(160 37)
\lpatt () \move (160 36.5) \lvec (165 35)
\clvec (175 40)(180 40)(190 38)
\move (35 40) \fcir f:0 r:1.3
\move (75 15) \fcir f:0 r:1.3
\move (88 32) \fcir f:0 r:1.3
\move (190 38) \fcir f:0 r:1.3
\move (165 35) \fcir f:0 r:1.3
% captions
%%%%%%%
\move (35 45) \htext{${}_{p_0}$}
\move (75 10) \htext{${}^{p_1}$}
\move (89 36) \htext{${}^{p_2}$}
\move (167.5 31) \htext{${}^{p_{2g-4}}$}
\move (195 33.5) \htext{${}^{p_{2g-3}}$}
\move (55 40) \htext{${}_{\hat B^\flat_1}$} 
\move (85 21) \htext{${}_{\hat B^\flat_2}$} 
\move (110 46) \htext{${}_{\hat B^\flat_{3}}$}
\move (180 46) \htext{${}_{\hat B^\flat_{2g-3}}$}
}}
%%%%
\noindent As we will see, depending on the posed geometric question either the basis $B^\flat_\ell$ or the group-theoretical basis $\hat B^\flat_\ell$ will turn out to be more convenient.

%%%
\subsec{Classification of $G$-flux on the local geometries}
%%%
After having described the local geometry of the transition, the next important step is to determine the consistent $G$-fluxes on top of it. Since the conifold transition represents a local surgery operation, the boundary of the local fourfold is the same in all phases
\eqn\Xlbdry{ \partial\widetilde X\,\equiv \, \partial\Xlsharpi a \,=\, \partial\Xlflat \ . }
Similarly the flux on the boundary does not change under a transition and must match throughout the different phases. In a global embedding, the geometry of the boundary $\partial\widetilde X$ and the flux on it will be further restricted by the requirement that one can consistently extend the local data to the global fourfolds $X^\sharp$ and $X^\flat$.

The relevant concepts to determine the flux choices on a non-compact fourfold have been described in \GukovYA\ and we review here the key results. Neglecting the shift in \Gflux\ for the moment, the background flux $G$ is classified by an element of $H^4(\widetilde X,\IZ)$. This group has two parts of different origin,  which are visible in the long exact sequence
\eqn\FluxSequence{ \cdots \longrightarrow 
  H^3(\partial\widetilde X,\IZ) \longrightarrow
  H^4_c(\widetilde X,\IZ) 
  {\buildrel \iota \over \longrightarrow} H^4(\widetilde X,\IZ) 
  \longrightarrow H^4(\partial\widetilde X,\IZ)\longrightarrow \cdots  \ , }
\def\K{G_c}The first part comes from the integral four-form cohomology with compact support $H^4_c(\widetilde X,\IZ)\simeq H^4(\widetilde X,\partial\widetilde X,\IZ)$, which is supported in the interior of $\widetilde X$. This interior part may change due to the local dynamics and will be denoted by $\K$. The second part arises from the homology of the boundary, $H^4(\partial\widetilde X,\IZ)$, cannot be changed by the dynamics in the interior and hence the flux in this part will be fixed under the phase transitions. In particular $H^4(\partial\widetilde X,\IZ)$ may include torsion classes, which capture at infinity flat -- but nevertheless topologically non-trivial -- configurations of the three-form gauge field $C$ \GukovYA. 

\vskip12pt\noindent{\it Resolved phase}\hfil\break
First consider the resolved conifold phase. Here $\partial\widetilde X$ arises as the boundary of the normal bundle $N\Ssharp \simeq \Xlsharp$, which is an $S^3$-bundle fibered over $\Ssharp$. From the Gysin long exact sequence we infer\foot{The torsion piece $\IZ_{2g-2}$ in $H^4(\partial\widetilde X,\IZ)$ arises in the Gysin sequence due to the second Chern class \XlCclass, which is the Euler class of the $S^3$-fibration in $\partial\widetilde X$. See Appendix~A\yyy{} for more information on the (co)homology groups of the local geometries that is used below.}
\eqn\XlbdryH{ H^3(\partial\widetilde X,\IZ)\,\simeq\, \IZ^{2g} \ , \qquad
  H^4(\partial\widetilde X,\IZ)\,\simeq\, \IZ^{2g}\oplus \IZ_{2g-2} \ , }
which implies by Poincar\'e duality
\eqn\XlbdryHom{ H_3(\partial\widetilde X,\IZ) \,\simeq\, \IZ^{2g} \oplus \IZ_{2g-2} \ , \qquad
 H_4(\partial\widetilde X,\IZ) \,\simeq\, \IZ^{2g} \ . } 
The torsion of $H_3(\partial\widetilde X,\IZ)$ is generated by a fiber $S^3_{tor}$ of $\partial\widetilde X$, which is the boundary of the conifold fiber \ConSing, and it obeys in homology $(2g-2)\,S^3_{tor}\,\simeq\,0$.

The long exact sequence \FluxSequence\ embeds the interior fluxes $\K^\sharp$ into the flux background $\GS$ according to 
\eqn\FMapFl{ \iota : \ H^4_c(\Xlsharp,\IZ) \hookrightarrow H^4(\Xlsharp,\IZ)\ , \qquad e^\sharp \mapsto (2g-2)e^{\sharp\,*} \ . }
Here $e^\sharp$ is the generator of $H^4_c(\Xlsharp,\IZ)$, which may be identified with $[\Ssharp]$, whereas the generator $e^{\sharp\, *}$ of $H^4(\Xlsharp,\IZ)$ may be identified with the volume form $\Svform$. It is dual to $e^\sharp$ via the intersection pairing \intSelfi.\foot{Poincar\'e duality for non-compact (complex four-dimensional) manifolds $\widetilde X$ associates $H^q_c(\widetilde X,\IZ)\simeq H_{8-q}(\widetilde X,\IZ)$ to $H^{q}(\widetilde X,\IZ)\simeq H_{q}(\widetilde X,\IZ)$ by (the non-degenerate part of) the intersection pairing $\cI:\ H_q(\widetilde X,\IZ)\otimes H_{8-q}(\widetilde X,\IZ)\rightarrow \IZ$.} Due to the intersection pairing $T^\sharp.S^\sharp=1$ of the (algebraic\foot{We continue to refer to complex submanifolds as ``algebraic'' cycles in the local case, even though -- strictly speaking -- we do not have a tool such as Chow's theorem \Chow\ which guarantees that they are algebraic in the compact case.}) four-cycle $\Ssharp$ with the non-compact (algebraic) four-cycle
\eqn\Tsharp{ T^\sharp \,=\, \pi^{-1}(p) \cap \left\{ x_1=x_3=u=0 \right\} \ , }
where $\pi^{-1}(p)\subset\Xlsharp$ is a resolved conifold fiber \bMat\ over some point $p$ of $\Cg$, the generator $e^{\sharp\, *}$ is dual to the non-compact four-cycle $T^\sharp$. As a consequence, the background flux may be written as\foot{Since the second Chern class \XlCclass\ is even in the local geometry, there is no half-integral shift in the quantization condition.}
\eqn\GSflux{ {\GS \over 2\pi}\,=\, {k^\sharp \over {2g-2}} e^{\sharp} \,=\, \left(k^\sharp\ {\rm mod}\ (2g-2)\right)[T^{\sharp}] + {\K^\sharp\over 2\pi} \ ,
 \qquad k^\sharp \in \IZ \ . }
In the second expression, we have separated the background flux $\GS$ into two contributions with non-compact and compact support. The first term characterizes the topologically non-trivial three-form $C$-field at infinity $\partial\widetilde X$ specified by the torsion $k^\sharp\in\IZ_{2g-2}$, which is given by the intersection $(k^\sharp \, T^\sharp) \cap \partial\widetilde X \simeq k^\sharp \, S^3_{tor}$. The second term is attributed to an interior background flux $\K^\sharp$. Note that a change $k^\sharp\rightarrow k^\sharp \pm (2g-2)$ keeps the torsion class invariant, but changes $\K^\sharp$ in agreement with \FMapFl.

\vskip12pt\noindent{\it Deformed phase}\hfil\break
In the deformed phase $\Xlflat$, the relevant part of the long exact sequence \FluxSequence\ is
\eqn\FSFlat{
  0 \longrightarrow 
  H^3(\partial \widetilde X,\IZ) {\buildrel {\al} \over \longrightarrow} H^4_c(\Xlflat,\IZ)
  {\buildrel \iota \over \longrightarrow} H^4(\Xlflat,\IZ) {\buildrel {\be} \over \longrightarrow}
   H^4(\partial \widetilde X,\IZ) \longrightarrow 0 \ . }
The part $\K^\flat$ of the flux comes from $(2g-3)$ generators in the cokernel of the map $\al$, which are Poincar\'e dual to the four-cycles $B^\flat_\ell$ and span a $(2g-3)$-dimensional integral lattice $ \Ga^\flat$ with intersection form \IPairing\ (or, in the group basis, \IntCartan):
\eqn\Fflattic{{\K^\flat\over 2\pi}\in \Ga^\flat \,=\, H^4_c(\Xlflat,\IZ)/{\al(H^3(\partial \widetilde X,\IZ))} \simeq  
  \langle\!\langle B^\flat_1, \ldots B^\flat_{2g-3}  \rangle\!\rangle \ . }
The background fluxes $G^\flat$ lying in the non-compact part of $H^4(\Xlflat,\IZ)$ can be further divided into two parts, depending on whether they are mapped under $\be$ onto a {\it non-torsion} four-form in $H^4(\partial\widetilde X,\IZ)$ or not. The fluxes $G^\flat_0$ in the second part span the $(2g-3)$-dimensional lattice $\Ga^{\flat\,*}$ dual to $\Gamma^\flat$ in \Fflattic:
\eqn\Fflattic{\Ga^{\flat\,*} \,\simeq\, \IZ^{2g-3}\,\subset\,H^4(\Xlflat,\IZ)  \ ,\qquad {G_0^\flat \over 2\pi} \in \Ga^{\flat\,*}  }
The torsion class in $H^4(\partial\widetilde X,\IZ)$ reflects again the fact that the lattice $\Ga^\flat$ may be viewed as a sublattice of its dual lattice $\Ga^{\flat\,*}$ of index $(2g-2)$, i.e.,
\eqn\TorLattice{ \Ga^{\flat\,*} / \Ga^\flat \,\simeq\,\IZ_{2g-2} \ . }
Background fluxes that map under $\beta$ onto the non-torsion four-forms on the boundary $\partial\widetilde X$ and that are orthogonal to $\Ga^\flat_0$ are denoted as $G^\flat_\perp$. Then the background fluxes $\GF$ can be written as
\eqn\GFflux{ 
  {\GF \over 2\pi} \,=\, {G^\flat_\perp \over 2\pi} + {G^\flat_0 \over 2\pi} \ , \qquad 
  {G^\flat_0 \over 2\pi} \,=\, \sum_{\ell=1}^{2g-3} \bq\ell\, e^{\flat\,*}_\ell \ \,=\, \sum_{\ell=1}^{2g-3} \lambda_\ell^\flat\, \hat e^{\flat\,*}_\ell \ , }
where $e^{\flat\,*}_\ell$ are the lattice generators of $\Ga^{\flat\, *}$ dual to the generators $e^\flat_\ell$ of $\Ga^\flat$, i.e., $e^\flat_\ell.e^{\flat\,*}_m=\delta_{\ell m}$ and $e^{\flat\,*}_m.e^{\flat\,*}_n = {\cal I}^{-1}_{mn}$, with $\cal I$ given in \IPairing. Alternatively, one may use the $SU(2g-2)$ basis $\hat e^\flat_\ell$ with $\hat e^{\flat\,*}_m.\hat e^{\flat\,*}_n = \hat{\cal I}^{-1}_{nm}$, as indicated in the second expression.

Note that the term $G_0^\flat$ describes both, the interior part $\K^\flat$ and the torsion classes in terms of  the lattice $\Ga^{\flat\,*}$. The latter is exactly the cohomology lattice of the $A$-type  local singularity in ref.~\GukovYA. A flux decomposition into elements in $\K^\flat$ and the torsion fluxes on $\partial\widetilde X$ corresponds to a decomposition of a lattice vector into root and weight vectors of $SU(2g-2)$, respectively. 

We can again express the lattice generators $e^{\flat\,*}_\ell$ of $\Ga^{\flat\, *}$ in terms of dual non-compact (algebraic) four-cycles. To this end we define the non-compact four-cycles
\eqn\Tflat{ T^\flat_\ell \,=\, \pi^{-1}(p_\ell) \cap \{ x_1 = x_3 = 0 \} \ , \quad \ell=0,\ldots,2g-3 \ , }
with intersection numbers $T^\flat_\ell.B^\flat_k = \delta_{lk}$ and $T^\flat_0.B^\flat_k = -1$ for $\ell,k=1,\ldots,2g-3$.\foot{Note that $B^\flat_\ell=T^\flat_\ell - T^\flat_0$ up to the ambiguity of adding four-cycle classes $A^\flat_n$, which, however, do not affect the intersection numbers. The representation of the cycles $B^\flat_\ell$ as differences of non-compact ``algebraic planes'' $T^\flat_\ell$ has been discussed in detail in ref.~\EguchiFM\ in the context of identifying the chiral fields of the dual Kazama-Suzuki model. In particular, the integral of $\Omega$ over the non-compact algebraic cycles $T^\flat_\ell$ is not zero due to contributions at infinity. }
Due to the shift in the second Chern class \ctwochang, the quantization condition \Gflux\ requires us to put half-integral fluxes on all the non-compact cycles $T^\flat_\ell$. Hence, expressed in terms of the four-cycles $T^\flat_\ell$ the background flux $G^\flat_0$ reads
\eqn\GFfluxii{ {G^\flat_0 \over 2\pi} \,=\, \sum_{\ell=0}^{2g-3} t^\flat_\ell\,[T^\flat_\ell] \ , \qquad t^\flat_\ell \in \IZ + {1\over 2} \ , }
where the half-integral flux quanta $t^\flat_\ell$ are related to the integral flux quanta $b^\flat_\ell$ according to
\eqn\btrel{ \bq\ell\,=\,t^\flat_\ell - t^\flat_0 \ , \qquad \ell=1,\ldots,2g-3 \ . }
The torsion part contains again information about the flat topological non-trivial three-form $C$-field on the boundary $\partial\widetilde X$, classified by the torsion element $k^\flat\in\IZ_{2g-2}$. Since the non-compact four-cycles $T^\flat_\ell$ intersect $\partial\widetilde X$ in the generator $S^3_{tor}$ of the torsion subgroup of $H_3(\partial\widetilde X,\IZ)$, the torsion $k^\flat$ becomes
\eqn\kftor{ k^\flat \,=\, \sum_{\ell=0}^{2g-3} t^\flat_\ell \,=\, \sum_{\ell=1}^{2g-3} \bq\ell + (2g-2) t_0^\flat \,=\,
 \sum_{\ell=1}^{2g-3} \bq\ell + (g-1) \quad  \mod\ (2g-2) \ , }
where we have used in the last step that $t^\flat_0$ is half-integrally quantized. Note that a change in the flux quanta $b^\flat_\ell\longrightarrow{\bq\ell}'$, such that $\sum_\ell \bq\ell = \sum_\ell {\bq\ell}'\ {\rm mod}\ (2g-2)$ affects $\K^\flat$, but not the torsion class $k^\flat$ on the boundary $\partial\widetilde X$.

%%%
\subsec{Non-dynamical flux constraints for the phase transitions}
%%%
A conifold transition between an M-theory compactification on $X^\sharp$ and $X^\flat$ can occur only if the fluxes $G^\sharp$ and $G^\flat$ in the two phases match certain conditions. A universal constraint comes from the flux $\Phi$ at infinity, defined in ref.~\GukovYA\ as:
\eqn\Phiis{\Phi \,=\, M + {1\over 2}  \int {G\over 2\pi} \wedge {G\over 2\pi} -  \int_X X_8(R)\ .}
We must require that this flux is constant through phase transitions among Calabi--Yau fourfolds, i.e., 
\eqn\PhiTrans{ \Phi\,\equiv\, \Phi_1^\sharp \,=\, \Phi_2^\sharp \,=\, \Phi^\flat \ , } 
Explicitly
\eqn\PhiDefs{\eqalign{
  \Phi_a^\sharp \,&=\,
 M^\sharp_a + {1\over 2} \int_{\Xlsharpi a} {\GSi a\over 2\pi} \wedge {\GSi a\over 2\pi} - \int_{\Xlsharpi1} X_8(R_a^\sharp) \ , \quad a=1,2 \ ,\cr
\Phi^\flat \,&=\,
 M^\flat+{1\over 2} \int_{\Xlflat}{\GF\over 2\pi} \wedge {\GF\over 2\pi} - \int_{\Xlflat} X_8(R^\flat) \ , }}
in terms of the M2 brane, flux and curvature contributions in the different phases.

Let us first derive the flux constraints for extremal conifold transitions, i.e., between two local Calabi--Yau fourfolds  $\Xlsharp$ and $\Xlflat$. First we observe that the background fluxes $G^\flat_\perp$ of the deformed phase in \GFflux\ do not have a counterpart in the resolved phases in \GSflux. As a consequence there is no dynamical phase transition between $\Xlsharp$ and $\Xlflat$ in the presence of non-trivial background fluxes $G^\flat_\perp$.
Setting $G^\flat_\perp\equiv 0$, the condition $\Phi^\sharp=\Phi^\flat$ can be written as 
\eqn\extConst{
4(g-1)(M^\flat-M^\sharp)=(g-1)^2-(k^\sharp)^2-2(g-1)\, \bq{} \cdot {\cal I}^{-1}\cdot \bq{} \ ,
}
where $({\cal I}^{-1})_{mn}=\delta_{mn}-(2g-2)^{-1}$ is the inverse of the intersection form \IPairing\ and we used eqs.~\SEuler, \XlflatEuler, \GSflux\ and \GFflux. From \extConst\ it follows, that the torsion classes must match at the common boundary, i.e.,
\eqn\intFluxconsti{
  k^\sharp \,=\, \sum_\ell \bq\ell + (g-1) \,=\, k^\flat \quad \mod\ (2g-2) \ , }
in terms of the torsion class $k^\flat$ of eq.~\kftor. Note that this torsion condition -- derived from the constraint~\PhiTrans\ -- agrees with the requirement to maintain the torsion class at the boundary $\partial\widetilde X$ throughout the extremal transition.
  
The basic transitions with minimal flux are geometric transitions without a change in the number M2 branes, i.e., $M^\sharp=M^\flat$. In this case eq.~\extConst\ simplifies to
\eqn\intFluxconstii{(g-1)^2 - (k^\sharp)^2\,=\, 2(g-1)\, \bq{} \cdot {\cal I}^{-1}\cdot \bq{} \,=\, 2(g-1)\, \lambda^\flat \cdot \hat {\cal I}^{-1}\cdot \lambda^\flat\ \ ,}
where the shift of flux on the l.h.s. comes from the gravitational contribution in \Phiis.
Note that the r.h.s. of eq.~\intFluxconstii\ is always positive, as follows e.g. from the last expression and \IntCartan. As a consequence a transition without M2 brane participation is only possible, if the flux quantum $k^\sharp$ of the resolved phase is in the charge window $-(g-1) \le k^\sharp \le (g-1)$. The solutions to these constraints are given by
\eqn\IntSol{\eqalign{
  0\le k^\sharp \le (g-1) \ , \quad &\bq\ell \in \{-1,0\} \ , \quad 
  k^\sharp - (g-1) \,=\, \sum_\ell \bq\ell \ , \cr
  &\qquad {\rm or} \cr
  -(g-1)\le k^\sharp \le 0 \ , \quad &\bq\ell \in \{0,1\} \ , \quad 
  (g-1)+k^\sharp \,=\, \sum_\ell \bq\ell \ . }}
The fluxes $G^\flat$ and $G^\sharp$ determined by \IntSol, \GSflux, \GFflux\ and $G^\flat_\perp=0$  are then the minimal flux choices on $\widetilde X^\sharp$ and $\widetilde X^\flat$ that allow for a transitions.

Rewriting the solutions \IntSol\ for $\bq\ell$ in the group theory basis ${G^\flat \over 2\pi} =\bq\ell e^{\flat\,*}_\ell =\lambda^\flat_\ell \hat e^{\flat\,*}_\ell$, one observes that the solution vectors $\lambda^\flat_\ell$ are simply the {\it miniscule weights} $\lambda^\flat$ of $SU(2g-2)$. These  have been already determined in refs.~\refs{\GukovYA,\EguchiFM} as the minimal flux choices on an $A$-type singularity which lead to a (Kazama-Suzuki coset) CFT with a mass gap after a circle compactification to 2d. Here we have found that the minimal fluxes (and states) of the $A$-type singularity are at the same time the minimal fluxes (and states) relevant for the fourfold conifold transition, where the group is $SU(2g-2)$ for the conifold fibration over a genus $g$ curve. In this context, the flux constraints \intFluxconsti\ and \intFluxconstii\ amount to partitioning the shifted flux on the $X^\sharp$ side with $(g-1)-|k^\sharp|$ flux quanta into single flux quanta $\bq\ell$ of charge $\pm 1$ in the deformed phase $X^\flat$. \vskip0.5cm

Next we consider the flux constraint for a flop transition between the two resolved conifold phases $\Xlsharpi 1$ and $\Xlsharpi 2$, which  is simpler. From eq.~\PhiTrans\ it already follows that $k_1^\sharp = \pm k_2^\sharp$.  Matching the torsion class at the common boundary $\partial\widetilde X$ through the flop gives
\eqn\FlopTorConstraint{ k_1^\sharp \,=\, -k_2^\sharp \ . }
We summarize the possible geometric phase transitions without the participation of M2 branes in the following phase diagram:
\eqn\PhaseDiag{
  \xymatrix{ 
  \Xlsharpi1 
    \ar@{<->}[rr]_{\rm flop}^{k_1^\sharp\,=\,-k_2^\sharp} 
    \ar@{<->}[rdd]^{\rm extr.\, trans.}_{ 
    {k^\sharp\equiv k_1^\sharp\,{\buildrel \IntSol \over \longleftrightarrow}\,\bq\ell}}&& 
   \Xlsharpi2 \ar@{<->}[ldd]^{ {k^\sharp\equiv -k_2^\sharp\,{\buildrel \IntSol \over \longleftrightarrow}\,\bq\ell}} \cr \cr
   &\Xlflat }
  }
The flux constraints discussed above and displayed in the diagram reflect only the necessary boundary conditions for a transition to exist. In addition, the transitions may be obstructed dynamically by the flux induced scalar potential for the deformations. Naturally, these obstructions cannot solely be phrased in terms of topological data, but depend on the actual values of the deformation 'moduli'. In the next section we therefore examine the flux-induced scalar potentials and exemplify their role in the context of these geometric phase transitions.

%%%
\subsec{M-theory three-form $C$-field and Cheeger-Simons cohomology}
%%%
In the previous section we determined the torsion classes of the $C$-field at the boundary by intersecting the non-compact (algebraic) four-cycles dual to the four-form flux with the boundary $\partial\widetilde X$. In order to characterize in greater detail the $C$-field at the boundary, we need a refined description of the M-theory three-form $C$-field together with its four-form flux $G$. As explained and spelled out in ref.~\MooreJV, in order to get a handle on the $C$-field in the presence of non-trivial background fluxes, we consider the pair $(C,G)$ as an element of Cheeger-Simons cohomology \refs{\CheegerSimons,\HopkinsSinger}
\eqn\ChSiFlux{ \left(g_\IZ, C, {G \over 2\pi} \right) \,\in\, C^4(X,\IZ)\times C^3(X,\IR) \times \Omega^4(X) \ , }
This triple consists of a closed integral four-cocycle $g_\IZ$, a real three cochain $C$ and a closed four-form ${G \over 2\pi}$ such that
\eqn\ChSireli{ dg_\IZ\,=\,0 \ , \quad dG\,=\,0 \ , \quad {G \over 2\pi} - g_\IZ \,=\, dC \ , }
modulo
\eqn\ChSimod{ \left(g_\IZ, C \right) \,\sim\,  \left(g_\IZ + d\Lambda, C-\Lambda-d\rho \right) \ , }
with $(\Lambda,\rho) \in C^3(X,\IZ) \times C^2(X,\IR)$.\foot{This description has to be appropriately adjusted for four-form background fluxes with shifted quantization conditions according to eq.~\Gflux.}

In this triple the integral cocycle $g_\IZ$ contains the topological information of the background flux, while the four-form $G$ is a solution to the M-theory equations of motion, which to leading order is a harmonic four-form. Finally, the real three-cocycle $C$, which corresponds to the expectation value of three-form M-theory gauge field, captures the deviation of the dynamical flux $G$ from the rigid topological integral cocycle flux $g_\IZ$. 

By integrating the four-form ${G\over 2\pi}$ or equivalently the integral four-cocycle $g_\IZ$ over four-cycles, we extract the integral quanta of the background flux, while M2 branes wrapped on three-cycles $\Sigma$ probe the three-form gauge field $C$ in terms of the holonomy phase \MooreJV
\eqn\ChSiPhase{ \phi(C,\Sigma) \,=\, \exp\left( 2\pi i \int_\Sigma C \right) \ . }
Note that this holonomy factor is well-defined, as it is invariant with respect to the transformations \ChSimod.

We now apply the Cheeger-Simons cohomology description to measure the $C$-field at the boundary $\partial\widetilde X$ of the discussed non-compact fourfolds. The four-form flux $\GS$ on the local fourfold $\Xlsharp$ is represented (at leading order) by a ${\bf L}^2$ harmonic form with compact support, as determined by the equations of motion for the real four-form $\GS$, whereas the topological flux $g^\sharp_\IZ$ -- representing the {\it integral} four-cocycle of the flux -- reaches out to the boundary $\partial\widetilde X$. At the boundary the deviation of the dynamical flux $G$ form the topological flux $g_\IZ$ is characterized by the $C$-field, which we analyze by computing the holonomies $\phi(C,\Sigma_a)$ over a set of generators of $H_3(\partial\widetilde X,\IZ)$
\eqn\ChSiBdryi{
   \phi(C,\partial\widetilde X) \,=\, \left(
   \exp\left( 2\pi i \int_{S^3_{tor}} C \right);\
   \exp\left( 2\pi i \int_{\Sigma_1} C \right), \ \ldots , \
   \exp\left( 2\pi i \int_{\Sigma_{2g}} C \right) \right) \ , }
where $S^3_{tor}$ and $\Sigma_n, n=1,\ldots,2g,$ are the generators of the torsion and non-torsion subgroups of the homology group \XlbdryHom, respectively.

For the local fourfold $\Xlsharp$ the holonomy phase factors \ChSiBdryi\ become
\eqn\ChiSiBdryii{ \phi(C,\partial\widetilde X) \,=\, \left( e^{2\pi i k^\sharp\over 2g-2};\, e^{2\pi i \nu_1^\sharp},\, \ldots,\, e^{2\pi i \nu_{2g}^\sharp} \right) \ , }
where the first factor measures the torsion of the $C$-field and the subsequent factors the holonomies $\nu_n^\sharp$ with respect to the non-torsion three-cycles $S^1\times S^2$. Note that the latter phase factors are continuous periodic moduli of M-theory on the local fourfold $\Xlsharp$. 

The flux $G^\flat$ on the fourfold $\Xlflat$ gives rise to similar phases at the boundary $\partial\widetilde X$
\eqn\ChiSiBdryiii{ \phi(C,\partial\widetilde X) \,=\, \left( e^{2\pi i k^\flat\over 2g-2};\, e^{2\pi i \nu_1^\flat},\, \ldots,\, e^{2\pi i \nu_{2g}^\flat} \right) \ , }
which capture again the torsion class $k^\flat$ of the $C$-field and the non-torsion holonomies $\nu_n^\flat$. 

Note that in the context of the local geometry $\Xlflat$ the phase factors $\nu_n^\flat$ may be interpreted as non-trivial four-form fluxes with compact support, which are Poincar\'e dual to the cycles $A^\flat_n$. Such fluxes are trivial in the cohomology with compact support, i.e., they are in the kernel of the map $\al$ in the sequence~\FSFlat. Therefore, the local geometry $\Xlflat$ does not impose a quantization condition on the parameters $\nu_n^\flat$. If, however, we couple to gravity -- by embedding the local geometry $\Xlflat$ into a global compact Calabi--Yau fourfold $\Xf$ -- a quantization condition may be imposed on the phase factors $\nu_n^\flat$. Therefore, by having a particular global compactification $\Xf$ in mind, we may impose even in the local setting $\Xlflat$ a particular quantization condition on the parameters $\nu_n^\flat$. 

Thus, in order to realize an extremal M-theory transition between the local geometries $\Xlflat$ and $\Xlsharp$, in addition to the constraints summarized in \PhaseDiag, we also need to ensure that the non-torsion holonomies match according to:
\eqn\ChiSiMatch{ e^{2\pi i \nu^\sharp_n} \,=\, e^{2\pi i \nu^\flat_n} \ , \qquad n=1,\ldots,2g }
%

%%%
\subsec{Flat directions of the superpotential and Abel-Jacobi map}
%%%
Having established the topological conditions on background fields, which must be fulfilled independently of any further details for a phase transition to exist, we now examine the dynamical conditions, i.e., the unobstructed directions of the scalar potential. Along these directions, the K\"ahler and complex structure moduli adjust such that $d\Wtot=0$ and the harmonic background fluxes are both primitive and of Hodge type $(2,2)$ \refs{\BeckerGJ,\GukovYA}. The conditions on the complex structure and the K\"ahler moduli can be study separately on the two parts $W(\Omega)$ and $\widetilde W(J)$ in eq.~\Wpots.
\vskip0.2cm

\noindent{\it Resolved phase}\hfil\break
In the resolved local fourfold $\Xlsharp$ the possible background fluxes $G^\sharp$ in eq.~\GSflux\ are Poincar\'e dual to (a rational multiple of) the algebraic surface $S^\sharp \subset \Xlsharp$, i.e., $e^\sharp \simeq [S^\sharp]$. As a consequence the superpotential $W(\Xlsharp)$ vanishes identically and there are no constraints on the complex structure. On the other hand, due to eqs.~\Wpots, \VolS\ and \GSflux\ a non-vanishing flux $G^\sharp$ gives rise to a twisted superpotential. Thus, in the phase $\Xlsharp$ we find
\eqn\Wsharp{ \widetilde W({\Xlsharp}) \,=\, {k^\sharp \over 4(g-1)} \left( {n\over 2} (J^F)^2 + J^F J^C \right) \ , \qquad W(\Xlsharp)\,=\, 0 \ . }
and the resolved phases $\Xlsharp$ tend to be dynamically lifted for non-zero flux $G^\sharp$ if $n\neq 0$.

\noindent{\it Deformed phase}\hfil\break
We have already argued that a transition requires $G^\flat_\perp=0$, and, therefore, we concentrate on non-vanishing fluxes of type $G^\flat_0$ as given in eqs.~\GFflux\ and \IntSol. These fluxes are represented by harmonic ${\bf L}^2$ four-forms. 

The flat directions of the flux-induced twisted superpotential $\widetilde W(\Xlflat)$ correspond to primitive ${\bf L}^2$ fluxes $G^\flat_0$, namely
\eqn\Wflattwist{ d\widetilde W(\Xlflat)\,=\, 0 \quad {\rm for} \quad G^\flat_0 \wedge J(\Xlflat) \,=\,0 \ , }
where $J(\Xlflat)$ is the K\"ahler form of the non-compact fourfold $\Xlflat$. The six form $G^\flat_0\wedge J(\Xlflat)$ is again ${\bf L}^2$ harmonic and {\it a priori} needs not to vanish. For compact manifolds the Hodge--DeRham theorem identifies harmonic forms with cohomology groups. However, for non-compact K\"ahler manifolds a similar relationship between ${\bf L}^2$ harmonic forms and cohomology groups is only established in special cases. Therefore -- despite of the vanishing cohomology group $H^6(\Xlflat,\IR)$ -- a non-trivial flux $G^\flat_0$ may still fail to be primitive. 
In the context of explicit global embeddings of $\Xlflat$ into a compact fourfold $\Xf$, we observe that the fluxes $G^\flat_0$ are primitive for the torsion class $k^\flat=0$ and tend to be imprimitive for other torsion classes $k^\flat\ne0$ (c.f., Section~4.1\yyy{}).

Before starting with analyzing the detailed conditions on the complex structure moduli from $W(\Xlflat)$, it is instructive to remember the structure of the result found in refs.~\refs{\CachazoJY,\CachazoPR} for the closely related problem of geometric engineering of $N=1$ supersymmetric four-dimensional Yang-Mills theory on Calabi--Yau threefold with flux. The non-zero flux potential describes a $N=1$ superpotential which drives the theory to loci in the deformation space with extra massless dyon states, which condense in the $N=1$ vacuum \refs{\SeibergRS}. The spectrum of massless states corresponds to a particular factorization of the Seiberg-Witten curve in the parent $N=2$ theory, which supports these states on vanishing cycles. The beautiful interplay between the factorization of the defining equations of the effectively one-dimensional geometry, a complex curve, and the minimization of the quantum superpotential computed by periods integrals is provided by the Abel-Jacobi theorem, which links the zeros of a linear combination of periods to the existence of a certain meromorphic function in the factorized geometry \CachazoPR. 

In the following we find a similar structure for the present minimization problem by reducing the superpotential for the complex structure moduli to Abel-Jacobi integrals on the curve $\Cg$.\foot{With a slight extension necessary to describe the $C$-fields discussed in the previous section.} The Abel-Jacobi theorem then relates the minima of the superpotential to the existence of line bundles with global holomorphic sections, with the latter associated to the massless 3d states that are the building blocks of the meson operators. The existence of these sections amounts to the split of the canonical divisor on $\Cg$ into two residual special divisors in the sense of \GriffithsPAG, and this condition can be written as a factorization condition on the polynomial $\eps$ in \DefConSing\ representing the deformations of $\Xlflat$.

In order to explicitly find the flat directions of the flux-induced superpotential $W(\Xlflat)$, we start by evaluating \Wpots\ in the presence of the background flux $G^\flat_0$
\eqn\FluxPot{ 
  W(\Xlflat)  \,=\, \sum_{\ell=1}^{2g-3} \bq\ell 
            \left( \int_{B_\ell^\flat} \Omega - {1\over 2g-2}\sum_{m=1}^{2g-3} \int_{B_m^\flat} \Omega \right) \ . }
Here we use the relations~\GFflux\ and \Fflattic\ to express the superpotential in terms of the integers $\bq\ell$, restricted by the topological condition \IntSol. 
For ease of notation we focus on the window $0\le k^\flat \le (g-1)$ and label the points $p_0$ through $p_{2g-3}$ in \lfig\Basisi\ such that the flux quanta \IntSol\ are distributed according to $\bq1=\ldots=\bq{(g-2)+k^\flat}=0$ and $\bq{(g-1)+k^\flat}=\ldots=\bq{2g-3}=-1$. Integrating over the $S^3$ fibers we obtain
\eqn\HiggsPotii{
W(\Xlflat)  \,=\, 
  {g-1-k^\flat\over 2g-2} \cdot \sum_{p_\ell\in Z_+} \int_{p_0(z)}^{p_\ell(z)}\!\!\!\!\om(z) 
   -{g-1+k^\flat\over 2g-2} \cdot \sum_{p_\ell \in Z_-} \int_{p_0(z)}^{p_\ell(z)}\!\!\!\!\om(z) \ . }
where $Z_\pm$ denote the two point sets 
\eqn\DefZ{
Z_+=\{ p_0,p_1,...,p_{g-2+k^\flat}\}\ , \qquad 
Z_-=\{p_{g-1+k^\flat},...,p_{2g-3}\} \ ,
}
with $g-1\pm k^\flat$ elements and $Z_+\cup Z_-$ is the divisor of $K_\Cg$, by construction. Moreover, for generic moduli, $\om(z)$ is a holomorphic one-form on the genus $g$ curve $\Cg$, which depends on both the parameters of deforming section $\eps$ of the canonical bundle $K_\Cg$ and the complex structure moduli of the base curve $\Cg$.\foot{Explicit examples will be considered in the next subsection.} Both types of moduli furnish complex structure deformations of the local Calabi--Yau fourfold $\Xlflat$, which we collectively denote by $z$. The line integrals are taken over paths as schematically depicted in \lfig\Basisi. For criticality of this superpotential we arrive at the condition
\eqn\CritCi{ dW(\Xlflat) = 
     \left({d_+ \over \Delta} \! \sum_{p_\ell\in Z_+}\int_{p_0}^{p_\ell}\!\!\!\!\partial_z\om(z) 
          - {d_-\over \Delta} \!\! \sum_{p_\ell\in Z_-} \int_{p_0(z)}^{p_\ell(z)}\!\!\!\!\partial_z\om(z) \right) \!dz = 0 \ , }
where we used that the differential $\om$ vanishes at the points $p_\ell$ and we defined the positive integers
\eqn\Defds{ d_\pm = {g-1\mp k^\flat \over \gcd(g-1-k^\flat,g-1+k^\flat)} \ , \quad 
  \Delta=d_+ + d_-={2g-2 \over  \gcd(g-1-k^\flat,g-1+k^\flat)} \ . }
As the derivatives $\partial_z\om(z)$ for all $z$ generate a basis of holomorphic one-forms $\om_\al, \al=1,\ldots,g$, we can formulate the criticality constraint~\CritCi\ in terms of the map
\eqn\tmumap{ 
  \tilde\mu: \{ p_\ell \} \mapsto \left( \int_{p_0}^{p_\ell} \om_1, \ldots , \int_{p_0}^{p_\ell} \om_{g} \right) \ , }
as 
\eqn\CritCii{ {d_+\over\Delta} \, \tilde\mu(p_0+ \ldots +p_{g-2+k^\flat}) - {d_-\over\Delta} \, \tilde\mu(p_{g-1+k^\flat}+ \ldots + p_{2g-3}) \,=\,0 \ , }
where we used linearity of $\tilde \mu$. The map $\tilde\mu$ is related to the Abel-Jacobi map
\eqn\AJmap{ \mu: \Cg \rightarrow {\rm Jac}(\Cg), \ q \mapsto \left( \int_{p_0}^q \om_1, \ldots , \int_{p_0}^q \om_{g} \right) \ , }
by the commuting diagram
\eqn\Lift{\xymatrix{ \{ p_\ell \} \ar[r]^{\tilde\mu} \ar[rd]^{\mu} & \IC^g \ar[d]^{P} \cr & {\rm Jac}(\Cg)} \  .}
The definition of the map $\tilde\mu$ includes a specific path of integration $p_0$--$p_\ell$, which is determined by the minimal volume condition when integrating over the $S^3$ fibers in eq.~\FluxPot. On the other hand, modding out by integral cycles in $H_1(\Cg,\IZ)$ defines the projection $P$ and one obtains the Abel-Jacobi map $\mu$, which is well-defined for a point $q$ on $\Cg$. Inversely, the lift from $\mu$ to $\tilde\mu$ requires specifying the path $p_0$--$p_\ell$.
 
To include also the $C$-fields, we have to consider in addition a contribution to the superpotential from background fluxes Poincar\'e dual to the four-cycles $A^\flat_n$. These fluxes have compact support and become exact in the cohomology $H^4(\Xlflat)$ according to $\FSFlat$. Nevertheless, they enter the superpotential due to their contributions at the boundary $\partial\widetilde X$. Integrating over the $S^3$ fibers, this flux-induced superpotential becomes
\eqn\WAfluxesii{ W(\Xlflat,\nu^\flat_n) \,=\, \sum_{n=1}^{2g} \nu^\flat_n \oint_{a^\flat_n} \omega(z) \ , }
with a basis of one-cycles $a^\flat_n$, $n=1,\ldots, 2g$, of $H_1(\Cg,\IZ)$. The flux parameters $\nu^\flat_n$ give rise to the (periodic) holonomy phases $e^{2\pi i \nu^\flat_n}$ in the Cheeger-Simons cohomology.\foot{Note that the continuous values of the parameters $\nu^\flat_n$ become only meaningful with respect to an explicitly chosen basis of four-cycles $B^\flat_\ell$ and $A^\flat_n$.} As discussed at the end of section~3.5\yyy, these flux parameters are quantized  in a global embedding of $\Xlflat$. The generalized criticality condition for the combined superpotential from eqs.~\HiggsPotii,\WAfluxesii\ can then be written as
\eqn\CritCii{ {d_+\over\Delta} \, \tilde\mu(p_0+ \ldots +p_{g-2+k^\flat}) - {d_-\over\Delta} \, \tilde\mu(p_{g-1+k^\flat}+ \ldots + p_{2g-3}) + \tilde\mu_{\nu}(\nu_n^\flat)\,=\,0 \ , }
with
\eqn\Exttildemu{ \tilde\mu_\nu: 
\nu_n^\flat \mapsto
  \left(\sum_n\nu_n^\flat \oint_{a^\flat_n} \om_1,\ldots,\sum_n\nu_n^\flat \oint_{a^\flat_n} \om_g\right) \ , } 
which, analogously to \Lift, also projects onto the intermediate Jacobian ${\rm Jac}(\Cg)$. Finally, by linearity of the map $\tilde\mu$, we may rewrite the criticality condition~\CritCii\ into the two equivalent constraints
\eqn\CritCiii{\eqalign{
 \tilde\mu(p_0+ \ldots +p_{g-2+k^\flat}) \,&=\, {d_-\over\Delta} \tilde\mu(p_0+ \ldots + p_{2g-3})-\tilde\mu_\nu(\nu^\flat_n) \ , \cr
 \tilde\mu(p_{g-1+k^\flat}+ \ldots + p_{2g-3}) \,&=\, {d_+\over\Delta} \tilde\mu(p_0+ \ldots + p_{2g-3})+\tilde\mu_\nu(\nu^\flat_n) \ . }}

The projection of the map $\tilde\mu$ onto the Abel-Jacobi map $\mu$ gives a nice geometric interpretation of the supersymmetry conditions~\CritCiii: Firstly, the zero of the Abel-Jacobi map establishes a one-to-one correspondence with holomorphic line bundles. Namely, we assign to the effective divisors appearing as the arguments of the map~$\tilde\mu$ on the left-hand side of the two relations in eq.~\CritCiii\ the line bundles $\cE_\pm$ 
\eqn\DivDev{ 
  \cE_+ \,=\, \cO_\Cg(p_0+ \ldots +p_{g-2+k^\flat}) \ , \qquad 
  \cE_-\,=\, \cO_\Cg(p_{g-1+k^\flat}+ \ldots +p_{2g-3}) \ . }
Then the supersymmetry conditions \CritCiii\ tell us that the two line bundles must be given by
\eqn\CritCiv{ \cE_+ \, \simeq \, K_\Cg^{d_{-}/\Delta} \otimes \cL_0 \ ,  \qquad
 \cE_-\,\simeq\,K_\Cg^{d_{+}/\Delta} \otimes \cL^*_0 \ . }
where $\cL_0$ is the degree zero line bundle associated to the point $-\mu_\nu(\nu^\flat_n)$ of the intermediate Jacobian ${\rm Jac}(\Cg)$. Here the root of the canonical bundle appears because the effective divisor $p_0+\ldots+p_{2g-3}$ corresponds to the canonical bundle $K_\Cg$ of the curve $\Cg$. As a result, we observe that the line bundles $\cE_\pm$ fulfill
\eqn\DivCan{ \cE_+ \otimes \cE_- \simeq K_\Cg \ .}
For later reference, we alternatively write the line bundles $\CritCiv$ as
\eqn\CritCv{ \cE_+ \, \simeq \,K_\Cg^{1/2} \otimes \cL \ , \qquad \cE_-\,\simeq \, K_\Cg^{1/2} \otimes \cL^* \ , }
in terms of the spin structure $K^{1/2}_\Cg$ and the degree-$k^\flat$ line bundle 
\eqn\DefBL{ \cL \,=\, K_{\Cg}^{k^\flat\over {2g-2}} \otimes \cL_0 \ . }
The spin structure $K^{1/2}_\Cg$ and the $(2g-2)$-th root of $K_\Cg$ in eq.~\DefBL\ must be chosen in accord with eqs.~\CritCiii. 

To recapitulate at this point, the flux superpotential $W(\Xlflat)$ is specified by the torsion class $k^\flat$ and the $C$-field parameters $\nu_n^\flat$ and imposes the constraint \CritCiii\ on the complex structure deformation space of ${\rm Def}(\Xlflat)$ parametrized by the global sections $\eps$ of the canonical bundle $K_\Cg$ on the family of curves $\Cg$. Via the above argument, the condition $dW(\Xlflat)=0$ translates into the condition, that the canonical divisor splits into two residual divisors associated to two line bundles $\cE_\pm$ with holomorphic sections $\eps_\pm$. The flat directions in ${\rm Def}(\Xlflat)$ are therefore of the factorized form 
\eqn\DefFactor{ x_1\,x_2 - x_3\,x_4 \,=\, \eps_+\eps_- \quad {\rm with}\quad \eps_\pm \,\in\, H^0(\Cg,\cE_\pm) \ .  }
The above conditions \DivCan\ and \CritCiv\ do not uniquely specify the line bundles $\cE_\pm$, as there is an ambiguity in taking the $\Delta$-th root in eq.~\CritCiv. The different choices of roots of the canonical bundle are distinguished by the $C$-field parameters $\nu_n^\flat$ in eq.~\CritCiii, and these have to be matched as well in an extremal transition. 

As the space of flat directions is parametrized by the deformations of $\eps$ that arise as the product of two global sections $\eps_\pm$ of the line bundles $\cE_\pm$, the dimension of the unobstructed deformation space is 
\eqn\FlatDef{ \dim {\rm Def}(\Xlflat,\cE_+\otimes \cE_-) \,=\, N_+ + N_- -1 \ . }
Here $N_\pm=h^0(\Cg,\cE_\pm)$ is the number of global sections of $\cE_\pm$ and the $-1$ accounts for a rescaling $(\eps_+,\eps_-) \rightarrow (\la \eps_+, \la^{-1} \eps_- )$ with $\la\in \IC^*$. Note that the individual number of global sections $N_\pm$, and thus the dimension of the deformation space \FlatDef, depends on the complex structure of the curve $\Cg$ and the holonomy phase factors $\nu^\flat_n$. Only the index
\eqn\Indexi{ N_+ - N_- \,=\, h^0(\Cg,\cE_+) - h^1(\Cg,\cE_+)\,=\, h^1(\Cg,\cE_-) - h^0(\Cg,\cE_-)\ , }
is a topological invariant.

In summary the non-vanishing superpotential $W(\Xlflat)$ obstructs a generic deformation into the deformed phase $\Xlflat$. The flux $G^\flat_0$ is still consistent with the topological constraints \PhaseDiag\ as we move along non-flat directions of the superpotential $W(\Xlflat)$ where the zero set $Z_+\cup Z_-$ of the canonical bundle does not split into the zero sets of holomorphic sections of the two line bundles $\cE_\pm$. On the other hand, the superpotential $W(\Xlflat)$ is identically zero on the flat directions determined by the factorization condition~\DefFactor. Combining the above results for the superpotential and the twisted superpotential, we can find dynamically unobstructed phase transitions at least for vanishing torsion class $k^\sharp=k^\flat=0$.

%%%
\subsec{Local transitions for special configurations}
%%%
To make some of the previous findings more concrete, we consider now some special configurations. Since the flux superpotential $W(\Xlsharp)$ in eq.~\Wsharp\ generically prevents a phase transition for $k^\sharp\neq 0$ (at least in the local case) we first focus on the most promising case of vanishing background flux $\GS$ in the resolved geometry $\Xlsharp$. This corresponds to the torsion class $k^\sharp=k^\flat=0$ and a flux 
\eqn\FluxForSec{
  {\GS\over 2\pi}\,=\, 0 \ , \qquad 
  {G^\flat_0\over 2\pi}\,=\, {1\over 2} \left( 
  \sum_{\ell:\, p_\ell\in Z_+} [T^\flat_{\ell}] - 
  \sum_{\ell:\, p_\ell\in Z_-} [T^\flat_{\ell}] \right) \ , }
where each set $Z_\pm$ defined in \DefZ\ contains $g-1$ points.
While there is no potential in the resolved phase $\Xlsharp$, the flux-induced superpotential $W(\Xlflat)$ is given by \HiggsPotii\ for $k^\flat=0$ and the flat directions correspond to the split \DivCan\ with $\cE_\pm=K_{\Cg}^{1/2}\otimes \cL^{\pm 1}$.

A particular interesting case, motivated by the existence of a global embedding, is the factorization $\cE_\pm\simeq K_{\Cg}^{1/2}$. The flat directions of the superpotential then correspond to the holomorphic global sections $N_+=N_-$ of a chosen spin structure $K^{1/2}_\Cg$ on $\Cg$. On the genus $g$ curve $\Cg$ there are $2^{2g}$ inequivalent spin structures or, in other words, $2^{2g}$ inequivalent square roots of the canonical bundle $K_\Cg$ \HitchinAA. These are in one-to-one correspondence with the $2^{2g}$ half-integral choices for the phases $\nu_n^\flat$
\eqn\PhaseSpinRel{ \nu^\flat_n \in \big\{ 0,{1\over2} \big\}\ ,\ n=1,\ldots,2g \quad {\buildrel{1:1}\over\longleftrightarrow} \quad {2^{2g}\ {\rm spin\  structures}\ K^{1/2}_\Cg} }
By a classical result due to Riemann (and proved in modern algebraic language by Mumford \MumfordAA) these spin structures (also called ``theta characteristics'') can be divided into two classes: for $2^{g-1}(2^g+1)$ of the spin structures, the dimension of the space of global sections is even, while for the remaining $2^{g-1}(2^g-1)$ spin structures, the dimension of the space of global sections is odd, hence non-vanishing.

As a first example consider the case with a single global section $\varepsilon=\eps_\pm$.\foot{For hyperelliptic Riemann surfaces there is always a spin structure with a single global section, whereas for the generic Riemann surface such a spin structure is conjectured in ref.~\BaerAA.} Then the unobstructed deformation space is one-dimensional with its flat direction parametrized by $\eps=\varepsilon^2$. As the deformation $\eps$ is a square, it yields $(g-1)$ double zeros. For explicitness we label the individual zeros in such a way that the pairs $(p_0,p_1), (p_2,p_3),\ldots,(p_{2g-4},p_{2g-3})$ correspond to the $(g-1)$ double zeros. In this way we naturally define a basis of four-cycles $\hat B^\flat_\ell$ as in \lfig\Basisii. Then the four-cycles $\hat B^\flat_{2k-1}, k=1,\ldots,g-1$ are associated to the vanishing paths $p_{2k-2}$ -- $p_{2k-1}$, and hence are shrunken to zero size. With these conventions and according to eq.~\FluxForSec\ the deformation $\eps=\varepsilon^2$ corresponds to a flat direction for the flux configuration
\eqn\FluxConf{ {G^\flat_0\over 2\pi} \,=\, {1\over 2} \sum_{k=0}^{g-2} \left([T^\flat_{2k+1}] - [T^\flat_{2k}]\right) \,=\,
  \hat e^{\flat\,*}_{1} - \hat e^{\flat\,*}_3 + \ldots +(-1)^g \hat e^{\flat\,*}_{2g-3} = {1\over 2} \sum_{k=0}^{g-2} [\hat B^\flat_{2k+1}] \  }
expressed in terms of dual forms of the non-compact four-cycles $T^\flat_\ell$ or, alternatively, in terms of a (particular choice of) basis $\hat e^{\flat\,*}_\ell$ Poincar\'e dual to the duals of the four-cycles $\hat B^\flat_\ell$. Due to the vanishing of the four-cycles $\hat B^\flat_{2k-1}, k=1,\ldots,g-1$, along the deformation direction $\eps=\varepsilon^2$, the superpotential $W(\Xlflat)= {1\over 2} \sum_{k=0}^{g-2}\int_{\hat B^\flat_{2k+1}}\Omega $ associated to the flux configuration \FluxConf\ is identically zero. Note that the deformed Calabi--Yau $\Xlflat$ remains singular at the vanishing cycles $\hat B^\flat_{2k-1}$ corresponding to the $(g-1)$ double points. A five-brane wrapped on a vanishing cycle $\hat B^\flat_{2k-1}$ gives rise to a tensionless domain wall which connects two vacua distinguished by a sign flip of the coefficient of [$\hat B^\flat_{2k-1}$] in \FluxConf.
\vskip0.2cm

%%%%%
\subsubsecc{Hyperelliptic curves}
%%%%% 
For a concrete class of phase transitions with higher dimensional deformation spaces we consider next the case of a hyperelliptic curve $\Cg$, where the relevant spaces of holomorphic sections can be studied quite explicitly. A hyperelliptic curve can be described as branched double covers of $\IP^1$, which can be realized as the locus
\eqn\CgHyperelliptic{ y^2 \,=\, \prod_{i=1}^{2g+2}(x-w_i) \ . }
in $\IC^2$. Hyperelliptic curves $\Cg$ come with an involution
\eqn\CgInv{ \iota:\ (x,y) \mapsto (x,-y) \ , }
which fixes the $(2g+2)$ Weierstrass points $w_i$, $i=1,...,2g+2$. The divisors $2 w_1\sim 2 w_2 \sim \ldots \sim 2w_{2g+2}$ are linearly equivalent, and the canonical bundle may be represented by \BaerAA
\eqn\ConHyper{ K_\Cg\,\simeq\,\cO_\Cg( (2g-2) w_i ) \quad {\rm for\ any} \quad i=1,\ldots,2g+2 \ . }
As described above, the flat directions of the superpotential arise from global holomorphic sections of the spin structures which have been classified in ref.~\BaerAA. Any spin structure of a hyperelliptic curve $\Cg$ can be expressed in terms of divisors built out of Weierstrass points 
\eqn\SStruct{ K^{1/2}_{\Cg} \,=\, \cO_\Cg(c_1 w_1 + \ldots + c_{2g+2} w_{2g+2}) \ , \qquad c_i\in\IZ \ , \qquad \sum_i c_i = g-1\ . }
The zeros of the holomorphic sections of $K_{\Cg}$ and $K^{1/2}_{\Cg}$, whose interplay led to the derivation of the factorization condition \DivCan,  are subject to the following general conditions: 
\item{(i)} The global holomorphic sections of the canonical bundle $K_\Cg$ are odd with respect to the involution $\iota$, which implies that their $(2g-2)$ zeros group into $(g-1)$ pairs of zeros $(p,\hat p)$ such that $p$ can be put at an arbitrary position while $\hat p=\iota(p)$. If a zero of $\eps$ coincides with a Weierstrass point $w_i$, than it is at least a double zero.
\item{(ii)} A global section of a spin structure has (at least) a simple zero at every Weierstrass point that appears with an odd coefficient $c_i$ in eq.~\SStruct\ . The remaining zeros come again in pairs $(p,\hat p)$ with $\iota(p)=\hat p$, with $p$ at an arbitrary position.
\vskip0,2cm

\noindent The classification of spin structures in \BaerAA\ distinguishes two classes, namely hyperelliptic curves $\Cg$ of odd and of even genus $g$. We examine these two situations separately and start first with odd genus curves $\Cg$. For convenience we reproduce the results of this classification here (with $i_1<i_2<\ldots<i_{g+1}$):
\eqn\SpinStOdd{
\hbox{\vbox{\offinterlineskip
\halign{\strut\vrule~~\hfil$#$\hfil~~\vrule&~~\hfil$#$\hfil~~\vrule&~~\hfil$#$\hfil~~\vrule&~~\hfil$#$\hfil~~\vrule\cr
\noalign{\hrule}
\#&K_\Cg^{1/2}&\dim H^0(\Cg,K_\Cg^{1/2})&\dim {\rm Def}(\Xlflat,K_\Cg^{1/2})\cr
\noalign{\hrule}
\noalign{\hrule}
1 & \cO_\Cg((g-1)w_1) & {g+1\over 2} & g\ \  ({\rm odd})\cr
\noalign{\hrule}
{2g+2\choose2} & \cO_\Cg((g-2)w_{i_1}+w_{i_2}) & {g-1\over2} & g-2 \cr
\noalign{\hrule}
{2g+2\choose4} & \cO_\Cg((g-4)w_{i_1}+w_{i_2}+w_{i_3}+w_{i_4}) & {g-3\over2} & g-4 \cr
\noalign{\hrule}
\vdots & \vdots & \vdots & \vdots \cr
\noalign{\hrule}
{2g+2\choose g-1} & \cO_\Cg(w_{i_1}+\ldots+w_{i_{g-1}}) & 1 & 1\cr
\noalign{\hrule}
{2g+1\choose g} & \cO_\Cg(-w_1+w_{i_2}+ \ldots +w_{i_{g+1}}) & 0 & 0 \cr
\noalign{\hrule} 
}}}}
The columns denote the number of distinct spin structures of a given type, the spin structure expressed by divisors, the number of global holomorphic sections, and the dimension of the unobstructed deformation space \FlatDef. By the above arguments, this space is supposed to parametrize the flat directions of the flux superpotential given by the two contributions \HiggsPotii\ and \WAfluxesii. To match these two descriptions we write the superpotential as a sum of two contributions
$$W(\Xlflat)=W_++W_-=\int g_+ \wedge \omega +\int g_- \wedge \omega \ ,$$
where $g_\pm$ is the part of the (reduction to $\Cg$ of the) flux which is even/odd under the involution. Since $\omega$ is odd, $W_+$ vanishes identically, while $W_-\neq 0$ and puts a non-trivial restriction on the deformation space.

For $k^\flat=0$ the superpotential \HiggsPotii\ can be expressed as a sum $W=\sum_\ell t^\flat_\ell [\hat T^\flat_\ell]=\sum_\ell \lambda^\flat_\ell [\hat B^\flat_\ell]$, with $t^\flat_\ell,\lambda^\flat_\ell\in \{{1 \over 2},-{1 \over 2}\}$. The poles $p_\ell$ of the four-cycles $\hat B^\flat_n$ are paired by the involution $\iota$ and there are three different possibilities to locate the four-cycles $\hat B^\flat_n$ relative to the two sheets, as shown in \lfig\WBone.

%%%%
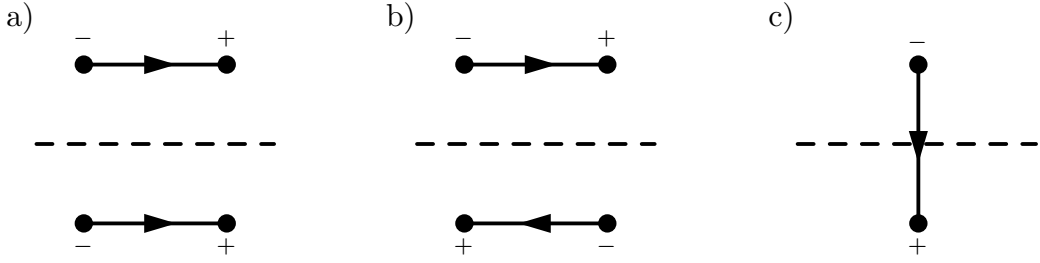
\figboxinsert\WBone{The diagram depicts the three different local flux configurations appearing on hyperelliptic curves $\Cg$. The dashed line indicates a branch cut separating the two sheets of the branched covering of $\IP^1$. As explained in the text, the circles and the connecting solid lines show the poles $p_\ell$ together with fluxes along four-cycles $\hat B^\flat_n$.}
{ 
\DrawDiag{
\move (0 60) % offset for upper margin
\move (0 -40) % offset for lower margin
%%%
% labels
\move (-5 40) \htext{a)}
\move (115 40) \htext{b)}
\move (235 40) \htext{c)}
%%%
% dashed lines
\linewd 1.2
\lpatt(5 5)
\move (0 0) \lvec(75 0)
\move (120 0) \lvec(195 0)
\move (240 0) \lvec(315 0)
%%%
% solid lines with dots
\arrowheadsize l:10 w:6
\lpatt()
\move(15 32) \htext{${}^-$} \move (15 25)\fcir f:0 r:3  \lvec(60 25) \fcir f:0 r:3 \move(60 32) \htext{${}^+$}
\move(15 -32) \htext{${}_-$}\move (15 -25)\fcir f:0 r:3 \lvec (60 -25)\fcir f:0 r:3 \move(60 -32) \htext{${}_+$}
\move (42 25) \avec(44 25) \move (42 -25) \avec(44 -25)
\move(135 32) \htext{${}^-$} \move (135 25)\fcir f:0 r:3 \lvec (180 25)\fcir f:0 r:3 \move(180 32) \htext{${}^+$}
\move(135 -32) \htext{${}_+$}\move (135 -25)\fcir f:0 r:3 \lvec (180 -25)\fcir f:0 r:3 \move(180 -32) \htext{${}_-$}
\move (162 25) \avec(164 25) \move (154 -25) \avec(152 -25)
\move(278 +32) \htext{${}_-$} \move(278 -32) \htext{${}_+$} \move (278 25)\fcir f:0 r:3 \lvec(278 -25)\fcir f:0 r:3
\move (278 -5) \avec (278 -6)
}}
%%%%

\noindent Here the solid circles denote the points $p_\ell$, the arrows the projection to $\Cg$ of the cycles $\hat B^\flat_n$ with the orientation determined by the sign of the coefficients $t^\flat_\ell,\lambda^\flat_\ell$ and the dashed line the branch cut separating the two sheets of $\Cg$. In the first configuration $a)$ the zeros $p_\ell$ and $\iota(p_\ell)$ appear with the same coefficient $t^\flat_\ell$ and both lie in a single factor $\eps_\pm$. The combined contribution of the two $\hat B^\flat$ cycles on the two sheets is $W_+=\int_{\hat B}\omega + \int_{\iota(\hat B)}\omega =\int_{\hat B}(\omega+\iota^*(\omega))=0$. For the other two configurations the zeros $p_\ell$ and $\iota(p_\ell)$ appear with opposite coefficients and lie in different factors $\eps_\pm$. The superpotential for these configurations is non-zero for generic position of $p_\ell$, but becomes critical if the zeros $p_\ell$ approach a Weierstrass points $w_i$ as shown in \lfig\WBtwo.

%%%%
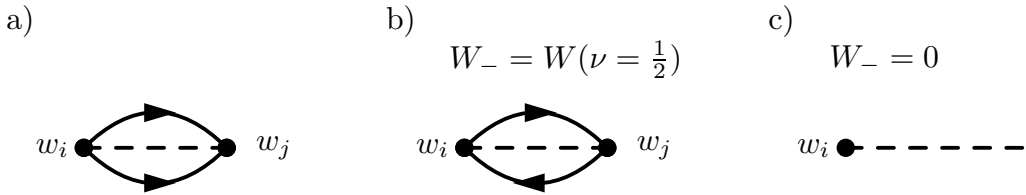
\figboxinsert\WBtwo{The diagram shows the flux configurations as the poles $p_\ell$ in \lfig\WBone\ approach Weierstrass points $w_i$ and $w_j$. Moving the poles onto the Weierstrass points realizes a flat direction in case a) whereas the superpotential for the flux configuration becomes critical in case b) and c).}
{ 
\DrawDiag{
\move (120 60) % offset for upper margin
\move (120 -20) % offset for lower margin
%%%
% labels
\move (-5 40) \htext{a)}
\move (115 40) \htext{b)}
\move (235 40) \htext{c)}
%%%
% dashed lines
\linewd 1.2
\lpatt(5 5)
\move (15 0) \lvec(60 0)  % diag a)
\move (135 0) \lvec(180 0)  % diag b)
\move (255 0) \lvec(315 0)  % diag c)
%%%
\arrowheadsize l:10 w:6
\lpatt()
% diag a)
\move (5 0)\htext{$w_i$} \move (75 0) \htext{$w_j$}
\move (15 0)\fcir f:0 r:3 
\move (39 11) \avec (44 11)
\move (15 0)\fcir f:0 r:3
\move (39 -11) \avec (44 -11)
%\lpatt(1 3)
\move (15 0)\clvec (30 15)(45 15)(60 0) \fcir f:0 r:3
\move (15 0)\clvec (30 -15)(45 -15)(60 0) \fcir f:0 r:3
%\lpatt()
% diag b)
\move (125 0)\htext{$w_i$} \move (195 0) \htext{$w_j$}
\move (135 0)\fcir f:0 r:3 \clvec (150 15)(165 15)(180 0) \fcir f:0 r:3
\move (159 11) \avec (164 11)
\move (135 0)\fcir f:0 r:3 \clvec (150 -15)(165 -15)(180 0) \fcir f:0 r:3
\move (151 -11) \avec (150 -11)
% diag c)
%%%
\move (245 0)\htext{$w_i$} \move (255 0) \fcir f:0 r:3
% text
\textref h:L v:C
\move (130 28) \htext{$W_-=W(\nu={1\over 2})$}
\move (250 28) \htext{$W_-=0$}
}}
%%%%

\noindent The superpotential for case $b)$ in \lfig\WBtwo\ is equal to a half-integer flux on a cycle $a^\flat_{ij}$ encircling the points $w_i,w_j$ and can be cancelled by adding a flux $\nu=-{1\over 2}$ in \WAfluxesii. For case $c)$, the flux is dual to vanishing cycles and gives a zero superpotential, similarly as in the case $N_\pm=1$ discussed above. The number of deformations is reduced by -2 and by -1 in these two cases, respectively.

To compare with the results of \BaerAA\ note that there is a unique maximal spin structure $K_\Cg^{1/2}$, corresponding to the first line of  table~\SpinStOdd, for which the unobstructed deformation space ${\rm Def}(\Xlflat,K_\Cg^{1/2})$ realizes the entire geometric deformation space ${\rm Def}(\Xlflat)$. This spin structure has only even coefficients, and as a result of point (ii), we can pick two sections $\eps_\pm$ each with ${1\over2}(g-1)$ pairs of zeros $(p,\hat p)$ where the zeros $p$ are at arbitrary positions. This maximal spin structure assigns to any pair of points $(p,\hat p)$ half-integral flux quanta of the {\it same} sign which contribute only to the configurations $a)$ with $W_+=0$, in accord with the above analysis of the flux superpotential. Moreover $\nu^\flat_\ell=0$ for all $\ell$ as adding a flux \WAfluxesii\ would induce a non-zero obstruction.

On general grounds, the other spin structures can be obtained from the maximal one by switching on half-integral $C$-field fluxes $\nu^\flat_\ell$ along cycles $a^\flat_\ell$. The critical points of the superpotential are then of the type $b)$ with flux coefficients $t^\flat_\ell$ of opposite signs on two points $p_\ell$ and $\iota(p_\ell)$. Indeed all the other spin structure in \SpinStOdd\ have global sections with zeros at $2\le 2k \le 2g-2$ Weierstrass points according to (ii). Let us denote these Weierstrass points by $w_{i_1}$ to $w_{i_{2k}}$. Thus the deformations $\eps=\eps_+\eps_-$ realized in terms of such a spin structure has (at least) a double zero at these Weierstrass points. This constrains the deformation space ${\rm Def}(\Xlflat)$ by $2k$ conditions, explaining the dimension of the deformation spaces ${\rm Def}(\Xlflat,K^{1/2}_\Cg) \subset {\rm Def}(\Xlflat)$ in \SpinStOdd. Since the deformations $\eps$ exhibit these $2k$ double points, there are $2k$ vanishing $B$-cycles in the deformed geometry and the fourfold $\Xlflat$ remains singular at the fibers over these Weierstrass points. The sections $\eps_\pm$ assign now half-integral flux quanta $[T^\flat_\ell]$ of {\it opposite} sign to the two distinct zeros in the double zeros of the $2k$ Weierstrass point according to eq.~\FluxForSec, resulting in one unit of flux for each vanishing $B$-cycles at the Weierstrass points $w_{i_1}$ and $w_{i_{2k}}$. 

The third critical configuration $c)$ describes a half-integral flux on a vanishing cycle $\hat B^\flat$, representing a root of the $A_M$ lattice \IntCartan\ with $M=2g-3$.  A Weyl reflection of $A_M$ sends $G/2\pi={1\over 2} [\hat B^\flat]\to -G/2\pi$. The two configurations are related by a fivebrane domain wall wrapping $\hat B$ and have the same number of M2 branes $\delta M(\hat B^\flat)=0$. Indeed, for a fivebrane wrapped on a four-cycle $D$, we should consider the averaged flux $\bar G =(G_1+G_2)/2 = G_1+2\pi [D]/2$ of the fluxes $G_{1/2}$ on both sides of the domain wall \GukovYA\  
and integrate over $D$ to obtain 
\eqn\deltaM{
\delta M(D) = \int_D {\bar G\over 2\pi} = \int \left( {G_1 \over 2\pi}\wedge [D]+ {1\over 2} [D]^2 \right) \ .}
This is zero for the above case as $\bar G=0$.\foot{The configurations of type $c)$ require $X$ to have an $A_n$ singularity with $n\geq 1$ in the sense of refs.~\refs{\GukovYA,\EguchiFM} and appear to be more general in that they need not be related to a factorization in the non-hyperelliptic case.}

We know briefly turn to the geometries based upon hyperelliptic curves of even genus. In this case the classification of spin structures reads (with $i_1<i_2<\ldots<i_{g+1}$) \BaerAA:
\eqn\SpinStEven{
\hbox{\vbox{\offinterlineskip
\halign{\strut\vrule~~\hfil$#$\hfil~~\vrule&~~\hfil$#$\hfil~~\vrule&~~\hfil$#$\hfil~~\vrule&~~\hfil$#$\hfil~~\vrule\cr
\noalign{\hrule}
\#&K_\Cg^{1/2}&\dim H^0(\Cg,K_\Cg^{1/2})&\dim {\rm Def}(\Xlflat,K_\Cg^{1/2})\cr
\noalign{\hrule}
\noalign{\hrule}
2g+2 & \cO_\Cg((g-1)w_1) & {g\over 2} & g-1 {\rm \ (odd)}\cr
\noalign{\hrule}
{2g+2\choose3} & \cO_\Cg((g-3)w_{i_1}+w_{i_2}+w_{i_3}) & {g-2\over2} & g-3 \cr
\noalign{\hrule}
{2g+2\choose5} & \cO_\Cg((g-5)w_{i_1}+\ldots+w_{i_5}) & {g-4\over2} & g-5 \cr
\noalign{\hrule}
\vdots & \vdots & \vdots & \vdots \cr
\noalign{\hrule}
{2g+2\choose g-1} & \cO_\Cg(w_{i_1}+\ldots+w_{i_{g-1}}) & 1 & 1\cr
\noalign{\hrule}
{2g+1\choose g} & \cO_\Cg(-w_1+w_{i_2}+ \ldots +w_{i_{g+1}}) & 0 & 0 \cr
\noalign{\hrule} 
}}}}
The analysis of spin structures for the even genus hyperelliptic curves proceeds analogously to the odd genus case. For even genus there are $2g+2$ maximal spin structures (as opposed to a single maximal spin structure for odd genus). However, no spin structure -- not even the maximal spin structures -- can dynamically realize the entire deformation space ${\rm Def}(\Xlflat)$, because any spin structure has at least one odd coefficient $c_i$ in \SStruct\ and hence at least one double zero along a Weierstrass point according to (ii). As a consequence the deformation space ${\rm Def}(\Xlflat,K^{1/2}_\Cg)$ is always a true subspace of ${\rm Def}(\Xlflat)$ for any spin structure $K_\Cg^{1/2}$. The individual spin structures are again distinguished by the phase factors $\nu^\flat_n$ as in \PhaseSpinRel. This can be worked out explicitly by determining integral vs. half-integral flux quanta on the four-cycles $A^\flat_n$.

To summarize we have seen how the Abel-Jacobi theorem applied to the flat directions of the superpotential \HiggsPotii\ and \WAfluxesii\ reproduces the classification of spin structures on hyperelliptic curves obtained in ref.~\BaerAA\ by different means. Moreover the above argument explicitly illustrates the correspondence~\PhaseSpinRel\ between spin structures and holonomy factors $\nu^\flat_n$ in the context of hyperelliptic curves. The above arguments based on the superpotential are however not restricted to the hyperelliptic case (and not even to spin structures, as they apply to more general line bundles, i.e., tensor products with the flat bundle $\cL_0$). It would be interesting to study the classification of spin structures on non-hyperelliptic curves from this perspective.

In this context note that the $\IZ_2$ symmetry, which asserts the cancellation of the two contributions to $W_+$ in the configuration $a)$, arises here from the hyperelliptic involution, but can be interpreted more generally as a $\IZ_2$ symmetry acting on the $SU(2g-2)$ lattice \Fflattic, which asserts the coincidence of the volumes of two subgroups of four-cycles (roots) with identical intersections. A natural ansatz for a systematic construction of critical subsets is therefore to classify and study those loci in the deformation space of $SU(2g-2)$, which are invariant under discrete symmetries acting on the group lattice.

%%%%%
\bigskip
\subsubsecc{Genus $6$ curves with a maximal spin structure}
%%%%% 
We now discuss the dynamics of the phases $\Xlsharp$ and $\Xlflat$ built upon a genus $6$ curve that arises as the zero locus of a (generic) homogeneous degree five polynomial $p_5$ in $\IP^2$
\eqn\gSix{ \Cg \,=\, \left\{ p_5(x_1,x_2,x_3) \,\equiv\, 0 \right\} \, \subset \, \IP^2 \ , }
with homogeneous coordinates $x_1$ to $x_3$.

In order to have dynamical transitions we focus on a scenario with vanishing flux $\GS=0$ in the resolved phase $\Xlsharp$. Then, as discussed, the flat directions of the deformed phase $\Xlflat$ are controlled by a spin structure $\cE_\pm\simeq K_\Cg^{1/2}$ arising as the square root of the canonical bundle $K_\Cg$. The canonical bundle of the curve \gSix\ is the line bundle $\cO(2)$ restricted to the curve $\Cg$, namely $K_\Cg \simeq \cO(2)|_\Cg$. Its  sections are homogeneous degree two polynomial
\eqn\DefSec{ q_2 \,=\, z_1\,x_1^2 + z_2\,x_2^2 + \ldots + z_6\,x_2 x_3 \ , }
with six parameters $z_1$ to $z_6$ parametrizing the deformation space ${\rm Def}(\Xlflat)$.

An obvious square root of the canonical bundle is given by $K_\Cg^{1/2} \simeq \cO(1)|_\Cg$ with three global sections $x_1, x_2, x_3$, i.e., $N_\pm=3$, which gives rise to five unobstructed deformations (c.f., eq.~\FlatDef). This five-dimensional unobstructed deformation space ${\rm Def}(\Xlflat,K_\Cg^{1/2})$ is a codimension one slice in the six-dimensional deformation space ${\rm Def}(\Xlflat)$ parametrized by sections of the canonical bundle $K_\Cg$. The unobstructed deformations correspond to those polynomials $q_2$, which factorize into two linear polynomials $q_2 = l_1 l_2$ with the linear factors $l_{1/2}$ representing sections of the spin structure $\cO(1)|_\Cg$.

As a side remark, we observe that the discussed spin structure $\cO(1)|_\Cg$ is {\it maximal}.  A spin structure of a genus $g$ curve is called maximal, if it has $\left[{g+1\over 2}\right]$ global sections, which is the maximal number of sections for a spin structure of a curve \HitchinAA. However, a spin structure can only be maximal if the curve $\Cg$ is either hyperelliptic or it is of genus four or genus six \refs{\MartensAA,\BaerAA}. The curve \gSix\ is {\it not} hyperelliptic, and therefore we encounter here an example of a maximal spin structure for the exceptional case $g=6$. In the subsequent examples, we come back to maximal spin structures of hyperelliptic curves.

%%%%%
\subsubsecc{Vanishing $G$-flux on the local fourfold $\Xlflat$}
%%%%% 
In our first scenario, we take a generic genus $g$ curve $\Cg$ ($g>1$) with the background fluxes 
\eqn\FluxExOnei{ {\GS\over 2\pi}\,=\, \pm {1\over 2} [S^\sharp] \ , \qquad
  {G^\flat \over 2\pi}\,=\, 0 \ ,  }
in the resolved and in the deformed phase, respectively. Such fluxes correspond to the torsion classes $k^\sharp=k^\flat=g-1$ with vanishing flux quanta $\bq\ell$ on the deformed geometry $\Xlflat$. Dynamically, the resolved phase $\Xlsharp$ is lifted due to the presence of the twisted superpotential~\Wsharp, while -- because of the absence of background fluxes $G^\flat$ -- the deformed phase $\Xlflat$ is unobstructed, realizing the entire deformation space of the deformed phase. Formally, we may identify the unobstructed phase $\Xlflat$ with the factorization bundles $\cE_+ \simeq K_\Cg$ and $\cE_- \simeq \cO_\Cg$ with $N_+=g$ and $N_-=1$ global holomorphic sections, giving rise to the $g$-dimensional deformation space $\FlatDef$.

%%%
\def\Sis{\Sigma^\sharp}
\def\Cch{\widehat C^\sharp}
\def\Cf{C^\flat}
\def\Pis{\Pi^\sharp}
\def\Pich{\widehat\Pi^\flat}
%%%%%%%%
\newsec{Conifold transition in global Calabi--Yau fourfolds}%
\seclab\secTransGlobal
%%%%%%%%
In this section we study the embedding of the local transitions into global fourfolds. The globalization of the locally consistent fluxes identified above leads us to consider fluxes supported on integral homology cycles of a mixed type which give a new class of solutions to the local anomaly conditions. We study the flat directions of the superpotential for these solutions and show that these are often the only solutions which give rise to supersymmetric vacua and dynamically realized phase transitions.

At the critical points the local condition captured by the Abel-Jacobi theorem translates to the appearance of new algebraic four-cycle classes at special complex structure where the polynomial of the global hypersurface allows for reducible algebraic cycles. Reducible algebraic four-cycles supporting $G$-flux have been already studied in the context of F-theory and elliptic fibrations in ref.~\BraunZM.\foot{A dual spectral cover description for $G$-fluxes supported on algebraic four-cycles has been given in refs.~\refs{\CurioBVA,\JMW,\DonagiCA,\MarsanoIX}.} The correspondence between reducible algebraic cycles in Calabi--Yau threefolds and its dual fourfolds and the critical points of $N=1$ superpotentials via an Abel-Jacobi argument is also prominent in the works~\refs{\MorrisonBM,\WalcherCalculations,\JMW,\AlimZA} on $N=1$ mirror symmetry.

To avoid being too technical from the beginning, we first consider the sextic fourfold as a simple example, as it illustrates already some of the distinct features of the new class of mixed fluxes emerging from the local discussion.

%%%
\def\Sef{X^\flat}
\def\Ses{X^\sharp}
\def\Tf{\cT^\flat}
%%%
\subsec{A simple example: Extremal transition for the sextic}
%%%
Before we delve into the general discussion of extremal M-theory transitions in Calabi--Yau fourfolds, we first present an instructive example. Let us consider the sextic hypersurface Calabi--Yau fourfold $\Sef\equiv\IP^5[6]$ in the projective space $\IP^5$, given in terms of a general homogeneous polynomial of degree six in the projective coordinates $x_1,\ldots,x_6$ of $\IP^5$. The fourfold $\Sef$ has Hodge numbers and Euler characteristic
\eqn\Sefh{ h^{1,1}_{\Sef} \,=\, 1 \ , \quad h^{2,1}_{\Sef} \,=\, 0 \, \quad h^{3,1}_{\Sef} \,=\, 426 \ , \quad h^{2,2}_{\Sef} \,=\, 1\,752 \ , \quad \chi(\Sef)\,=\,2\,610 \ . }
As the defining sextic polynomial degenerates to $x_5\, g(x) + x_6\, h(x)$ with two degree-five polynomials $g(x)$ and $h(x)$, it develops along the codimension two locus $x_5\,=\,x_6\,=\,g(x)\,=\,h(x)\,=\,0$ a genus $76$ curve $\Cg$ of conifold singularities. A small resolution of this singular curve yields the Calabi--Yau fourfold $\Ses$ with Hodge numbers and Euler characteristic\foot{We obtain the small resolution as follows (c.f., \refs{\CandelasUG,\GreeneHU}). The curve $\Cg$ lies within $\IP^3$ defined by $x_5=x_6=0$. We blow up $\IP^3$ within $\IP^5$, which may be realized in $\IP^5 \times \IP^1$ as the locus $s_1x_5 -s_2x_6=0$ with the projective coordinates $s_1,s_2$ of $\IP^1$. The resolved Calabi--Yau fourfold $\Ses$ is the locus $0=s_1x_5 -s_2x_6=s_1g(x)+s_2h(x)$ within $\IP^5\times\IP^1$. Alternatively, we will give a (related) toric hypersurface realization of $\Ses$ in Section~4.4\yyy{}.}
\eqn\Sesh{ h^{1,1}_{\Ses} \,=\, 2 \ , \quad h^{2,1}_{\Ses} \,=\, 0 \ , \quad h^{3,1}_{\Ses} \,=\, 350 \ , \quad h^{2,2}_{\Ses} \,=\, 1\,452 \ , \quad \chi(\Ses)\,=\,2\,160 \ . }

We observe that the Euler characteristic changes by $\delta\chi=\chi(\Sef)-\chi(\Ses)=450=6-6\,g(\Cg)$, in agreement with the change in Euler characteristics deduced from eqs.~\SEuler\ and \XlflatEuler\ in the context of the local extremal transitions. Furthermore, the Hodge numbers change as anticipated in eq.~\hnchange, which we will show in general in Section~4.2\yyy{} and Section~4.3\yyy{} for such extremal transitions.

To describe dynamical phase transition between $\Sef$ and $\Ses$, let us now turn to the background fluxes. Firstly, we observe that the Euler characteristic $\chi(\Ses)$ is divisible by $24$ and the second Chern class $c_2(\Ses)$ is even in $H^4(\Ses,\IZ)$. As consequence, the Calabi--Yau fourfold $\Ses$ together with $M_{\Ses}={2\,160\over 24}=90$ space-time filling M2 branes fulfills both the tadpole constraint \Acon\ and the quantization condition~\Acon\ in the absence of background fluxes $G^\sharp=0$. Thus, such a scenario represents a consistent M-theory background. Due to the absence of background fluxes, there are no potentials \Wpots\ and we arrive at a supersymmetric M-theory vacuum on $\Ses$ with
\eqn\Sesi{ M_{\Ses}\,=\,90 \ , \quad G^\sharp\,=\,0 \ . }
From a local perspective in the vicinity of the curve $\Cg$ -- that is to say from the vantage point of the local fourfold $\Xlsharp$ of a resolved genus $76$ curve $\Cg$ as discussed in Section~\secTransLocal\ -- the vacuum \Sesi\ describes a local configuration with vanishing torsion class $k^\sharp=0$. Furthermore, due to the absence of homologically non-trivial three-cycles, i.e, $h^{2,1}_{\Ses}=0$, the phase factors $\nu_n^\sharp$ are set to zero due to the embedding into the global geometry $\Ses$.  

Let us now examine dynamical phase transitions into the fourfold $\Sef$. The Euler characteristic of $\chi(\Sef)$ is {\it not} divisible by $24$ and the second Chern class $c_2(\Sef)=15 H^2$ is odd, where the class $H$ is induced from the hyperplane class of the ambient $\IP^5$.\foot{Since $\int_{\Sef} H^2 \wedge H^2=6$, the four form class $H^2$ is {\it not} divisible by two in $H^4(\Sef,\IZ)$. Hence, $c_2(\Sef)$ is {\it not} divisible by two.} Thus, consistency of M-theory on the fourfold $\Sef$ requires a half-integrally quantized background flux $G^\flat$ such that the quantization condition~\Gflux\ is met. Then the tadpole condition~\Acon\ can be fulfilled with an integral number of space-time filling M2 branes $M_{\Sef}$.  

For a dynamical phase transition into $\Sef$, the required background flux $G^\flat$ must be supersymmetric for an unobstructed deformation into $\Sef$. In the discussed global setting these deformations are parametrized by homogeneous polynomials of degree six $\eps(x)$ according to
\eqn\Sefdef{ x_5\, g(x) + x_6\, h(x)\,=\, \eps(x) \ . }
The deformation $\eps(x)$ restricts on the genus $g$ curve $\Cg$ to a section of the canonical bundle $K_\Cg$, as discussed in the context of the local fourfold $\Xlflat$. The analysis of the superpotential in the local geometry of Section~3.6\ \yyy{} told us that -- for vanishing phase factors $\nu_n^\sharp$ and for vanishing torsion classes $k^\sharp=k^\flat=0$ -- the section of the canonical bundle $\eps$ must factor into sections of an appropriate spin structure $K^{1/2}_\Cg$ as $\eps=\eps_+ \eps_-$.\foot{Except for the critical configurations related to $A_n$ singularities.}

When the local geometry is embedded into a global fourfold $\Sef$, 
the two local algebraic cycles defined by $x_5=x_6=\epsilon_i(x)=0$ 
must extend to globally defined algebraic cycles.
The easiest way to achieve this is to assume that the factorization condition globalizes,  i.e., the sextic deformation $\eps(x)$ in eq.~\Sefdef\ has to factor into two homogeneous polynomials of degree three, which is a constraint on the complex structure of $\Sef$. The new aspect in the global geometry is that for this complex structure there is a new integral algebraic four-cycle class. Indeed if we set $x_5=x_6=0$ in eq.~\Sefdef, then -- contrary to the generic sextic fourfold $\Sef$ near the transition -- we get a {\it reducible} algebraic four-cycle with two components $\Tf_\pm:\ x_5=x_6=\eps_\pm(x)=0$. By construction the integral four-form classes $[\Tf_\pm]$ add up to the square of the hyperplane class, namely $H^2=[\Tf_+]+[\Tf_-]$. 

A global version of the local result \FluxForSec\ obtained for $k^\sharp=0$ in Sect.~3.6\yyy{} is the flux
\eqn\SefFlux{ {G^\flat\over 2\pi} \,=\, {1\over2}\left([\Tf_+] - [\Tf_-] \right) \ . }
As expected from the local analysis, this satisfies the quantization condition \Gflux:
\eqn\SefQuant{  {G^\flat\over 2\pi} - {c_2(\Sef)\over2} \,=\,{1\over2}\left([\Tf_+] - [\Tf_-] \right) -{15\over 2}H^2 \,=\, -7 [\Tf_+] - 8 [\Tf_-] \ . }
Furthermore, with the help of the intersection numbers $\Tf_+.\Tf_+=\Tf_-.\Tf_-=39$ and $\Tf_-.\Tf_+=-36$ of the four-cycles $\Tf_\pm$,\foot{We calculate the Euler number of the normal bundle $N\Tf_\pm$, which determines the self-intersections of $\Tf_\pm$ to be $39$. Then we infer $\Tf_-.\Tf_+=-36$ from $6=\int_{\Sef} H^2\wedge H^2=(\Tf_++\Tf_-)^2$.} we determine for the flux \SefQuant\ a tadpole-free M-theory configuration with space-time filling M2 branes that is unchanged from \Sesi
\eqn\Sefi{ M_{\Sef}\,=\, {2\,610\over 24} -{1\over 2\cdot 4}\left(2 \cdot 39 +2 \cdot 36\right) \,=\, 90 \ . }
Note that the classes $[\Tf_\pm]$ continue to exist, and provide a solution to the local anomaly cancellation on each integral four-cycle, for any complex structure away from the factorization locus. For generic complex structure the representatives for $[\Tf_\pm]$ are neither algebraic nor special Lagrangian and the flux \SefFlux\ is of a mixed type. For the special complex structure the lattice vectors in $H^4(X,\IZ)$ associated with the classes $[\Tf_+]$ and $[\Tf_-]$ become orthogonal to the class represented by $\Omega$ (see also the discussion in App.~A\yyy.) 

The flux discussed above gives rise to an unobstructed  phase transition between $\Sef$ and $\Ses$. As for the K\"ahler moduli, the mixed $G$-flux $G^\flat$ in eq.~\SefFlux\ is primitive because of $J\wedge G^\flat\sim (\Tf_++\Tf_-)(\Tf_+-\Tf_-)=0$. The twisted superpotential $\widetilde W$ is zero. This feature of the new mixed solution to the anomaly constraint should be compared to the other obvious solution of the split type: 
\eqn\SefFluxs{ {G_{split}^\flat\over 2\pi} \,=\, {1\over2}H^2 \,=\, {1\over2}\left([\Tf_+] + [\Tf_-] \right) \ . }
In distinction to the flux \SefFlux, this choice of flux generates a twisted superpotential for the K\"ahler modulus of the sextic. Note that the two configurations \SefFlux\ and \SefFluxs\ are connected by a five-brane domain wall wrapped on $T_-$.

On the other hand, the mixed $G$-flux generates the superpotential $W$ for the complex structure deformations studied in Section~3.6\yyy. As argued there, the critical points of this superpotential are precisely the complex structures for which $\eps$ factorizes as $\eps(x)=\eps_+(x)\eps_-(x)$. These deformations will keep both cycles $\Tf_\pm$ as cycles of type $(2,2)$, as expected from the results of \refs{\BeckerGJ,\GukovYA}. Hence, moving along factorized deformation directions, we find a dynamically unobstructed phase transition from $\Ses$ into the deformed phase $\Sef$.

More precisely, the 150 zeros of $\eps(x)$ on the genus $76$ curve $\Cg$ split into the $g-1=75$ zeros $p_\ell\in Z_\pm$ of $\eps_\pm(x)$, defining the four-cycles $T^\flat_\ell$ that appear with positive/negative coefficients in \FluxForSec. In the global embedding these two sets of four-cycles in the local geometry add up to the four-cycle classes $\Tf_\pm$ in $\Sef$. Thus the analyzed phase transition represents a global embedding of the local transition discussed in eq.~\IntSol.

Extremal transitions for other torsion classes correspond to other classes of factorizations of $\eps(x)$ and these again give rise to (different) new algebraic four-cycle classes in the global fourfold $\Sef$ for special complex structures. For the sextic, a globally consistent factorization requires $\eps_\pm$ to be of degree $3,2,1$, corresponding to the torsion classes $k^\sharp=0,25,50$ and a net number of M2 branes $90,92,98$, respectively. Note that the flux contribution to the M2 brane charge on $X^\sharp$ has the wrong sign for $k^\sharp>0$ (see \Acon) and indicates an obstruction towards the transition to $X^\sharp$. These obstructed cases are similar to the transitions in elliptic fourfolds considered in \BraunZM, where the flux contribution on the resolved phase appears to have always the wrong sign.

As a concrete example consider the factorization of $\eps(x)=\eps_5(x)\eps_1(x)$ into a degree five and a degree one polynomial. In this case the $G$-flux becomes
\eqn\SefFluxii{ {G^\flat\over 2\pi}\,=\,{1\over 2}\left( [\Tf_5] - [\Tf_1] \right) \ , }
in terms of the corresponding algebraic cycles with $H^2=[\Tf_5] + [\Tf_1]$. By similar arguments, this flux is again consistent with the quantization constraint~\Gflux\ and yields $M^\flat=98$ due to the tadpole condition~\Acon\ (because $\Tf_5.\Tf_5=25$, $\Tf_1.\Tf_1=21$ and $\Tf_5.\Tf_1=-20$). From the $125$ and $25$ zeros of $\eps_5(x)$ and $\eps_1(x)$, respectively, we obtain $125$ non-vanishing flux-quanta $b^\flat_\ell$ of charge one, which yield the torsion class $k^\flat = 50$ according to eq.~\kftor. However, the flux $G^\flat$ is {\it not} primitive anymore, because the intersection $\int_{\Sef} G^\flat \wedge H^2 = 2$ gives rise to a twisted superpotential
\eqn\SefTwist{ {\widetilde W}^\flat \,=\, {1\over 2} \int_{\Sef} {G^\flat\over 2\pi} \wedge J\wedge J \,=\, J_H^2 \ , }
in terms of the K\"ahler form $J=J_H\,H$ with K\"ahler parameter $J_H$.
In the resolved fourfold $\Ses$ the corresponding background flux $G^\sharp$ is given by\foot{Despite the overall factor of ${1\over 3}$, a closer analysis reveals that ${G^\sharp\over 2\pi}$ is actually integral and thus compatible with the quantization condition~\Gflux.}
\eqn\SesFluxii{ {G^\sharp\over 2\pi}\,=\,{1\over 3}[\Ssharp] + {1\over3}\tilde H^2 \, \in \, H^4(X^\sharp,\IZ) \ , }
expressed in terms of the four-form $[\Ssharp]$ dual to the surface \Zfib\ and the two-form class $\tilde H$ induced from the hyperplane class $H$ in $\IP^6$. As $[\Ssharp]$ is orthogonal to $\tilde H^2$, the tadpole condition~\Acon\ yields again the M2 brane number $M^\sharp=98$. Furthermore, the flux \SesFluxii\ generates a twisted superpotential of the form 
\eqn\SesTwist{ {\widetilde W}^\sharp \,=\, {1\over 2} \int_{\Ses}{G^\sharp\over 2\pi}\wedge J \wedge J \,=\, J_H^2 + 10 J_H J_F\ , }
with the K\"ahler form $J=J_H \tilde H +J_F [F]$, where $J_F$ is the volume of the $\IP^1$ fiber $F$ of the fibration $S^\sharp$. The first term corresponds to the twisted superpotential~\SefTwist\ also appearing in the fourfold $\Sef$. The second term arises in the extremal transition in the vicinity of the genus $76$ curve $\Cg$, as discussed in the context of the local geometries in Section~\secTransLocal. This is the contribution in \Wsharp\ for the torsion class $k^\sharp=50$.

Hence, the factorization $\eps(x)=\eps_5(x)\eps_1(x)$ realizes a local extremal M-theory transition scenario along a genus $76$ curve $\Cg$ with torsion class $k^\flat=k^\sharp=50$. However, in addition to the potential terms arising in the vicinity of the local transition geometries $\Xlflat$ and $\Xlsharp$, we find an additional overall twisted superpotential term, which we attribute to the chosen realization of the global embedding fourfolds $\Sef$ and $\Ses$.  As we review and discuss in section 5, twisted superpotentials yield Chern-Simons terms in the 3d field theory, and mixed terms, involving one dynamical and one background field, have the interpretation of a FI term in the low-energy field theory.  Then, in \SesTwist, the first term $\sim J_H^2$ is a genuine global effect present on both sides of the transition, while the second term has an interpretation as a non-zero FI-term in the 3d field theory. For the above choice of global flux, the transition will therefore be obstructed, even in the field theory sense. This is in accord with the fact that again one needs 8 {\it more} M2 branes on $X^\sharp$ with flux \SesFluxii\ as in the vacuum with zero flux. The FI-term can be however removed by an additional integral flux on $X^\sharp$, i.e., $G^\sharp = {1\over 3}([S^\sharp]+\tilde H^2)-2\tilde H^2$ with twisted superpotential $\widetilde W^\sharp=-5J_H^2$ on both sides of the transition.

%%%
\subsec{M-theory transitions via topological surgery}
%%%
As we have exemplified in the previous section, M-theory transitions among non-compact geometries $\Xlsharp$ and $\Xlflat$ furnish a local description of M-theory conifold transitions among compact Calabi--Yau fourfolds $\Xs$ and $\Xf$. For a global conifold transition, we envision the local geometries $\Xlsharp$ and $\Xlflat$ to be topologically glued to a common complementary space $\Xc$, so as to give rise to the global geometries $\Xs$ and $\Xf$ via topological surgery according to
\eqn\XcSurg{ 
  \Xs\,=\,\Xc \cup \Xlsharp \ , \quad 
  \Xf\,=\,\Xc \cup \Xlflat \quad {\rm with} \quad
  \partial\widetilde X\,=\,\partial\Xlsharp\,=\,\partial\Xlflat\,=\,-\partial\Xc \ . }
The structure of the participating cycles and chains along the extremal conifold transition are schematically depicted in \lfig\TransCycles\ and are explained in detail below.

%%%%
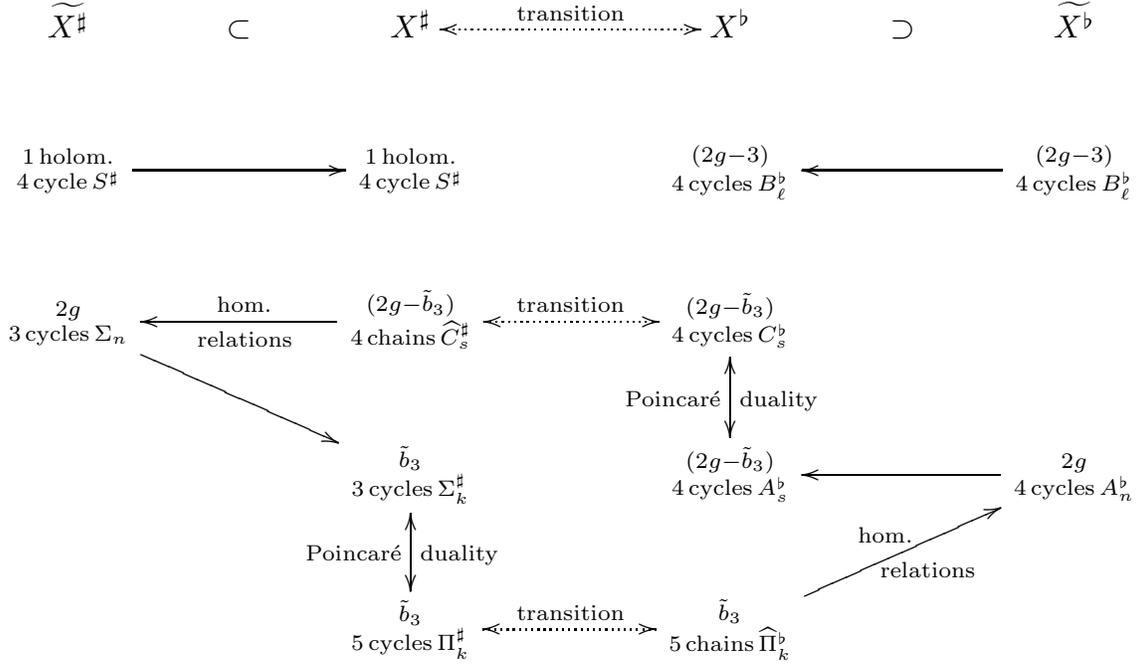
\figboxinsert\TransCycles{The diagram summarizes the topological properties of the conifold transition between the compact fourfold $\Xs$ and $\Xf$. It shows the homological structure of cycles inherited from the local fourfold $\Xlsharp$ and $\Xlflat$ and their resulting homological relations in the embedding compact Calabi--Yau fourfolds.}{
\hbox{\xymatrix{
\Xlsharp &\subset& \Xs \ar@{.>}[rr]^{\rm transition} && \Xf \ar@{.>}[ll] &\supset& \Xlflat \cr
{{\rm 1\, holom.}\atop{{\rm 4\,cycle}\, \Ssharp}} \ar[rr] && 
   {{\rm 1\, holom.}\atop{{\rm 4\,cycle}\, \Ssharp}}  &&
      {(2g-3)\atop{{\rm4\,cycles}\,B^\flat_\ell}}  &&
          {(2g-3)\atop{{\rm4\,cycles}\,B^\flat_\ell}} \ar[ll]  \cr
 {2g\atop{{\rm3\,cycles}\,\Sigma_n}} \ar[drr]&&
   {(2g-\tilde b_3)\atop{{\rm4\,chains}\,\Cch_s}}\ar[ll]_{\rm\ hom.}^{\rm\ relations}  \ar@{.>}[rr]^{\rm transition}&&
       {(2g-\tilde b_3)\atop{{\rm4\,cycles}\,\Cf_s}}\ar[d]_{\rm\strut Poincar\acute{e}}^{\rm\strut duality} \ar@{.>}[ll]\cr
&& {\tilde b_3\atop{{\rm3\,cycles}\,\Sis_k}} \ar[d]_{\rm\strut Poincar\acute{e}}^{\rm\strut duality} &&
   {(2g-\tilde b_3)\atop{{\rm4\,cycles}\,A^\flat_s}} \ar[u] &&
       {2g \atop {{\rm4\,cycles}\,A^\flat_n}} \ar[ll] \cr
&& {\tilde b_3\atop{\rm5\,cycles}\,\Pis_k} \ar[u] \ar@{.>}[rr]^{\rm transition} &&
   {\tilde b_3\atop{\rm5\,chains}\,\Pich_k} \ar[urr]^{\rm hom.}_{\rm relations} \ar@{.>}[ll]
}}}
%%%%

In the compact K\"ahler fourfold $\Xs$ the algebraic four-cycle $\Ssharp \subset \Xlsharp \subset \Xs$ represents a non-trivial class in $H_4(\Xs,\IZ)$. In addition, there are $0\le\tilde b_3\le 2g$ three cycle classes $\Sis_k, k=1,\ldots,\tilde b_3$, arising from the $2g$ non-torsion three-cycles $\Sigma_n$ in $H_3(\widetilde\partial X,\IZ)$ together with $(2g-\tilde b_3)$ relations in homology. These relations are geometrically realized by $(2g-\tilde b_3)$ four-chains, $\Cch_s, s=1,\ldots,2g-\tilde b_3$, with $\partial\Cch_s \subset\left(\bigcup_n \Sigma_n\right)$. Finally, the cycle classes in $H_5(\Xs,\IZ)$, which are Poincar\'e dual to the non-trivial three cycles $\Sis_k$, are denoted by $\Pis_k,k=1,\ldots,\tilde b_3$. 

As we go through the conifold transition to the fourfold $\Xf$, the algebraic four-cycle $\Ssharp$ disappears, and instead we obtain $(2g-3)$ new four-cycle classes $B^\flat_\ell, \ell=1,\ldots,2g-3$, which correspond to the local $B$-cycle classes \fourcyclesXlf\ in $H_4(\Xlflat,\IZ)$. Furthermore, all the three-cycles classes $\Sigma_n \in H_3(\widetilde\partial X,\IZ)$ become homologically trivial in $\Xlflat$ (because $H_3(\Xlflat,\IZ)\simeq0$). Thus, there are $(2g-\tilde b_3)$ three cycle classes in $H_3(\partial\widetilde X,\IZ)$ that are trivial in both $H_3(\Xc,\IZ)$ and $H_3(\Xlflat,\IZ)$. Then -- due to the long exact homology sequence $\ldots\rightarrow H_4(\Xf,\IZ) \rightarrow H_3(\partial\widetilde X,\IZ) \rightarrow H_3(\Xc,\IZ)\oplus H_3(\Xlflat,\IZ)\rightarrow\ldots$ -- we find that there are $(2g-\tilde b_3)$ non-trivial four-cycle classes, $\Cf_s,s=1,\ldots,2g-\tilde b_3$, in $H_4(\Xf,\IZ)$, which in the transition come from closing off the $(2g-\tilde b_3)$ four-chains $\Cch_s$. The Poincar\'e dual four-cycle classes to $\Cf_s$ correspond to appropriate (linear combinations of) four-cycles, $A^\flat_s,s=1,\ldots,2g-\tilde b_3$, of the $2g$ local four-cycles $A^\flat_n,n=1,\ldots,2g$, in eq.~\fourcyclesXlf, while the remaining local $A$-type cycles become trivial in the global geometry $\Xf$, as they are bounded by $\tilde b_3$ five chains $\Pich_k,k=1,\ldots,\tilde b_3$. These chains come from the five cycles $\Pis_k$, which open up as we go through the conifold transition.

%%%
\subsec{M-theory transitions via the Clemens--Schmid exact sequence}
%%%
We will now compare the topology and Hodge structures on $\Xs$ and $\Xf$ using the Clemens--Schmid exact sequence, which is reviewed in Appendix D.1\yyy{} and demonstrated for Calabi--Yau threefold conifold transitions in Appendix D.2\yyy{}. To apply this to our fourfold extremal transition, we must first construct a {\it semistable degeneration}. Here, such a degeneration is a certain (smooth) fibration of Calabi--Yau fourfolds over a disk $\Delta$, where a normal-crossing component of the singular central fiber is birational equivalent to the resolved fourfold $\Xs$, while the non-central (smooth) fourfold fibers describe the deformed Calabi--Yau fourfold $\Xf$. With the constructed semistable degeneration the Clemens--Schmid exact sequence allows us to calculate the change in Hodge structure as we go through the extremal fourfold transition. The details of this computation are relegated to Appendix~D.3\yyy{}, and we now summarize the result of this analysis.

By transitioning from the central fiber of the constructed semi-stable degeneration~(D.10)\yyy{} to deformed Calabi--Yau geometry $\Xf$ of the generic smooth fiber, we arrive at the Hodge diamond~(D.14)\yyy{} of the fourfold $\Xf$:
\eqn\HDXfii{
  \dim H^{p,q}(\Xf)\,=\!\!\!
    \vcenter{\offinterlineskip
    \halign{\hbox to 7.5ex{#}&\hbox to 7.5ex{#}&\hbox to 7.5ex{#}&\hbox to 7.5ex{#}&\hbox to 7.5ex{#}&\hbox to 7.5ex{#}&\hbox to 7.5ex{#}&\hbox to 7.5ex{#}&\hbox to 7.5ex{#}&\hbox to 7.5ex{#}\cr
    % 0-forms
    &&&& \sentry{1} \cr\phantom{X}\cr
    % 1-forms
    &&& \sentry{0} && \sentry{0} \cr\phantom{X}\cr
    % 2-forms
    && \sentry{0} &&  \sentry{h^{1,1}_{\Xs}\!-\!1} &&  \sentry{0} \cr\phantom{X}\cr
    % 3-forms
    & \sentry{0} &&  \sentry{\hskip-2exh^{2,1}_{\Xs}\!-\!\tilde h^{2,1}\hskip-2ex} &&  \sentry{\hskip-2exh^{2,1}_{\Xs}\!-\!\tilde h^{2,1}\hskip-2ex} &&  \sentry{0}  \cr\phantom{X}\cr
    % 4-forms
    \sentry{1} && \sentry{h^{3,1}_{\Xs}\!-\!\tilde h^{2,1}\!\!+\!g} &&  \sentry{\hskip-3exh^{2,2}_{\Xs}\!-\!2\tilde h^{2,1}\!\!+\!4g\!-\!4\hskip-3ex} &&  \sentry{h^{3,1}_{\Xs}\!-\!\tilde h^{2,1}\!\!+\!g} &&  \sentry{1} \cr\phantom{X}\cr
    % 5-forms
    & \sentry{0} &&  \sentry{\hskip-2exh^{2,1}_{\Xs}\!-\!\tilde h^{2,1}\hskip-2ex} &&  \sentry{\hskip-2exh^{2,1}_{\Xs}\!-\!\tilde h^{2,1}\hskip-2ex} &&  \sentry{0}  \cr\phantom{X}\cr
    % 6-forms
    && \sentry{0} &&  \sentry{h^{1,1}_{\Xs}\!-\!1} &&  \sentry{0} \cr\phantom{X}\cr
    % 7-forms
    &&& \sentry{0} && \sentry{0} \cr\phantom{X}\cr
    % 8-forms
    &&&& \sentry{1} \cr
    }}}
Here we express the Hodge diamond of $\Xf$ in terms of the Hodge numbers $h^{p,q}_{\Xs}$ of the resolved Calabi--Yau fourfold $\Xs$. The number $\tilde h^{2,1}$ refers to the number of harmonic $(2,1)$-forms participating in the extremal transition. They are in the image of the canonical map $H^{2,1}(\Xs) \to H^{2,1}(\Ssharp)$ and therefore disappear with the cycle $\Ssharp$, as we go through the extremal transition form $\Xs$ to $\Xf$. By comparing with the topological properties of the conifold transition we can now also identify the number $\tilde b_3$ of non-trivial three-cycles $\Sigma^\sharp_k$ of $\Xs$ in \lfig\TransCycles\ with the number $\tilde h^{2,1}$ of participating $(2,1)$-forms according to
\eqn\RelTopHodge{ \tilde b_3 \,=\, \tilde h^{2,1}+ \tilde h^{1,2}\,=\, 2\, \tilde h^{2,1} \ . } 
Furthermore, we can associate the remaining $(h^{2,1}_{\Xs}-\tilde h^{2,1})$ three-forms as part of the complementary space $X^c$ (c.f., eq.~\XcSurg). Hence, they arise as non-trivial three-forms in both fourfold geometries $\Xf$ and $\Xs$. Finally, we note that the determined Hodge diamond \HDXfii\ yields the characteristic change of Hodge numbers advertised in eqs.~\hnchange.

%%%
\subsec{$G$-flux quantization condition}
%%%
%
In the following we consider the change of $G$-flux quantization during an extremal transition $\Xs$ to $\Xf$, where the four-cycle $\Ssharp$ with self-intersection $\Ssharp.\Ssharp = 2 - 2g < 0$ shrinks and disappears from $H_4(\Xs,\IZ)$.\foot{For a discussion of the quantization condition in F-theory on elliptic fibrations, see refs.~\refs{\MarsanoIX,\CollinucciGZ,\MarsanoHV,\CollinucciAS}.}  The algebraic cycle $S^\sharp$ need not be a generator of the homology lattice $H_4(\Xs,\IZ)$, but instead it could be homologous to a linear combination of generators of $H_4(\Xs,\IZ)$. However, for the considered transition $\Ssharp$ is the only four-cycle vanishing in the resolved geometry $\Xs$ (see  \lfig\TransCycles). As a result, for the extremal transitions under consideration $\Ssharp$ can only be a multiple of generator $\Ssharp_{1/\ell}$ with $\Ssharp \sim \ell\,\Ssharp_{1/\ell}$ in homology. As result the self-intersection of $\Ssharp$ must factor as $2-2g=\ell^2 m$, and we have intersection numbers $\Ssharp_{1/\ell}.\Ssharp_{1/\ell}=m$. Furthermore, as the intersection pairing of $H_4(\Xs,\IZ)$ is unimodular, there is always a dual cycle $\cT^\sharp$ with $\cT^\sharp.\Ssharp_{1/\ell}=1$, and we can write the dual integral four-form as
\eqn\cTi{ [\cT^\sharp]\,=\,{1\over m} ([\Ssharp_{1/\ell}] + \Theta ) \, \in \, H^4(\Xs,\IZ) \ , }
where $\Theta$ is a four-form such that $[\cT^\sharp]$ is integral. Firstly, $\Theta$ is integral itself because both $m [\cT^\sharp]$ and $[\Ssharp_{1/\ell}]$ are integral. Secondly, $\Theta$ is orthogonal to $\Ssharp_{1/\ell}$ because $\int_{\Ssharp_{1/\ell}} [\cT^\sharp] = \cT^\sharp.\Ssharp_{1/\ell} = 1$. 

In order to determine consistent $G$-flux, we need to look at the second Chern class of $\Xs$.  The second Chern class of $\Xs$ takes the following form
\eqn\ctwoXs{ c_2(\Xs)\,=\,-\ell\,m\,[\cT^\sharp] + \Delta c_2\,=\,-\ell\,[S^\sharp_{1/\ell}]-\ell\,\Theta + \Delta c_2 \ , }
where the integral piece $\Delta c_2$ must be orthogonal to $\Ssharp$ to yield $\int_{\Ssharp} c_2(\Xs) = \int_{\Ssharp} c_2(\Xlsharp) = 2g-2$ in agreement with eq.~\XlCclass. Then we make for the $G$-flux $G^\sharp$ the ansatz
\eqn\Gsi{ {G^\sharp\over 2\pi}\,=\, \kappa\,[\cT^\sharp] + {\Delta G\over 2\pi} \ , \qquad \kappa \in \IZ \ , }
with $\Delta G$ orthogonal to $\cT^\sharp$ such that the quantization condition \Gflux\ is fulfilled. Note that $\kappa$ is integrally quantized because $\ell\,m$ in eq.~\ctwoXs\ is even.

In the local geometry $\Xlsharp$ the cycle $\cT^\sharp$ reduces to $\ell$ copies of the non-compact cycles \Tsharp. Therefore, the flux $G^\sharp$ corresponds to the local torsion class $k^\sharp=\kappa\,\ell$, whereas the flux $\Delta G$ is attributed to the complement $X^c$ of the non-compact local geometry $\Xlsharp$. As a side remark, we observe here that the global embedding geometry $\Xs$ restricts the globally possible torsion classes to multiples of $\ell$ where $\ell$ must obey $\ell^2\, |\, \chi(\Cg)$. 

As we go through the extremal transition to $\Xf$ the second Chern class \ctwoXs\ becomes
\eqn\ctwoXf{ c_2(\Xf)\,=\, -\ell\, \Theta + \Delta c_2 \ , }
because the contributions $\Theta$ and $\Delta c_2$ are associated to the complementary geometry $X^c$, and, as we transition to $\Xf$, we (generically) do not generate any new holomorphic four-forms that could contribute to the the second Chern class $c_2(\Xf)$. As a check we find that
\eqn\ctwosquare{ \int_{\Xf} c_2^2(\Xf) - \int_{\Xs} c_2^2(\Xs)\,=\, 2g-2 \ , }
which, according to ref.~\SethiES, agrees with ${\delta\chi\over 3}={\chi(\Xs)\over 3}-{\chi(\Xf)\over 3}$. 

Similarly, the flux $\Delta G$ just carries over to $\Xf$, and we arrive at the $G$-flux
\eqn\Gfi{ {G^\flat\over 2\pi}\,=\,  {G^\flat_0\over 2\pi} + {\Delta G\over 2\pi} \ . }
For the four-form flux $G^\flat_0$, we make the ansatz
\eqn\Gfii{ {G^\flat_0\over 2\pi} \,=\, {1\over 2}\left( [\cT^\flat_+] - [\cT^\flat_-] \right) \ , }
with the four-forms 
\eqn\cTi{ [\cT^\flat_\pm]\,=\, {1\over m} \left(\pm[\cB^\flat] + \left({\ell\,m\over2}\pm\kappa\right) \Theta  \right) \ . }
Here $[\cB^\flat]$ arises from the local geometry $\Xlflat$ and is chosen such that the four-forms $[\cT^\flat_\pm]$ become integral.\foot{Using long exact Mayer-Vietoris singular homology sequences for the topological surgeries $\Xs=X^c\cup \Xlsharp$ and $\Xf=X^c\cup \Xlflat$, we can argue that we can always find a form $\cB^\flat$ such that $\cT^\flat_\pm$ becomes integral.}

As $\cT^\flat_\pm$ is integral, we note that the quantization condition \Gflux\ is fulfilled
\eqn\cGfiii{  {G^\flat_0\over 2\pi} - {c_2(\Xf)\over2}\,=\,  [\cT_+^\flat] + \left({\Delta G\over 2}- \Delta c_2 \right) \, \in \, H^4(\Xf,\IZ) \ . }
Furthermore, the flux \Gfii\ arises from the flux component $\kappa [T^\sharp]$ in eq.~\Gsi, as both $G$-fluxes \Gsi\ and \Gfi\ have the (fractional) four-form part ${\kappa\over m}\Theta$ in common. 

Due to the flux contribution ${2\over m}[\cB^\flat]$, the $G$-flux $G^\flat_0$ becomes in the local geometry $\Xlflat$ the local flux \GFfluxii. Requiring that the four-forms $[\cT^\flat_\pm]$ are integral, does not entirely fix the four-form component $[\cB^\flat]$. Its structure, however, is constraint by the local tadpole condition~\kftor\ examined in detail in Section~\secTransLocal. In particular, requiring no change in the number of space-time filling M2 branes along the transition, we get further constraints on the choice of $[\cB^\flat]$.

%%%%%%
\subsec{Non-Abelian gauge groups and relation to F-theory}
%%%%%%
The topology changing transitions considered in the previous sections proceed via Higgsing of a $U(1)$ factor in the gauge group, with the gauge field arising from the three-form reduced on  the $\IP^1$ fiber over the curve $\Cg$ on the resolved side. This can be generalized in a straightforward way to phase transitions that involve singularities with several intersecting $\IP^1$, leading to non-Abelian gauge groups with various matter representations.\foot{Parallel discussions for the case of Calabi--Yau threefold singularities can be found e.g. in refs.~\refs{\KatzHT,\KlemmKV,\KlemmBJ,\KatzXE,\KatzFH,\KatzEQ,\IntriligatorPQ}}

A comprehensive study of the relevant fourfold singularities in the context of F-theory and twisted 8-dimensional SYM has appeared in refs.~\refs{\BeasleyDC,\BeasleyKW,\DonagiCA}. The essential local geometry is that of an ADE singularity (possibly with monodromy) over a surface $S_G \in X^\sharp$, which is enhanced over a matter curve $\Cg\subset S_G$. For the F-theory compactification on a fourfold $X$ to four dimensions, $X$ has to be elliptically fibered and this is related to M-theory on $X$ by $S^1$ compactification. The existence of an elliptic fibration in the compactification of the normal bundle to $S_G$ restricts the possible gauge and matter content,\foot{As is expected from the fact that the constraints from four-dimensional anomalies are more restrictive then those from $\IZ_2$ anomalies in three dimensions.} but is otherwise inessential to the local analysis of the gauge theory engineered by the local singularity. The results of our M-theory analysis is therefore slightly more general, but directly applicable to the F-theory compactification to four dimensions, if the 3d spectrum fulfills the four-dimensional anomaly constraints and an elliptic fibration exists. In particular the M-theory picture must reproduce the results of refs.~\refs{\BeasleyDC,\BeasleyKW} in this case and we will indeed see explicitly in Section~5 that this is the case for the spectrum obtained from a topological twist in M-theory. 

As discussed above, a topological transition to a deformed manifold $X^\flat$ in the presence of consistently quantized fluxes describes a motion along a flat direction in the parameter space of the non-Abelian space-time gauge theory associated with this local geometry. Alternatively, this can be viewed as a microscopic engineering of a $G$ bundle over the surface $S_G$ in the internal Calabi--Yau space. 

As is clear from the analysis of \refs{\BeasleyDC,\BeasleyKW}, the spectrum and the superpotential couplings will depend very much on the details of a concrete geometry, and in particular on the choice of quantized $G$-flux, which captures topological data of the gauge bundle on $S_G$ in the F-theory picture. Instead of trying to be general we restrict here to illustrate the application of the results of the M-theory analysis on topological transitions in a simple non-Abelian example. 

Our non-Abelian example is a $SU(6)$ gauge theory that repoduces the sextic compactification of Section~4.1\yyy\ as the end point of a chain of topology changing transitions. For the engineering of the $SU(6)$ gauge group, we need an $A_5$ surface singularity with local equation
\eqn\afive{
xy+z^6=0\ .
}
The engineering of global elliptic fibrations with the appropriate singularities is well-understood, but the requirement of $X$ being elliptic leads to slightly more complicated geometries then needed for our purposes. To study the 3d physics associated with the transition it suffices to study the case without elliptic fibration. Since the difference is not essential for the local physics we will anyway often comment on the F-theory picture and heavily borrow from the results of \refs{\BeasleyDC,\BeasleyKW}.\foot{It is self-evident, that the Coulomb branches of the resolved singularities can arise in F-theory only after compactification on $S^1$.}

To make contact with the sextic example, we identify $x$ and $z$ with the homogeneous coordinates $x_5,x_6$ on $\IP^6$ and $y$ with a degree 5 polynomial $p_5(x_i)$ depending only on the other coordinates $x_i$, $i=1,...,4$. Equation \afive\ then describes an $A_5$ singularity over a quintic hypersurface $S_G:\ p_5(x_i)=0$ in $\IP^3(x_1,x_2,x_3,x_4)$ with Hodge numbers 
\eqn\hodgesg{
h^{0,0}(\SZ)=1\,,\quad
h^{1,0}(\SZ)=0\,,\quad  
h^{1,1}(\SZ)=45\,,\quad 
h^{2,0}(\SZ)=4\,,\quad 
\chi(\SZ) = 55\,.
}
More generally, an $A_{k-1}$ singularity over $\SZ$ gives rise to a $G=SU(k)$ gauge theory in 3d with $h^{2,0}=4$ chiral multiplets in the adjoint representation \KatzTH\BeasleyDC. Adding additional monomials allowed for the general sextic in $\IP^6(x_1,...,x_6)$ to \afive\ describes (partial) resolutions of the singularity with $k\leq 5$. As a concrete model we consider a chain of hypersurfaces $X_k$, $k=0,...,6$ that arise as the zero of the polynomial 
\eqn\SecticPk{
P_k=\sum_{a,b}p_{6-a-b}^{(b)}x_5^{b}x_6^{a}\prod_{n=1}^kx_{6+n}^{a+n(b-1)}\,,
}
in the toric ambient spaces $\IP[\Delta_k]$ with homogeneous coordinates $x_i$, $i=1,...,6+k$. The toric ambient spaces $\IP[\Delta_k]$ describe a $k$-fold blow up of $\IP^6$ and can be described as in \refs{\BatyrevHM,\CoxVI} by a series of polyhedra $\Delta_k$, $k=0,\ldots, 6$, specified by the vertices of the dual polyhedra $\Delta_k^* \,=\, \{ \nu_0^* , \ldots \nu_{6+k}^*\}$:
\eqn\SexticPoly{
\hbox{\valign{\vfil#\vfil\cr
\vbox{\offinterlineskip
\halign{\strut~$#$\hfil~\vrule&~\hfil$#$~&~\hfil$#$~&~\hfil$#$~&~\hfil$#$~&~\hfil$#$~~\cr
\multispan6{\strut\hfil\hskip5ex$\Delta_0^*$\hfil}\cr
\noalign{\hrule}
\nu_0^* &  0 &  0 &  0 &  0 &  0 \cr
\nu_1^* & -1 & -1 & -1 & -1 & -1 \cr
\nu_2^* &  1 &  0 &  0 &  0 &  0 \cr
\nu_3^* & 0 &  1 &  0 &  0 &  0 \cr
\nu_4^* & 0 &  0 &  1 &  0 &  0 \cr
\nu_5^* & 0 &  0 &  0 &  1 &  0 \cr
\nu_6^* & 0 &  0 &  0 &  0 &  1 \cr
}}\cr
\vbox{\hbox{\hskip1in}}\cr
\vbox{\offinterlineskip
\halign{\strut~~$#$\hfil~~\vrule&~~\hfil$#$~~&~~\hfil$#$~~&~~\hfil$#$~~&~~\hfil$#$~~&~~\hfil$#$~~\cr
\multispan6{\strut\hfil\hskip7ex Blowup vertices \hfil}\cr
\noalign{\hrule}
\nu_7^* & \-0 & \-0 & \-0 & \-1 & \-1 \cr
\nu_8^* & 0 & 0 & 0 & 2 & 1 \cr
\nu_9^* & 0 & 0 & 0 & 3 & 1 \cr
\nu_{10}^* & 0 & 0 & 0 & 4 & 1 \cr
\nu_{11}^* & 0 & 0 & 0 & 5 & 1 \cr
\nu_{12}^* & 0 & 0 & 0 & 6 & 1 \cr
}}\cr
}}
}
\noindent The vertices $\nu_i^*$ fulfill the relations $\sum l^{a}_i\nu_i^*=0$ with 
\eqn\moric{
l^1=(-6,1,1,1,1,1,1,0^6)\ ,\qquad l^2=(-1,0,0,0,0,1,1,-1,0^5)\ ,
}
and $l^a_i=-2\delta_{a+4,i}+\delta_{a+4,i-1}+\delta_{a+4,i+1}$ for $a=3,...,6$. Appropriate linear combinations of the $l^a$ define a phase in the K\"ahler moduli space of the toric hypersurfaces \refs{\WittenYC,\CoxVI}.

The sections $p^{(l)}_k$ in \SecticPk\ are generic degree $k$ polynomials in the coordinates $x_i$, $i=1,...,4$ and the defining quintic polynomial of $S_G$ is $p_5^{(1)}=0$. The coefficients of the homogeneous monomials in $P_k$ parameterize the complex structure moduli space of the Calabi--Yau fourfolds $X_k$ with independent Hodge numbers $h^{2,1}(X_k)=0$ and 
\eqn\CYHodge{
\hbox{\vbox{\offinterlineskip
\halign{\strut~~\hfil$#$\hfil~~\vrule&~~\hfil$#$~~&~~\hfil$#$~~&~~\hfil$#$~~&~~\hfil$#$~~&\vrule#&~~\hfil$#$~~&\hfil$#$\hfil&\vrule#&~~\hfil$#$\hfil~~&~~\hfil$#$\hfil&~~\vrule~~\hfil$#$~~\hfil\cr
&\multispan4{\hfil~Hodge numbers~\hfil}&&\multispan2{\hfil~Euler characteristic~\hfil}&&\multispan3{~Singularity structure~}\cr
{\rm CY}_4&\hfil h^{1,1}\hfil& \hfil h^{3,1}\hfil & \hfil h^{3,1}_{np}\hfil & \hfil h^{2,2}\hfil && \hfil\chi\hfil &\hfil\mod 24\hfil&&\Cg&g(\Cg)&S_G\cr
\noalign{\hrule}
X_0 & 1 & 426 &(0)& 1\,752  && 2\,610 & 18 && & &\cr
X_1 & 2 & 350 &(0)& 1\,452  && 2\,160 & 0 && A_0 & 76 & \cr
X_2 & 3 & 299 &(4)& 1\,252  && 1\,860 & 12&& A_2 & 51 &A_1 \cr
X_3 & 4 & 268 &(8)& 1\,132  && 1\,680 & 0 &&  A_3 & 31 &A_2\cr
X_4 & 5 & 252 &(12)& 1\,072  && 1\,590 & 6&& A_4 & 16 &A_3 \cr
X_5 & 6 & 246 &(16)& 1\,052  && 1\,560 & 0&& A_5 & 6 &A_4 \cr
X_6 & 6 & 246 &(20)& 1\,052  && 1\,560 & 0&& & &A_5 \cr
}}}}
For each of this partial resolutions with $2\leq k\leq 5$, there is an $A_k$ singularity over $S_G\subset X_{k+1}$ with local equation
\eqn\SexticSing{
x_5p_5^{(1)}+x_6^{k+1}p^{(0)}_{6-(k+1)}\ +\ x_6^{k+2}p^{(0)}_{5-(k+1)}=0\ . 
}
As specified in the rightmost column, this singularity enhances over the genus $g$ matter curve $X_{k+1}\supset \CCg {k+1} :\ p_5^{(1)}=0=p^{(0)}_{6-(k+1)}$ to $A_{k+1}$. The Euler characteristic and genus of the complete intersection curves $\CCg {k+1}$ are
\eqn\Cgenus{ \chi(\CCg {k+1}) \,=\, - 5 (5-k) (6-k) \ , \quad g(\CCg {k+1}) \,=\, 5\cdot {6-k \choose 2} + 1 \ , \quad k=0,\ldots,4 \ . }
The enhancement of the singularity on $\Cg $ gives rise to matter in the fundamental representation of $G$ \refs{\BerglundUY,\KatzXE,\BeasleyDC}. As alluded to above, the local geometry is almost identical to the matter curves representing intersecting 7-branes in an F-theory compactification, except for the absence of a global elliptic fibration. The topological twist and the spectrum for this local geometry has been computed in ref.~\BeasleyDC, and is reproduced by the results reported in Sections~5.3\yyy{} and 5.5\yyy.

In the M-theory compactification on $X_k$, a generic point in the K\"ahler moduli corresponds naively to a Coulomb branch of $G$, where the gauge symmetry is broken to $G=U(1)^{{\rm rk}\,G}$.  Moreover there is a common bare mass for the fundamentals proportional to the volume of the single extra $\IP^1$ in the resolution of the $A_k$ singularity over $\CCg k$, which again represents a Coulomb vev for the $U(1)$ associated with the single mass parameter. Except for a possible obstruction from the flux configuration, the moduli spaces of two manifolds $X_{k+1}$ and $X_{k}$ can be connected by a topology changing transition, where the extra $\IP^1$ over $\CCg {k+1} \subset X_{k+1}$ shrinks, each step giving rise to a local conifold transition of the type described in Section~3\yyy. The change in the Hodge numbers in \CYHodge\ in each step is of the expected form 
\eqn\SexticChangeHN{
\Delta h^{1,1} = +1\,,\qquad 
\Delta h^{3,1} = -g\,,\qquad 
\Delta h^{2,2} = -4(g-1)\,,\qquad 
\Delta \chi = -6(g-1)\,.
}
In F-theory language, transitions of this type describes a process, where a stack of parallel 7-branes is deformed to a set of intersecting branes, which recombine under addition of fluxes \refs{\BeasleyDC,\BraunZM}.

The spectrum of the 3d theory obtained from the local $A_5$ singularity over $\SZ$ in $X_6$ is that of an $SU(6)$ theory with 4 adjoint chiral multiplets. Giving a vev to scalars in the Cartan subalgebra breaks $SU(6)\to U(1)^5$. There are two different types of Coulomb branches depending on whether the scalars $J^a$ in the 3d vector multiplet or the scalars $z^\alpha$ in the chiral multiplet get a vev. The second branch is also available in the F-theory compactification (if the global embedding $X$ would be chosen to be elliptic) and then describes a deformation of parallel 7-branes to intersecting 7-branes \BeasleyDC.  

In the partial resolutions \CYHodge, the Cartan part of the adjoint fields correspond to the non-polynomial deformations $h^{3,1}_{np}$, which are frozen in the hypersurface representation and can not be represented by coefficients in $P_k$ in the given representation of $X_k$ as a hypersurface in $\IP[\Delta_k]$.  The number of non-polynomial deformations $h^{3,1}_{np}$ in \CYHodge\ matches the number $4\cdot {\rm rk}(G)$ of neutral components of the four adjoint hypermultiplets. Indeed the difference in the hypersurface equations for $X_6$ and $X_5$ is that the 4 coefficients of the polynomial $p_1^{(0)}\in \Gamma(K_\SZ)$ are set to zero in $X_6$. Since the adjoint chiral multiplet are sections of the canonical bundle $K_\SZ$, these coefficients should be identified with a vev for the 4 chiral multiplets lying in a $U(1)\subset SU(6)$ subgroup. Thus the moduli of the hypersurface $X_6$ with 20 frozen deformations describes a pure 3d Coulomb-branch $U(1)^5$ with vev's only of the 3d vector multiplet $J^a$, while the transition to $X_5$ describes moving from a pure 3d C-branch to a mixed $U(1)^4\times U(1)_z$ Coulomb-branch, where the subscript $z$ denotes a non-zero vev of a neutral scalar in the chiral multiplet.

The zero locus of the section $p_1^{(0)}$ defines a genus six  curve $\CC_5 \subset X_5$ above which the singularity enhances to $A_5$; the local deformation theory for the genus six case was one of the examples treated in detail in Section~3.7\yyy. M2 Branes wrapping the extra node account for the charged components of the adjoint chiral multiplets with a 3d mass proportional to the vev in the vector multiplet. In the F-theory context, a $SU(6)$ stack of parallel D7-branes is deformed to a $SU(5)$ stack intersected by a single brane \BeasleyDC.

Starting from $X_5$ the partial resolutions are related successively by conifold transitions $X_{k+1}\simeq X^\sharp \ \rightarrow X_{k}\simeq X^\flat$, upon condition of appropriate background flux. Some or all of the Coulomb and/or Higgs branches will be lifted, depending on the choice of consistent $G$-flux and $C$-fields analyzed in Section~3, leading to many components of the $N=1$ deformation space with different spectra and disconnected in the field theory limit. 
For a suitable triangulation the charges for the toric $\IC^*$ actions encoding the Mori cone of $\IP[\Delta_k]$ for $k=5$ are
\eqn\MConeThree{
\hbox{\vbox{\offinterlineskip
\halign{\strut~\hfil$#$\hfil~&\vrule~\hfil$#$\hfil~&\vrule~\hfil$#$\hfil~&~\hfil$#$\hfil~&~\hfil$#$\hfil~&~\hfil$#$\hfil~&~\hfil$#$\hfil~&~\hfil$#$\hfil~&~\hfil$#$\hfil~&~\hfil$#$\hfil~&~\hfil$#$\hfil~&~\hfil$#$\hfil~&~\hfil$#$\hfil~\cr
\IP[\Delta_5]& \,p & x_1 & x_2 & x_3 & x_4 & x_5 & x_6 & x_7 & x_8 & x_9 & x_{10} & x_{11} \cr
\noalign{\hrule}
\om_1 & -1 & 1 & 1 & 1 & 1 & -4 & \-0 & \-0 & \-0 & \-0 & \-0 & \-1 \cr
\om_2 & -1 & 0 & 0 & 0 & 0 & \-1 & \-0 & \-0 & \-0 & \-0 & \-1 & -1 \cr
\om_3 & \-0 & 0 & 0 & 0 & 0 & \-0 & \-0 & \-0 &  \-0 &  \-1 & -2 & \-1 \cr
\om_4 & \-0 & 0 & 0 & 0 & 0 & \-0 & \-0 &  \-0 &  \-1 & -2 & \-1 & \-0 \cr
\om_5 & \-0 & 0 & 0 & 0 & 0 & \-0 & \-0 &  \-1 & -2 & \-1 & \-0 & \-0 \cr
\om_6 & \-0 & 0 & 0 & 0 & 0 & \-0 & \-1 & -2 & \-1 & \-0 & \-0 & \-0 \cr}}}}
For $1\leq k<5$ the toric divisors $D_\ell: \{x_\ell=0\}$ are blown down for $\ell>k+6$ and the charge vectors in each step related by 
$\om_1(X_k)=\om_1(X_{k+1})+\om_2(X_{k+1})$, 
$\om_2(X_{k})=\om_2(X_{k+1})+\om_3(X_{k+1})$,
$\om_{2+\alpha}(X_{k})=\om_{3+\alpha}(X_{k+1})$, $\alpha = 1,...,k-1$. With these definitions, the classes of the surfaces $\Ssharp_{k+1} \subset X_{k+1}$, which are the $\IP^1$-fibrations over the curves $\CCg {k+1}$ \Zfib, are
\eqn\Sk{ \Ssharp_{k+1} \, \simeq \, (4\,D_{k+6}+5\,D_{k+7})\cap D_{k+7} \,\subset\, X_{k+1} \ , \qquad k=0,\ldots, 4 \ . } 

For $k$ even, $\chi=0\,\mod\,24$ on $X_{k+1}$ and a flux in the torsion class $k^\sharp=0$ gives rise to a flat direction for the conifold transition from $X_{k+1}\to X_k$ as shown in Section~3. For $1<k<4$ odd, $\chi\neq 0\mod 24$ on $X_{k+1}$ and a canonical flux solving the local quantization condition is in the torsion class $k^\sharp=\pm(g(\CCg{k+1})-1)$, i.e., 
\eqn\SeriesFlux{ {G^\sharp \over 2\pi} \,=\, \pm {1\over2} [S^\sharp_{k+1}] \ , }
which leads to the twisted superpotential
\eqn\SeriesWpot{ 
  \widetilde W(X_k) \,=\, \ \pm {5(6-k)\over 2} J_{1} J_{2} \pm  {n_k\over 4} J_{2}^2 \ , \quad
  n_k\,=\, 5(6-k)(k-1) \ , }
A full analysis of the different disconnected branches of the $N=1$ deformation space, distinguished by the fluxes consistent with the local quantization condition, is beyond the scope of this exposition. It would be interesting to work out more details for a concrete model with a phenomenological perspective. This will require also a further study of non-generic configurations. E.g. note that if we start with zero flux on $X_5$ and blow down the first $\IP^1$ fiber over $\CCg 5:\ p_5^{(1)}=0=p_1^{(0)}$, the critical points of the superpotential on $X_4$ describe the local singularity 
\eqn\fifo{
X_5\to X_4:\ xy+z^6+p_1z^5\to xy+z^6+p_1z^5+p_2z^4 \ ,
}
with $p_2$ factorized into two linear polynomials as $p_2=t_1.s_1$. On this locus, the genus 16 curve $X_4\supset \CCg 4:\ p_5^{(1)}=0=p_2^{(0)}$ degenerates to two genus 6 curves intersecting each other and the original genus 6 curve. The intersection points may induce further superpotential couplings, as described in \BeasleyDC.

%%%%%%%%%%%%%%%%%%%%%%%%%%%%%%%%%%
\newsec{M-theory phases in the effective $N=2$ three-dimensional field theory}%
\seclab\secFieldTheory
%%%%%%%%%%%%%%%%%%%%%%%%%%%%%%%%%%
%
In this section, we first review aspects of $N=2$ three-dimensional $U(1)$ gauge theories with charged matter fields. We discuss the interplay among global symmetries, phase structures, Chern-Simons terms and parity anomalies of such field theories. By constructing such theories by dimensional reduction of certain five-dimensional field theories with eight supercharges, we associate the resulting three-dimensional field theories with the analyzed conifold transitions in M-theory. While completing this manuscript, the paper \GrimmFX \ appeared, which has a certain overlap with the aspects discussed in this section.

%%%
\subsec{$N=2$ three-dimensional field theory}
%%%
%
We briefly review a few aspects, mostly following the notation of ref.~\AharonyBX. The algebra of three dimensional $N=2$ (four supercharges) admits a $U(1)_R$ symmetry\foot{For 3d, $N$-extended supersymmetry ($2N$ supercharges) it's an $SO(N)_R$ symmetry.} and a real central term $Z$,  $\{ Q_\alpha, \bar Q_{\beta}\} = 2\sigma ^\mu _{\alpha \beta}P_\mu +2i\epsilon _{\alpha \beta}Z$, with all other anticommutators vanishing.  The matter multiplets are in chiral\foot{The terminology is in analogy with 4d, even though there is no chirality in 3d, since there is no analog of $\gamma _5$ and all 3d fermions $\psi _\alpha$ are two-component.  Also similar to 4d, the coupling of the real scalar in the $N=2$ vector multiplet (the $A_4$ component in reducing from 4d) distinguishes between 3d chiral superfield matter in representations ${\bf r}$ vs ${\bf \bar r}$.  So, much as in 4d, we can distinguish between vector-like (real or ${\bf r\oplus \bar r}$) vs chiral matter representations in 3d $N=2$ theories.}  superfields $X$, similar to those of 4d $N=1$, $[\bar Q_\alpha, X]=0$, with CPT conjugate anti-chiral superfields $\bar X$, with $[Q_\alpha, \bar X]=0$.  

Considering, say a $U(1)^r$ Abelian\foot{The non-Abelian case is similar, since the gauge group is anyway broken to the Cartan $U(1)^r$ on the Coulomb branch.  We will briefly remark about the differences for the non-Abelian case.  One difference, for the case of non-Abelian $N=2$ pure Yang-Mills, with no matter, is that instantons generate a non-perturbative, runaway superpotential that lifts the Coulomb branch \AffleckAS.  For non-Abelian gauge theories with matter, instantons have too many fermion zero modes to generate a superpotential, though there can be other non-perturbative effects, see e.g. \AharonyBX.}  gauge theory, with vector multiplets $V^a$, ${a=1,\dots,r}$, one can form associated linear multiplets $\Sigma^a=\epsilon^{\alpha \beta}\bar D_\alpha D_\beta V^a$, with $D^2\Sigma^a=\bar D^2 \Sigma^a=0$, i.e. the same as for a conserved current, $D^2\JC=\bar D^2\JC=0$.  Indeed, abelian gauge fields lead to $U(1)_J$ global conserved currents, $j^\mu = \epsilon ^{\mu \nu \rho}F_{\nu \rho}$, that shifts the scalar dual of the photon, with $\Sigma$ the corresponding superspace conserved current.   The bottom components of $\Sigma^a$ are  the real scalars of the Coulomb branch, $\Sigma^a|=\LM^a$.\foot{Note that we refer to the bottom component of $\Sigma$ by the roman letter $\LM$, whereas we use the calligraphic letter $\JC$ for the conserved currents.}

The lagrangian terms involving the gauge multiplet include 
\eqn\lagterms{{\cal L}_{N=2}\supset \int d^4 \theta \left(\tau _{ab}\Sigma^a \Sigma^b + V^a \JC_a+ {k_{ab} \over 4\pi} \Sigma^a V^b\right) \ ,}
(summing repeated indices)  where $\tau \sim g^{-2}$ give the gauge kinetic terms, and $k_{ab}$ give the $N=2$ supersymmetric Chern-Simons terms:
\eqn\lagtermsii{
  {\cL}_{N=2}\supset -{1\over 4}\tau _{ab}F_{\mu \nu}^a F^{b\,\mu \nu}+\half \tau_{ab}\partial_\mu\LM^a \partial ^\mu\LM^b
    +{k_{ab}\over 4\pi}\epsilon^{\mu \nu\rho}A^a_\mu F^b_{\nu \rho}-{1\over 8\pi ^2}k_{ac}\,k_{bd}\,\tau^{cd}\LM^a\LM^b \ .}
The Chern-Simons terms give masses $\sim k_{ab}$ to the gauge fields, and the last term in \lagtermsii\ give supersymmetry-preserving superpartner masses to the $\LM^a$, lifting the Coulomb branch. The $N=2$ Chern-Simons terms in \lagterms\ can be expressed in terms of the ``twisted superpotential'' $\widetilde W(\Sigma)=\half k_{ab}\Sigma^a \Sigma^b$,\foot{Again, $\widetilde W(\Sigma)$ is not a superpotential in 3d, since $\Sigma$ is real, but reduces in 2d to a superpotential for twisted chiral superfields.} where the term in Lagrangian is
\eqn\lagtermst{{\cal L}_{N=2}\supset \int d^4\theta\, \partial_a \widetilde W(\Sigma)\, V^a \ . }

The $V^a \JC_a$ sum in \lagterms\ can include both dynamical gauge fields coupled to gauge currents, and also background gauge fields coupled to any global currents; for background gauge fields,  $\tau_{ai}\to \infty$, and $\Sigma^i\sim \tilde m$ coupling to a global symmetry gives the real mass term parameters, $\tilde m$.  Fayet-Iliopoulous terms $\int d^4\theta\,\xi_a V^a$ can be regarded as mixed Chern-Simons couplings between the dynamical field $V^a$ and a background field  $\Sigma^i$, $\xi_a\sim k_{ai}\Sigma^i$.   The central term $Z$ can get contributions from real mass terms or FI terms, 
\eqn\zis{Z\,=\,\sum_i q_i \tilde m^i,}
where $q_i$ is the charge of the field under a global $U(1)_i$ symmetry, $\tilde m^i$ is the real mass that can be regarded as a $U(1)_i$ background field, $\tilde m^i=\Sigma^i|$, and the sum includes $U(1)_J$, with $m_J=\xi$ the FI parameter.

The $N=2$ supersymmetric Chern-Simons term coefficients $k_{ab}$ in \lagterms\ to \lagtermst\ include both classical and one-loop contributions,
\eqn\ktotal{k^{total}_{ab}\,=\,k_{ab}^{cl}+{1\over2} \sum _f (q_f)_a (q_f)_b \,{\rm sign}(M_f)\ ,}
where $f$ runs over all fermions, $(q_f)_a$ is its charge under $U(1)_a$ and $M_f$ is its real mass, including the contribution from the $\LM^a$ expectation values on the Coulomb branch,  $M_f=\tilde m_f +\sum_a (q_f)_a \LM^a$. There is a quantization condition $k_{ab}^{total}\in \IZ$ or $\half \IZ$.\foot{For non-Abelian groups, gauge invariance quantizes $k^{total}$, independent of any details.  For Abelian groups, the quantization relies on having compact $U(1)s$, and considering the theory on a compact spacetime $X_3$, with the normalization of the gauge field specified via  $\oint A \in 2\pi \IZ$.  The quantization condition depends on $X_3$: if $X_3$ is restricted to be spin, as is required for compactification of $M$-theory (or having fermions for that matter), one gets $k\in \half \IZ$; the half-integer case is is referred to as Abelian spin-Chern-Simons \refs{\BelovZE,\KapustinHK}. This is similar to  compactification of M-theory to 5d, where the spin restriction on $X_5$ gives  $c_{total} \in \IZ$ (rather than $c\in 6\IZ$).}  

If the charged matter spectrum representation is vector-like, with all matter chiral superfields in real or ${\bf r\oplus \bar r}$ representations, the induced Chern-Simons term vanishes.  For example, for $U(1)$ with chiral matter of charge $+1$ and $-1$, the conjugate fields have opposite ${\rm sign}(M_f)$ and make canceling contributions to \ktotal. This can also be understood in terms of the non-renormalization theorems given in ref.~\AharonyBX\ related to the non-coupling of chiral vs linear multiplets: the Chern-Simons term cannot depend on chiral multiplets, so it cannot depend on complex masses which is a background chiral multiplet.  So vector-like matter can be decoupled with arbitrarily large complex mass $m_C$, with \ktotal\ unaffected.  In the non-Abelian case, the induced Chern-Simons term on the Coulomb branch is related to the cubic Casimir, essentially $k=k^{cl}+\half \sum _f d_3({\bf r_f})$ \AharonyBX, where the sum runs over all chiral superfields.  See ref.~\PoppitzHR\ for a detailed discussion and the more precise statement.   The upshot is a connection between the one-loop induced Chern-Simons terms of the 3d theory and the 4d ${\rm Tr}\,F^3$ gauge anomaly: the 3d induced Chern-Simons term vanishes precisely if the matter content would be gauge anomaly free in 4d.

%%%
\subsec{Effective $N=2$ field theory for M-theory on smooth fourfolds with flux}
%%%
%
M-theory compactified on a smooth Calabi--Yau fourfold (with or without flux) yields at low energies three-dimensional $N=2$ -- i.e., four supercharges --  supergravity coupled to matter and vector multiplets.  As shown in refs.~\refs{\BeckerGJ,\GukovYA,\HaackJZ}, reducing eleven-dimensional supergravity, along with the additional eleven-dimensional M-theory $C\wedge G\wedge G$ and $C\wedge X_8$ and purely gravitational interactions, leads to the $N=2$ three-dimensional supergravity Lagrangian for the $U(1)^{h^{1,1}}$ Coulomb branch moduli $\LM^a$, with gauge kinetic terms and Chern-Simons terms as in \lagterms.  In particular, three-dimensional Chern-Simons terms arise from the reduction of the M-theory eleven-dimensional Chern-Simons term $C\wedge G\wedge G$ on fourfold compactifications with non-trivial four-form background fluxes $G$ \refs{\MayrSH,\HaackJZ}
\eqn\CStermflux{ k_{ab} = \partial_a\partial_b \widetilde W=\int_X {G \over 2\pi} \wedge \om_a\wedge\om_b \ , }
where $\om_a$ is a basis for $H^{1,1}(X,\IZ)$.
The quantization condition is $k_{ab}\in Z$ or $k_{ab}\in Z+\half$ where the latter possibility is because, as discussed in the above footnote, the 3d spacetime is necessarily a spin manifold.
In comparing with the lagrangian derived in  ref.~\HaackJZ, we can restore the three-dimensional Planck mass scale $M_3$,\foot{The canonical scaling dimensions are assigned as $\Delta [z^\alpha]=\half$ the free scalar field dimension, the vector multiplet has $\Delta [\LM^a]=\Delta [A_\mu ^a]=1$, $\Delta [\tau_{ab}]=\Delta [1/e^2]=-1$, and $\Delta [\widetilde W]=2$.}  and consider the low-energy, gravity decoupling limit, $M_3\to \infty $.  For example ${\cal L}_{3d}^{SUGRA}\subset {M_3 \over 2 } R - {1\over M _3}\widetilde W^2$, which decouple in this limit.  The remaining terms are the twisted superpotential Chern-Simons terms \CStermflux.  In addition to the terms in  ref.~\HaackJZ, we have the charged matter contributions, from wrapped M2 branes, and their superpotential interactions.

\def\neut{\varphi}
%%%%%
\subsec{Singularities, charged matter, and the $5d\to 3d$ reduction}%
%%%%%
The smooth-fourfold theory of the previous subsection is a three dimensional abelian gauge theory, without charged matter.  Geometric singularities are needed to obtain non-Abelian groups or charged matter.  In our construction of the fourfold as coming from a threefold that is fibered over a genus $g$ curve ${\cal C}$, we can take the shrinking $\IP^1$ in the threefold fiber to be very small compared with the curve ${\cal C}$, i.e. writing the K\"ahler class $J=J_F+J_C$ for the sizes of the threefold fiber and the curve ${\cal C}$, respectively, we can consider the limit $J_F\ll J_C$, which leads to the mass hierarchy $M_{11d\to 5d}\gg M_{5d\to 3d}$.  The  three-dimensional matter spectrum can then be analyzed by first reducing M-theory on the threefold, which yields a low-energy five dimensional $N=1$ (eight supercharge) theory.  The five dimensional gauge theory is next reduced (fibered) over the curve ${\cal C}$ to reduce to the three-dimensional $N=2$ (four supercharge) theory.  

Let us briefly outline how the latter reduction works, for a general five-dimensional $N=1$ gauge theory, reduced to three dimensions on a genus $g$ curve $\Cg$.  The five-dimensional theory in flat spacetime would have $SO(4,1)\times SU(2)_R$ isometry group, and reduction on $C$ breaks $SU(4,1)\to SO(2,1)\times U(1)_L$, where the eight supercharges of five-dimensions transform in the $({\bf 2}_s, \pm \half,{\bf 2})$ of $SO(2,1)\times U(1)_L\times SU(2)_R$.  Four of these supercharges are preserved if the theory is twisted, modifying the $U(1)_L$ factor of the Lorentz group to $U(1)'_L$, with generator 
\eqn\Lgennew{ J'_L \,=\, J_L + J_3 \  }
where $J_L$ is the 2d $U(1)_L$ generator, $J_3=\half \sigma _3$ is the Cartan generator of $SU(2)_R$, and the preserved supercharges have $J'_L=0$.  The four preserved supercharges have a three-dimensional $U(1)_R$ symmetry, with generator given by $R^{3d}=2J_L$, so the four supercharges transform under $SO(2,1)\times U(1)_R$ as $({\bf 2}_s, \pm 1)$.    

The five-dimensional $N=1$ (eight supercharge) fields are 
\eqn\vhfive{
  \hbox{5d-vector}= \pmatrix{&A_\mu &\cr \lambda && \psi \cr &\phi _R&}\ , \qquad 
  \hbox{5d-hyper}=\pmatrix{&\psi _{\qp} &\cr \qp& & \qt^\dagger \cr &\psi ^\dagger _{\qt}&} \ ,}
where $SU(2)_R$ acts on the rows, with $J_3=\pm \half$.    All 5d spinors in \vhfive\ reduce as $\psi ^{d=5}\to (\psi ^{d=3}_\alpha \otimes \psi ^{d=2}_{+})\oplus (\psi ^{' d=3}_\alpha\otimes \psi ^{d=2}_{-})$, where $\psi ^{d=3}_{\alpha =1,2}$ is a two-component 3d spinor and $\psi ^{d=2}_{\pm}$ is a 2d spinor of $U(1)_L$ Lorentz charge $J_L=\pm \half$.  

Consider first reducing the 5d-vector multiplet in \vhfive\ on $\Cg$.  Collecting the fields according to their $U(1)_{L'}$ spin \Lgennew, there are fields with $J'_L=0$, which assemble into a 3d $N=2$ vector multiplet.  There are also fields with $J'_L=\pm 1$, which assemble into a 3d $N=2$, adjoint valued chiral multiplet.    Reducing on $\Cg$, the $U(1)_{L'}=0$ fields yield  a massless 3d, $N=2$ vector multiplet, from the constant mode on $\Cg$ (along with a massive tower from the other modes of the laplacian on $\Cg$).  The $J'_L=\pm 1$ fields are 1-forms on the curve $\Cg$, e.g. $A_{z,\bar z}=A_4\pm i A_5$, and thus the $\Cg$ laplacian zero modes are given by the $g$ holomorphic and $g$ anti-holomorphic  1-forms on $\Cg$.  We thus obtain, in addition to the 3d $N=2$ massless vector multiplet, $g$ additional massless, 3d $N=2$ chiral multiplets, $\neut^{i=1\dots g}$, in the adjoint representation of the gauge group $G$.    The scalar components of these $g$ chiral multiplets  $\neut^i$ come from the $g$ holomorphic 1-form $A_z$ components of the five dimensional gauge field on $\Cg$.\foot{Indeed, the $2g$ real scalars in these multiplets are the $2g$ Wilson loops $\oint _{\alpha _n}A $, where $\alpha _n$ are the $2g$ one-cycles associated with the four-cycles $A_n$ in Fig. 1.} The coupling of $A_z$ to charged matter implies that $\neut^i$ have superpotential couplings to charged matter in three dimensions.  For $g=1$, the supersymmetry is enhanced to three dimensional $N=4$ and $\Phi$ is the adjoint chiral superfield of the $N=4$ vector multiplet.   

Now consider reducing a five dimensional matter hypermultiplets in \vhfive\ on $\Cg$.  Collecting the fields according to their $U(1)_{L'}$ spin, we find a 3d $ N=2$ chiral superfield, $\qp$, in some representation ${\bf r}$ of the gauge group, with $U(1)_{L'}$ spin $J'_L=+\half$, and a 3d $N=2$ chiral superfield $\qt$, in conjugate representation ${\bf \overline r}$ of the gauge group, with $U(1)_{L'}$ spin $J'_L=\half$. Upon reducing on $\Cg$, we get massless 3d chiral superfields from the zero modes of the two-dimensional Dirac operator on $\Cg$ with $U(1)_{L'}$ spin $\pm \half$.  The five-dimensional hypermultiplet  thus reduces to massless three-dimensional chiral multiplets as:
\eqn\hyperred{({\rm 5d-hyper})\to \qp^{f=1\dots N_+}\oplus \qt^{\tilde f=1\dots N_-},}
where $N_\pm$ are the numbers of $U(1)_L$ spin $\pm \half$ fermion zero modes of the two-dimensional Dirac operator in representation $\bf r$:
\eqn\zeromodes{\slashedD _2\chi _+^{f=1\dots N_+}=0, \qquad \slashedD _2\chi _-^{\tilde f=1\dots N_-}=0.}
For example, for a $U(1)$ gauge group and a 5d hypermultiplet of charge $1$, the reduction on a genus $g$ curve $\Cg$ yields the three-dimensional spectrum given by:
\eqn\ThreeFields{
\hbox{\vbox{\offinterlineskip
\halign{\strut\vrule~~\hfil#\hfil~~&\vrule~~\hfil#\hfil~~\vrule&~~\hfil#\hfil~~\vrule&~~\hfil#\hfil~~\vrule&~~\hfil#\hfil~~\vrule&~~\hfil#\hfil~~\vrule\cr
\noalign{\hrule}
3d $N=2$ multiplet & 3d field & $SO(2,1)$ & $U(1)_g$ & $U(1)_R$ \cr
\noalign{\hrule}
Vector multiplet $V$
& $\phi $ & ${\bf 1}$ & $0$ & $\-0$ \cr
&$\lambda_0$ & $\ {\bf 2}_s$ & $0$ & $+1$ \cr
&$\lambda_0'$ & $\ {\bf 2}_s$ & $0$ & $-1$ \cr
&$A_\mu$ & ${\bf 3}$ & $0$ & $\-0$ \cr
\noalign{\hrule}
neutral chiral
& $\neut^i=A_z^i$ & ${\bf 1}$ & $0$ & $\-2$ \cr
multiplets $\neut^i$
& $\lambda_z^i$ & $\ {\bf 2}_s$ & $0$ & $\-1$ \cr
$i=1,\ldots,g$ 
& $\lambda_{\bar z}^i$ & $\ {\bf 2}_s$ & $0$ & $-1$ \cr
\noalign{\hrule}
charged chiral
& $q^f_+$ & ${\bf 1}$ & $+1$ & $\-0$ \cr
multiplets $q^{f}_+$
& $\psi _{q_+}^f$ & $\ {\bf 2}_s$ & $+1$ & $-1$ \cr
$f=1,\ldots,N_{+}$
& ${\psi'}^{f}_{q_+}$ & $\ {\bf 2}_s$ & $+1$ & $\-1$ \cr
\noalign{\hrule}
charged chiral
& $q ^{\tilde f}_-$ & ${\bf 1}$ & $-1$ & $\-0$ \cr
multiplets $q^{\tilde f}_-$
& $\psi _{q_-}^{\tilde f}$ & $\ {\bf 2}_s$ & $-1$ & $-1$ \cr
$\tilde f=1,\ldots,N_{-}$
& ${\psi'}^{\tilde f}_{q_-}$ & $\ {\bf 2}_s$ & $-1$ & $\-1$ \cr
\noalign{\hrule}
}}}}

The multiplicities $N_\pm$ of the charged chiral multiplets are governed by the zero-modes of the two-dimensional Dirac operator on $\Cg$ twisted by the background flux $F_\Cg$. Note also  that the gauge covariant derivatives in \zeromodes\ contain the gauge connection on $\Cg$, so the fermion zero modes are affected by the expectation values of the $g$, gauge adjoint chiral multiplets $\neut ^{i=1\dots g}$ that came from the 5d vector multiplet.  Non-zero expectation values of the $\neut ^{i=1\dots g}$ can give 3d complex masses to pairs $\qp^f$, $\qt^{\tilde f}$.  In terms of the 3d gauge theory obtained by dimensional reduction on $\Cg$, this effect is accounted for by a superpotential,
\eqn\Wvm{W\,=\,\sum _{i=1}^g \sum _{f=1}^{N_+}\sum _{\tilde f=1}^{N_-}c_{i f\tilde f}\,\Tr \  \qp^f \neut ^i \qt^{\tilde f}\ }
with $\Tr$ over the gauge indices.  For the case of $g=1$, where supersymmetry is enhanced, this is the expected superpotential of 3d $N=4$ supersymmetric gauge theories.  The constants $c_{if\tilde f}$ can be determined by a topological calculation (independent of the metrics), as we will exhibit shortly in the connection with M-theory.   The 3d complex masses obtained from \Wvm\ for non-zero $\vev{\neut ^i}$ affects the phase structure of the 3d field theory, for example lifting Higgs branch moduli.

 While the solutions \zeromodes\ and the numbers $N_\pm$ depend on the $\neut ^i$ moduli, the difference $N_+-N_-$ is the topological index that's independent of the $\neut ^i$ moduli (since $\vev{\varphi ^i}$ gives complex masses to pairs, $\qp$ and $\qt$ in pairs).  This index is that of the two-dimensional Dirac operator on $\Cg$ twisted by the background flux $F_\Cg$.  In particular, for the case of a $U(1)$ gauge theory, 
\eqn\indextwo{{\rm index}({\slashedD}_2)=N_+-N_-=\int _\Cg {F_\Cg\over 2\pi}=n(1-g) \ .}
The $F_\Cg$ flux in \indextwo\ corresponds to a magnetic monopole and is thus quantized (see e.g. section 14.4 in ref.~\GreenMN) as $A_\alpha =\half n \omega _\alpha$, where $\omega _\alpha$ is the $U(1)_{L}$ spin connection on $\Cg$ and $n$ is an integer unit of flux.  For $g=1$ there is enhanced supersymmetry and the theory is necessarily vector-like, fitting with the vanishing index \indextwo.  For $g\neq 1$ and $n\neq 0$ flux units, the three-dimensional theory has chiral matter, $N_+\neq N_-$.

%%%%%%%%%%%%%%%%%
\subsec{The dynamics of 3d $U(1)$ gauge theories for general matter fields of charge $\pm 1$}
%%%%%%%%%%%%%%%%%%%
The analysis in \AharonyBX\ focused on the vectorlike matter case, $N_+=N_-=N_f$, noting the branched moduli space of vacua, with the Higgs branch, and two distinct quantum Coulomb branches, all meeting at the origin.  We here generalize this to allow for $N_+\neq N_-$.  

Consider a  $U(1)$ gauge theory with $N_+$ chiral superfields $\qp^{f=1\dots N_+}$ of charge $+1$, and $N_-$ chiral superfields $\qt^{\tilde f=1\dots N_-}$ of charge $-1$.   The fields and global symmetries are
\eqn\nfgenui{
\matrix{  
\quad     & U(1)_R & U(1)_J& U(1)_A & SU(N_+) & SU(N_-)  \cr
          & & & &&\cr 
\qp & 0 & 0 & 1 & {\bf N_+} & {\bf 1}\cr 
\qt & 0 & 0 & 1 & {\bf 1} & {\bf \overline N_-} \cr	
M & 0 & 0 &2  & {\bf N_+} & {\bf \overline N_-} \cr
V_\pm & N_f & \pm 1  & -N_f &{\bf 1} &\bf 1 \cr}} 
where $M^{f\tilde g}=\qp^f \qt^{\tilde g}$ are the $N_+\times N_-$ gauge invariant mesons, and $V_\pm$ are chiral superfields labeling the Coulomb branch, which can be obtained from $\Sigma$ by dualizing the gauge field to a compact, real scalar.  The $U(1)_R$ is an R-symmetry and we could just as well have chosen a different basis of the $U(1)s$, e.g. $\widetilde U(1)_R=U(1)_R+rU(1)_A$.  The charges of $V_\pm$ under the global symmetries follow from a one-loop diagram coupling the global currents to the gauge fields \AharonyBX, and here $N_f\equiv \half (N_++N_-)$.  

For $N_+N_-\neq 0$, there is a $N_++N_--1$ complex dimensional Higgs branch moduli space of classical vacua when $W_{tree}=0$.  The Higgs branch can be parameterized by expectation values of the $N_-N_+$ meson gauge invariants, subject to their classical constraint coming from $M^{f\tilde g}=\qp^f \qt^{\tilde g}$:
\eqn\mesons{{\cal M}_{\rm Higgs}:\  \vev{M^{f\tilde g}}  \qquad \hbox{with}\qquad {\rm rank}(M)=1.}
The case $N_+=N_-$ was discussed in \refs{\AharonyBX, \deBoerKR}.  As in that case, the moduli space generally separates into three branches: The Higgs branch \mesons\ with $\vev{\LM}=0$, and two distinct Coulomb branches, Coulomb$_{\pm}$ with $\vev{M}=0$, and ${\rm sign}(\LM)=\pm 1$.  The quantum Coulomb branches are parameterized by the chiral superfields $V_\pm$ in \nfgenui. 

The $\qp$ and $\qt$ fields get real masses on the Coulomb branch, $\tilde m_+ +\LM$ and $\tilde m_--\LM$, respectively, where $\tilde m_{\pm}$ are in the adjoint of $SU(N_\pm)\times U(1)_A$.  For vanishing real masses, the Chern-Simons term on the branch Coulomb$_{\pm}$ is given by \ktotal\ to be
\eqn\kis{k_\pm^{total}=k^{cl}\pm {1\over2}(N_+-N_{-})} 
with $\pm \half (N_+-N_{-})$ the quantum contribution from integrating out the massive fields $\qp$ and $\qt$ fields.   When $N_+\neq N_-$, generally $k^{total}\neq 0$, lifting both Coulomb branches.  It is possible to tune $k^{cl}$ to leave one of the two Coulomb branches unlifted.  Note that $k^{cl}$ cancels if we consider the difference of the Chern-Simons coefficient on the two Coulomb branches, which is given by the topological index:
\eqn\kdiff{k^{total}_+-k^{total}_-=N_+-N_-.}

Recall first the case $N_+=N_-=N_f>0$, with $k_{cl}=0$ so the two Coulomb branches are unlifted and meet the Higgs branch at the origin.  The theory at the origin flows to an interacting SCFT \AharonyBX\ that can be described by a dual effective theory of the gauge invariant moduli fields in \nfgenui, with   
\eqn\wui{W=-N_f(V_+V_-\det M)^{1/N_f}.}
For $N_f=1$ the dual theory \wui\ is non-singular and provides a complete dual description of the low-energy theory and interacting SCFT at the origin.  For $N_f>1$, the superpotential \wui\ is singular, corresponding to the fact that additional degrees of freedom are needed to describe the interacting SCFT at the origin.   

SCFTs are generic in 3d, so anything pointing toward additional degrees of freedom beyond the moduli fields can be regarded as some evidence for an interacting SCFT.  In particular, singular moduli spaces or moduli spaces with branches are some general evidence pointing toward an interacting SCFT at the singularity.  

Let us now discuss the $N_+\neq N_-$ cases.  For $k_{cl}=0$, the Coulomb branches are lifted by the Chern-Simons terms $k_\pm ^{total}\neq 0$, so the moduli space consists only of the Higgs branch \mesons.  Whenever both  $N_+>1$ and $N_->1$, the Higgs branch \mesons\ is singular at the origin (the classical singularity can not be smoothed by quantum effects), which suggests an interacting SCFT there.  

On the other hand,  for $N_+>1$ and $N_-=1$, the $N_+$ dimensional Higgs branch is smoothly parameterized by the mesons $M^i=\qp^i\qt$, so it is possible that the theory with $k^{cl}=0$, and hence lifted Coulomb branches for $N_+\neq 1$, is an IR free smoothly confined theory of the mesons $M^i$, without additional SCFT degrees of freedom at the origin.   A weak test of such a scenario is the $\IZ_2$ parity anomaly matching \AharonyBX\ analog of {}'t Hooft anomaly matching for the global symmetries.  The microscopic fields ($\qp^i$ and $\qt$ and the $U(1)$ multiplet) in \nfgenui\ give $k_{RR}=\half N_+$, $k_{RA}=k_{AA}=\half (N_++1)$ mod integers.  The $M_i$ low energy fields give $k_{RR}=\half N_+$, $k_{RA}=0$, $k_{AA}=0$ mod integers.  So the parity anomalies involving $U(1)_A$ do not match for $N_+$ even, suggesting again an interacting SCFT at the origin in that case.  Only $N_+$ odd and $N_-=1$ might have an IR free theory of mesons, rather than a SCFT.    

If $N_+\neq N_-$ and $k^{cl}$ is tuned so that $k_+^{total}$ or $k_-^{total}$ in \kis\ vanishes, then there is an unlifted Coulomb branch, intersecting the Higgs branch \mesons\ at the origin, and there again we expect an interacting SCFT at the intersection point on the moduli space.  

%%%
\subsec{Comparing M-theory conifold transitions with 3d field theory phase structure}
%%%
%
Reducing M-theory on the ruled surface $\Ssharp$ -- as analyzed in Section~\secTransLocal\ -- actually leads to {\it two} three-dimensional $U(1)$ gauge fields, $U(1)_F$ and $U(1)_C$, coming from reducing $C_3$ on either $\IP^1$ or $\Cg$ in \Zfib.  In terms of the $11\to 5\to 3$ reduction, $C^{11d} \to A^{5d}\wedge J_F+C^{5d}\to  A^{3d}_F \wedge J_F+A^{3d}_C\wedge J_C$.  The $U(1)_F$ gauge field $A^{3d}_F$ comes from the five-dimensional gauge field $A^{5d}$, while $U(1)_C$ comes from the five-dimensional $C^{5d}$ 3-form gauge field.  We are interested in the $J_F\ll J_C$ limit where the interesting dynamics is in the $U(1)_F$ gauge theory, and we can take a low-energy limit where the $U(1)_C$ gauge theory essentially decouples, together with the gravity multiplet (it comes from $C^{5d}$, in the five-dimensional gravity multiplet). The point is that $U(1)_F$ has light charged matter, from M2 branes wrapping the small $\IP^1$, whereas all matter charged under $U(1)_C$ is much heavier for $J_C\gg J_F$, as it gets a large real mass $\tilde m\sim J_{C}$ on the Coulomb branch. It is in this limit that we make contact with the three-dimensional field theory associated to the spectrum \ThreeFields. Let us know discuss how this correspondence between the M-theory geometry and the discussed 3d field theory comes about.

As we discussed above, the numbers $N_\pm$ of massless 3d chiral superfields, coming from the solutions of the Dirac operator zero modes \zeromodes, are determined by the number of flux units, with index given by  \indextwo.  Lifting the 5d gauge theory results to $M$-theory on $\Ssharp$, the background flux $F_\Cg$ arises from integrating out the $\IP^1$-fibers.  We therefore identify the index~\indextwo\ with the torsion classes in eq.~\intFluxconsti
\eqn\indexone{
  \int_{\Ssharp} {G^\sharp\over 2\pi}\,=\,\int_{\Cg}{F_\Cg\over 2\pi} \quad \Rightarrow \quad 
  k^\sharp = k^\flat = N_+ - N_- = {\rm index}(\slashedD_2) \ . }
The moduli dependence of zero modes \zeromodes\ enters in the M-theory transition through the dynamical obstructions induced form the flux superpotential \FluxPot. In particular, the deformation sections \DefFactor, which are the global sections of the line bundles $\cE_\pm$ in \CritCv, are in one-to-one correspondence with the zero mode structure of the Dirac operators \zeromodes. Identifying the $U(1)$ line bundle $\cL$ with the line bundle $\cL$ in eq.~\DefBL\ arising in the M-theory transition and using the previous definitions $\cE_+=K^{1/2}_\Cg\otimes\cL$ and $\cE_+=K^{1/2}_\Cg\otimes\cL^*$ in eq.~\CritCv , we get a 1-1 correspondence
$$
  H^0(\cC,\cE_\pm) \ {\buildrel1:1\over\longleftrightarrow} \ \{ \,\chi_\pm^{f=1,\ldots,N_\pm}\, \} \ ,
$$  
which implies for the multiplicities $N_+$ and $N_-$ of the charged chiral fields $q_+$ and $q_-$
$$
  N_+ \,=\, h^0(\Cg,\cE_+) \ , \qquad N_- \,=\, h^1(\Cg,\cE_-) \ . 
$$

As discussed in the previous sections, interacting SCFTs are generic at the origin of 3d $N=2$ gauge theories with non-zero matter content, and in particular occur at the transition point in the moduli space between Higgs and Coulomb branches.  The interacting SCFT at the transition point has additional degrees of freedom.  In the M-theory geometry description, M2 branes give a natural source for the additional degrees of freedom.  Indeed, there are tensionless domain walls located at the singularity, which is another tell-tale sign of an interacting SCFT, where the number of M2 branes can change. This is because at the origin of the Higgs branch the $SU(N_+)\times SU(N_-)$ flavor symmetry is restored, which implies that there are vanishing cycles supporting tensionless domain walls with $\delta M\neq 0$ computed by eq.~\deltaM.

If the 3d theory is related to a 4d theory by a circle compactification, we recover the result ref.~\BeasleyDC\ in the four-dimensional theory from F-theory compactification on the fourfold.  As in that case, the index of the charged chiral matter fields $q_+$ and $q_-$ can be written, using 
Serre duality and the Riemann-Roch theorem, as 
\eqn\IndexMatter{\eqalign{
 N_{+} - N_{-}\,&=\, h^0(\Cg,K^{1/2}_\Cg\otimes\cL) - h^1(\Cg,K^{1/2}_\Cg\otimes\cL) \cr
 \,&=\,(1-g) + \int_\Cg c_1(\cL \otimes K^{1/2}_\Cg) \,=\,  \int_\Cg {F_\Cg \over 2\pi} \ . }}
The latter equality in the second line of \IndexMatter\ can again be directly understood as the statement of the index theorem for the two-dimensional twisted Dirac operator \indextwo.

The spectrum obtained form the dimensional reduction of the 5d multiplets \vhfive\ upon $\Cg$ and their associated geometric sections are summarized in the following table:
\eqn\FiveDCharges{
\hbox{\vbox{\offinterlineskip
\halign{\strut\vrule~#\hfil~\vrule&~\hfil#\hfil~\vrule&~\hfil#\hfil~\vrule&~\hfil#\hfil~\vrule&~\hfil#\hfil~\vrule&~\hfil#\hfil~\vrule&~\hfil#\hfil~\vrule\cr
\noalign{\hrule}
&fields& $SO(2,1)$ & $U(1)_L$ & $U(1)_g$ & $SU(2)_R$ &  sections of \cr
\noalign{\hrule}
Supercharge & $(Q,Q')$ & $\ {\bf 2}_s$ & $+1/2$ & $0$ & ${\bf 2}$ & $K^{1/2}_\Cg$ \cr
\noalign{\hrule}
Vector multiplet 
& $\phi$ & ${\bf 1}$ & $0$ & $0$ & ${\bf 1}$ & $\cO_{\Cg}$ \cr
& $(\lambda,\psi)$ & $\ {\bf 2}_s$ & $+1/2$ & $0$ & ${\bf 2}$ & $K^{1/2}_\Cg$ \cr
& $A_\mu$ & ${\bf 3}$ & $\-0$ & $0$ & ${\bf 1}$ & $\cO_\Cg$ \cr
& $A_z$ & ${\bf 1}$ & $+1$ & $0$ & ${\bf 1}$ & $K_\Cg$ \cr 
\noalign{\hrule}
Hypermultiplet
& $(\qp,\qt^\dagger)$ & ${\bf 1}$ & $0$ & $+1$ & ${\bf 2}$ & $\cL$ \cr
& $\psi_{\qp}$ & $\ {\bf 2}_s$ & $+1/2$ & $+1$ & ${\bf 1}$ & $K^{1/2}_\Cg\otimes\cL$ \cr
& $\psi_{\qt}$ & $\ {\bf 2}_s$ & $+1/2$ & $-1$ & ${\bf 1}$ & $K^{1/2}_\Cg\otimes\cL^*$ \cr
\noalign{\hrule}
}}}}
Here $(Q,Q')$ is the pair of pseudo-real spinorial supercharges of five-dimensional $N=1$ supersymmetry. Performing the topological twist \Lgennew, which corresponds to tensoring the sections of the fields in table~\FiveDCharges\ with $K^{J_3}_\Cg$ according to their quantum number of the Cartan generator ${J}_3$ of $SU(2)_R$, we obtain the 3d $N=2$ supercharges with $U(1)_{L'}$ spin $J'_3=0$ and arrive at the 3d spectrum of table~\ThreeFields.

In the geometry, the coefficients $c_{i{f\tilde f}}$ in \Wvm\ are determined by the integrals 
\eqn\Yukawa{ c_{i\,f\tilde f}=\int_{\Cg} (\eps_{+}^f \,\eps_{-}^{\tilde f}) \wedge  \bar\mu^i \ , }
where $\eps_\pm$ are the sections of $\cE_\pm$ associated with the matter fields, and $\bar\mu^i$ the $(0,1)$-form for the $i$-th Wilson line, see eq.~(4.105) in ref.~\BeasleyDC.

With these identifications at hand we can now compare the phase structure of the discussed 3d field theory with the local M-theory geometries of Section \secTransLocal. The two resolved phases $\Xlsharpi1$ and $\Xlsharpi2$ map to the two Coulomb branches with ${\rm sign}(\phi)>0$ and ${\rm sign}(\phi)<0$. Indeed, as computed explicitly in Appendix~B\yyy, the flop transition~\PhaseDiag\ induces the discontinuous jump in the Chern-Simons term according to the flux-induced twisted superpotentials integrated over the two flopped volumes \VolS\ and \VolSiii. This matches the expected jump as induced form the chiral spectrum \ktotal, in agreement with \kdiff 
$$
  \Delta k\,=\, {\rm index}(\slashedD_2) \,=\, N_+ - N_- \ .
$$
In the context of both the M-theory geometry and the associated 3d field theory, the presence of non-vanishing Chern-Simons couplings from twisted superpotentials lift those branches geometrically in the former and field theoretically the latter description.

The deformed M-theory geometry $\Xlflat$ is identified with the Higgs branch of the 3d field theory. The dimension of this Higgs branch~\mesons\ agrees with the dimension of the unobstructed deformation space \FlatDef. In particular the factorized deformations~$\eps=\eps_+\eps_-$ in eq.~\DefFactor\ get mapped to the gauge invariant mesonic operators $M^{f\tilde g}$.

Let us emphasize that this correspondence holds not only on the level of multiplicities and of dimensionality of moduli spaces, but also on the level of moduli spaces itself. In the local M-theory transition the moduli of the background fluxes $F_\Cg$ are mapped to the moduli of the line bundle \DefBL, while both in the 5d-to-3d reduction and in the geometric M-theory conifold transition, the moduli of the curve $\Cg$ enter through their dependence on the structure of global sections. The underlying reason for this correspondence is that the Albanese map of the ruled surface $\Ssharp$ gets identified with the Abel Jacobi map of the curve $\Cg$ \GriffithsPAG, which in turn encods both the dynamically unobstructed deformation directions in Section \secTransLocal\ and the field-theoretic spectrum \ThreeFields. 

Accordingly, as discussed in Section~3\yyy, non-zero $C$-field backgrounds correspond to non-zero Wilson lines $\neut^i$ and also lift massless fields via the couplings \Wvm,\Yukawa. If $h^{2,1}(\Xs)=h^{2,1}(\Xf)=0$, all the fields $\neut^i$ are non-dynamical and their expectation values are fixed by the global embedding geometry. A very explicit example is provided by the hyperelliptic cases studied in Section~3.7\yyy, where a choice of half-integer $C$-field values corresponds to a choice of a particular spin structure $K^{1/2}_\Cg$. The latter will be fixed in a global embedding and in turn determine the actual number of holomorphic sections/massless fields given in eqs~\SpinStOdd,\SpinStEven. Similarly, the fields $\neut^i$ associated to $(2,1)$-forms participating in the M-theory transition (c.f.~eq.~\RelTopHodge) are dynamical and further reduce the Higgs branch dimension by their equations of motion, e.g., setting $F_{\neut^i}=0$.

%%%%%%%%%%%%%%%%%%%%%%%%%%%%%%%%%%
\newsec{Conclusions}%
\seclab\secCon
%%%%%%%%%%%%%%%%%%%%%%%%%%%%%%%%%%
%
We examine topological changing transition for M-theory compactification on Calabi-Yau fourfolds, which give rise to three-dimensional theories with four supercharges. Compared to extremal transitions of M-theory/type II string compactifications on Calabi-Yau threefolds, which result in low energy effective theories with eight supercharges, a crucial new ingredient emerges due to the presence of non-trivial background $G$-fluxes. Therefore, the central theme of this work is the interplay among the quantization conditions of $G$-fluxes, the contribution of $G$-fluxes to the tadpole cancellation condition and the flat directions of the flux-induced scalar potential in the effective three-dimensional description.

To model the conifold transitions of interest, we first analyze non-compact local Calabi-Yau fourfolds with a genus $g$ curve of conifold singularities. Geometrically, we find three phases -- two small resolutions and one deformation -- that smooth the singular fourfold geometry.\foot{Note the interesting distinction from the familiar case of type II string theories on a Calabi--Yau threefold: in that case, the two small resolutions unify into a single branch of the moduli space \AspinwallNU.}
 By including the M-theory $G$-flux we find a rich structure of consistent flux configurations governing the dynamics along such a conifold transition. The flux configurations at the boundary constrains and determines the dynamics in the interior. We argue that finding the flat directions of the flux-induced superpotential in the M-theory description is mapped to the classical problem of studying global holomorphic sections of divisors on Riemann surfaces: 
Namely, the canonical line bundle of the genus $g$ curve factorizes into the two line bundles $\cE_\pm$ as determined by the background $G$-flux data. The global holomorphic sections of the factors $\cE_\pm$ then give rise to flat directions of the flux-induced superpotential. We illustrate our findings for particular curves of genus $g$ (mainly for hyperelliptic curves), but as our result holds more generally, it would be interesting to apply our techniques so as to study linear systems of line bundles on generic Riemann surfaces.

The geometrically derived factorization condition enjoys a beautiful interpretation in the associated three-dimensional $U(1)$ gauge theory. The M-theory phase structure matches with the phase structure of such gauge theories, i.e., the two resolved phases correspond to the two Coulomb branches whereas the deformed phase maps to a Higgs branch. Moreover, the global sections of the factored bundles $\cE_\pm$ are in one-to-one correspondence with the $\pm 1$ charged chiral spectrum of the $U(1)$ gauge theory. The products of such sections -- realizing the flat directions of M-theory in the deformed phase -- correspond in the field theory to gauge invariant mesonic condensates, which parametrize the Higgs branch of the three-dimensional $U(1)$ gauge theory. Moreover, in the two Coulomb branches these charged multiplets become massive, they are integrated out and generate (for a non-vanishing index of the chiral spectrum) a Chern-Simons coupling at one loop. We demonstrate that the characteristic structure of such one-loop Chern-Simons terms is reproduced by the M-theory phases attributed to the two small resolutions. 

At the transition point itself, the structure of the global symmetries together with their anomaly structure of the $U(1)$ gauge theory signal the emergence of new degrees of freedom giving evidence for a non-trivially interacting three-dimensional $N=2$ SCFT. Clearly, it would be interesting to further investigate such a $N=2$ SCFT at the transition point. To make direct contact with the M-theory description we believe that a detailed analysis of the quantum effects is necessary. 

By embedding the local Calabi-Yau fourfold geometries into compact Calabi-Yau fourfolds, we realize conifold extremal transitions in the context of global M-theory compactification. A consistent choice of $G$-flux also specifies the boundary conditions of the associated local M-theory geometry. The quantization condition imposed by the global Calabi-Yau fourfold imposes further constraints on the possible realization of $G$-flux in the local M-theory geometries. In order to realize a dynamically unobstructed conifold transition in this global setting, we find a consistently quantized background $G$-flux of ``mixed type''. As explained, these flux backgrounds of mixed type include $G$-flux quanta, which reside in both the vertical and horizontal cohomology of the Calabi-Yau fourfold. Such $G$-flux configurations reflect the factorization condition discovered for flat directions in the local Calabi-Yau fourfold transitions. In the global fourfold the factorization condition signals the appearance of non-generic algebraic four-cycle supported with $G$-flux and extremizing the flux-induced superpotential.

The class of local geometries for the conifold transitions studied here comprises geometries of ``matter curves'' in gauge theories from F-theory, studied in depth in refs.~\refs{\BeasleyDC,\BeasleyKW,\DonagiCA}. The local solutions to the quantization and flatness conditions for the $G$ flux (potential) encountered in our M-theory analysis thus carry over to solutions for the 7-brane dynamics in F-theory compactifications, provided the 3d spectrum is anomaly free in the 4d sense and the normal bundle allows for an elliptic fibration. The extension to non-Abelian $SU(n)$ gauge theories is demonstrated at the hand of an  explicit example of a chain of topologically distinct fourfolds with $A_{n-1}$  surface singularities, connected by extremal conifold transitions along curves of varying genera. We discuss consistency of $G$-fluxes for the individual fourfolds in the transition chain, but we do not examine the implications of the flux-induced potentials at the level of the underlying non-Abelian phase structure, and we hope to return to this analysis elsewhere. 

The M-theory analysis again parallels the F-theory discussion in \refs{\BeasleyDC,\BeasleyKW} and we hope the results on the M-theory classification of consistent fluxes and the flat directions of their potential also prove useful in the context of F-theory in the future.

A somewhat curious observation that deserves further study is the relation between three different objects, namely the phase transitions in (possibly interacting superconformal) 3d field theory on one hand, the fluxes for the associated extremal fourfold transition consistent with the local quantization condition on the other hand, and finally the close connection of these fluxes to 2d Kazama-Suzuki models based on the group $G=SU(M)/(SU(M-\ell)\times SU(\ell) \times U(1))$ described in refs.~\refs{\GukovYA,\EguchiFM}. In the present context, $M=2g-2$ was related to the genus $g$ of the curve $\Cg$ of conifold singularities and the integer $\ell = ||k^\sharp|-(g-1)|$ is related to the index $k^\sharp$ of the 3d field theory. This connection might be interesting from the point of a possible group theoretical classification of the components of the vacuum space as well as for a better understanding of the field theory spectrum at the 3d conformal fixed points.

%%%%%%%%%%%%%%%%%%%%%%%%%%%%%%%%%%
\bigskip
\noindent {\bf Acknowledgements:}\break
We would like to thank
Mina Aganagic,
Murad Alim,
Andr\'es Collinucci,
Thomas Grimm,
Sergei Gukov,
Jonathan Heckman,
Christoph Mayrhofer,
Sakura Sch\"afer-Nameki,
and
Timo Weigand
for useful discussions and correspondence.
H.J. and D.R.M. would like to thank Arnold Sommerfeld Center of the LMU Munich, the Banff International
Research Station, and the Simons Center for Geometry and Physics
for hospitality at various stages of this project.
H.J. would also like to thank the Kavli Institute for Theoretical Physics, where this project was initiated; D.R.M. would also like to thank the
Aspen Center for Physics for hospitality.
K.I. is supported in part by DOE-FG03-97ER40546. 
H.J. is supported by the DFG grant KL 2271/1-1. 
P.M. is supported by the program ``Origin and Structure of the Universe" of the German Excellence Initiative and the Deutsche Forschungsgemeinschaft.
D.R.M. is supported in part by NSF Grant DMS-1007414.
M.R.P. is supported in part
by NSF Grant DMS-0606578.

\bigskip
%%%%%%%%%%%%%%%%%%%%%%%%%%%%%%%%%%

%%%%%%%%%%%%%%%%%%%%%%%%%%%%%%%%%%%%%
\appendix{A}{Calabi--Yau fourfolds and collection of (co)homology data}
%%%%%%%%%%%%%%%%%%%%%%%%%%%%%%%%%%%%%
\noindent In this appendix we collect some information on the cohomology groups of the global and local fourfolds used in the text.

\noindent{\it Hodge numbers}\hfil\break
A compact Calabi--Yau fourfold $X$ has Hodge numbers $h^{p,q}$ 
satisfying the usual Hodge symmetry $h^{q,p}=h^{p,q}$, Poincar\'e
duality $h^{n-p,n-q}=h^{p,q}$ and the Calabi--Yau condition
$h^{0,0}=h^{4,0}=1$, $h^{1,0}=h^{2,0}=h^{3,0}=0$.
This apparently leaves four independent Hodge numbers
$h^{1,1}$, $h^{2,1}$, $h^{3,1}$, and $h^{2,2}$, but as shown in
\refs{\SethiES} they are not independent:
$$ -h^{1,1}+h^{2,1}-h^{3,1} = 8 - {\chi \over 6}$$
$$ -2h^{2,1} + h^{2,2} = 12 + {2\chi \over 3}$$
(where $\chi$ is the Euler characteristic of $X$), from which it follows that
$$h^{2,2}=44+ 4h^{1,1}-2h^{2,1}+4h^{3,1}.$$

The K\"ahler form $J$ of the Calabi--Yau fourfold also determines a
Lefschetz decomposition of the cohomology.  The second cohomology
has a one-dimensional imprimitive part spanned by $[J]$ and the orthogonal
complement  $(J^3)^\perp\subset H^2(X)$ is the primitive part
$H^2_{prim}(X)$.  The fourth cohomology has two imprimitive
parts: one spanned by $[J^2]$ and the other of the form
$J\wedge H^2_{prim}(X)$.

\noindent{\it Homology four-cycles and  $H^4(X,\IZ)$: global case}\hfil\break
For a compact Calabi--Yau fourfold $X$, Poincar\'e duality asserts that if $E_i$ is a basis of four-cycles for $H_4(X,\IZ)$, then there exists another basis of four-cycles $E^*_i$ such that $E_i\cap E^*_j=\delta_{ij}$. Thus $H^4(X,\IZ)\simeq H_4(X,\IZ)$ is an unimodular lattice of rank $b_4=2+2h^{1,3}+h^{2,2}$.
The signature of the lattice is the pair $(n_+,n_-)$, with the  (anti-)self-dual forms contributing to the positive (negative) part of dimension $n_+$ ($n_-$).
From the Hodge index theorem (working over $\IC$), the forms of type (4,0) and (0,4) and primitive (2,2)-forms contribute to $n_+$, forms  of type (3,1) and (1,3) to $n_-$.  Of the imprimitive (2,2) forms, those coming from $J\wedge H^{1,1}_{prim}$ (a space of dimension $h^{1,1}-1$) contribute to $n_-$ while the form $J^2$
contributes $1$ to $n_+$.  Putting these together, we find
$$(n_+,n_-) = (2+h^{2,2}-(h^{1,1}-1),2h^{3,1}+(h^{1,1}-1)).$$
On the other hand, from the Hirzebruch signature theorem it follows that
\GukovYA
\eqn\hirzeb{\eqalign{
\sigma&=n_+-n_-=8({\chi \over 24} + 4),
}}
which also follows from the relations above:
$$n_+-n_- = 4 + h^{2,2}-2h^{1,1}-2h^{3,1}=48+2h^{1,1}-2h^{2,1}+2h^{3,1}=32+{\chi \over 3}.$$

If $c_2(X)$ is even, $H^4(X,\IZ)$ is even \WittenMD, and the lattice $\Gamma_{n_+,n_-}\simeq H^4(X,\IZ)$ is unique up to isometry \Conway, with inner product
\eqn\inteven{
(e_i^*,e^*_j)\ \simeq \ H^{\oplus {b_4-\sigma \over 2}}\oplus E_8^{\oplus {\sigma \over 8}}\ ,\qquad  
H=\pmatrix{0&1\cr1&0}\ .
}
Here $E_8$ denotes the Cartan matrix of $E_8$ and we write the formula for $\sigma>0$, with the obvious changes for $\sigma<0$. If $c_2(X)$ is odd, then $H^4(X,\IZ)$ is odd \WittenMD\ and the lattice is again unique up to isometry
\Conway, this time taking the form
\eqn\intodd{
(e_i^*,e^*_j)\ \simeq \ (1)^{\oplus n_+} \oplus (-1)^{\oplus n_-}.
}

\noindent{\it Vertical and horizontal cohomology}\hfil\break
Slightly modifying the construction in \refs{\GreeneVM,\OoguriCK,\BeckerAY}, we
decompose the even cohomology of $X$ into ``vertical'' and ``horizontal''
parts.  The vertical cohomology is the subring of $H^*(X,\IZ)$
generated by $H^2(X,\IZ)$.  That is, the vertical cohomology consists
of all linear combinations of expressions $J_{i_1}\wedge \cdots \wedge
J_{i_k}$ for integral K\"ahler classes $J_{i_\alpha}$.  
These classes are always of type $(k,k)$, no matter what complex structure is
chosen on $X$, and can be thought of as ``complete
intersections'' of divisors on $X$.

On the other hand, the horizontal cohomology is the subset of $H^4(X,\IZ)$
which is orthogonal to $J_i\wedge J_j$ for every pair of K\"ahler classes
$J_i$ and $J_j$.  These classes are always primitive, no matter what
K\"ahler structure is chosen on $X$.

Finding appropriate supersymmetric four-cycles to represent a given 
integral cycle class is a challenging problem.  When the class has
type (2,2), the celebrated Hodge conjecture asserts that there should
be an algebraic cycle representing the class (which would provide a
supersymmetric cycle).  A class which is not in the vertical cohomology
can only be of type (2,2) for a proper subset of the complex structures 
on $X$.

On the other hand, when the class is primitive, it has all of the
cohomological properties one expects of a special Lagrangian cycle
(although unfortunately we have no general existence results which
would guarantee that it {\it is}\/ a special Lagrangian 
cycle.
A special Lagrangian representative would be supersymmetric.
Note that a class which is not in the horizontal cohomology can
only satisfy this primitivity assumption (the ``cohomological special
Lagrangian'' assumption) for a proper subset of the K\"ahler structures
on $X$.

There is a third possibility for a supersymmetric representative, described in
\BeckerAY: there could be a representative for the cohomology class
which is a Cayley cycle, which would give a 1/4-BPS cycle.

Note that the intersection form restricted to the horizontal cohomology
will not in general be unimodular.  Thus, the decomposition of
$H^4(X,\IZ)$ into vertical and horizontal pieces cannot in general
be done over the integers: one needs rational coefficients.  We refer
to a four-cycle with components in both spaces as a {\it mixed}\/
four-cycle.

Moreover, although a basis of integral cycles for these vertical/horizontal subspaces can be determined by fourfold mirror symmetry as described in \refs{\MayrSH,\KlemmTS}, these will {\it not}  generate $H^4(X,\IZ)$, which is the relevant group for the (appropriately shifted) $G$-flux. 

As a simple example consider the sextic $X$ in $\IP^5$ with $h^{1,1}(X)=1$ generated by the hyperplane $H$ with $\int_X H^4=6$. The generator of non-primitive four-forms in $H^{2,2}_V(X)$ is the dual of an irreducible sextic in $\IP^3$ of class $H^2$ with $\int_X (H^2)^2=6$. The primitive part is expected to be generated  (over the rationals) by duals of special Lagrangian cycles. The basis of algebraic and special Lagrangian cycles generates a finite index sublattice of $H_4(X,\IZ)$. A basis of $H^4(X,\IZ)$ necessarily includes a mixed class of the form  $e={1\over 6}H^2+\theta$, with $\theta$ a rational multiple of a form dual to a special Lagrangian cycle, possibly dual to a Cayley cycle. 

\vskip12pt

\noindent{\it Mirror symmetry and the Hodge conjecture}\hfil\break
We have identified certain cycles -- the integral $(p,p)$ cycles -- as being 
suitable for fluxes that minimize the superpotential, and other cycles
 -- the integral primitive cycles -- as being suitable for fluxes that
minimize the twisted superpotential.  Mirror symmetry between pairs
of Calabi--Yau fourfolds should exchange several things: the horizontal
and vertical cohomologies should be exchanged, the integral $(p,p)$ cycles
(which may include more cycles than just the vertical cohomology)
should be exchanged with the integral primitive cycles (which may include
more cycles than just the horizontal cohomology).

To get supersymmetric representatives of cycles, calibrated cycles should
be used, and the two natural calibrations for Calabi--Yau fourfolds -- the
K\"ahler calibration and the special Lagrangain calibration -- should also
be exchanged under mirror symmetry (since they lead to different types of branes).  Since the 
Hodge conjecture can be interpreted as asserting that any integral $(p,p)$
cycles is, up to torsion, a rational linear combination of K\"ahler-calibrated
cycles (i.e., algebraic cycles), it is tempting to formulate the following:

\vskip6pt

\noindent{\bf Mirror Hodge Conjecture}.
If $G$ is a middle-dimensional cycle on a compact Calabi--Yau manifold such that
$[G]\wedge J$ is an exact form, then, up to torsion, $G$ is a rational
linear combination of special Lagrangian cycles.

\vskip6pt

Although not as well motivated by mirror symmetry, one can go on to formulate
the:

\vskip6pt

\noindent{\bf Symplectic Hodge Conjecture}.
If $G$ is a middle-dimensional cycle on a compact symplectic  manifold such that
$[G]\wedge \omega$ is an exact form (where $\omega$ is the symplectic form), then, up to torsion, $G$ is a rational
linear combination of Lagrangian cycles.

\vskip12pt\noindent{\it Cohomology groups of local fourfolds}\hfil\break
As described in the text, the common boundary $\partial\widetilde X$ of the local fourfolds $\widetilde X^\sharp$ and $\widetilde X^\flat$ is a $S^3$ bundle with base $S^\sharp$. The integral cohomology and homology groups $H^3(\partial\widetilde X,\IZ)$ can be computed from the Gysin long exact sequence, or the Leray spectral sequence, to be 
$$
  H^q(\partial\widetilde X,\IZ) \simeq \cases{ \IZ & $q=0,7$ \cr \IZ^{2g} & $q=1,3,6$ \cr \IZ^{2} & $q=2,5$ \cr \IZ^{2g}\oplus\IZ_{2g-2} & $q=4$ } \ , \quad
  H_q(\partial\widetilde X,\IZ) \simeq \cases{ \IZ & $q=0,7$ \cr \IZ^{2g} & $q=1,4,6$ \cr \IZ^{2} & $q=2,5$ \cr \IZ^{2g}\oplus\IZ_{2g-2} & $q=3$ } \ .
$$
By exploiting the fibered structure of the local fourfolds $\Xlsharp$ and $\Xlflat$, we determine their cohomology groups to be
$$
 H^q(\Xlsharp,\IZ) \simeq \cases{ \IZ & $q=0,4$ \cr \IZ^{2g} & $q=1,3$ \cr \IZ^2 & $q=2$ \cr 0 & else } \ , \qquad
 H^q(\Xlflat,\IZ) \simeq \cases{ \IZ & $q=0,2,5$ \cr \IZ^{2g} & $q=1$ \cr \IZ^{4g-3} & $q=4$ \cr 0 & else } \ .
$$
Via the duality relations $H^q(\widetilde X^{\sharp/\flat},\IZ)\simeq H_q(\widetilde X^{\sharp/\flat},\IZ) \simeq H^{8-q}_c(\widetilde X^{\sharp/\flat},\IZ)$, we can also deduce the cohomology groups $H^{8-q}_c(\widetilde X^{\sharp/\flat},\IZ)$ of compact support, which are Poincar\'e dual to the homology groups $H_q(\widetilde X^{\sharp/\flat},\IZ)$.

\vskip12pt\noindent{\it Hodge structure of the stable degeneration components $X$ and $Y$}\hfil\break
As described in the section~4.3\yyy{} in the semi stable degeneration -- relevant to the extremal transition $\Xf$ to $\Xs$ -- the Calabi--Yau fourfold $\Xf$ degenerates into two four-dimensional component varieties $X$ and $Y$ intersecting transversely in the three-dimensional variety $E$. Since the variety $X$ is the blowup of $\Xs$ along the genus $g$ curve $\Cg$, its non-vanishing Hodge numbers read \GriffithsPAG
\eqn\HodgeX{\eqalign{ 
  &h^{0,0}(X)=h^{4,4}(X)=h^{4,0}(X)=h^{0,4}(X)=1 \ ,\cr 
  &h^{1,1}(X)=h^{3,3}(X)=h^{1,1}(\Xs)+1 \ , \cr
  &h^{2,1}(X)=h^{1,2}(X)=h^{3,2}(X)=h^{2,3}(X)=h^{2,1}(\Xs)+g \ , \cr 
  &h^{3,1}(X)=h^{1,3}(X)=h^{3,1}(\Xs) \ , \cr
  &h^{2,2}(X)=h^{2,2}(\Xs)+2 \ , }}
expressed in terms of the Hodge numbers $h^{1,1}(\Xs), h^{2,1}(\Xs), h^{3,1}(\Xs)$, and $h^{2,2}(\Xs)$ of the Calabi--Yau fourfold $\Xs$. 

The variety $Y$ is a quadratic hypersurface inside a $\IP^4$ bundle of the genus $g$ curve $\Cg$. For generic fibers over $\Cg$ the quadric fibers are of rank $5$, while there are $2g-2$ non-generic points on $\Cg$, where the quadric fibers drop to rank $4$. Using topological surgery techniques in the vicinity of the points of $\Cg$, where the quadric fibers degenerate to rank $4$, the Hodge numbers of the variety $Y$ can be derived
\eqn\HodgeY{\eqalign{ 
  &h^{0,0}(Y)=h^{4,4}(Y)=1 \ , \cr 
  &h^{1,0}(Y)=h^{0,1}(Y)=h^{4,3}(Y)=h^{3,4}(Y)=g \ , \cr
  &h^{2,1}(Y)=h^{1,2}(Y)=h^{3,2}(Y)=h^{2,3}(Y)=g \ , \cr
  &h^{1,1}(Y)=h^{3,3}(Y)=2 \ , \cr
  &h^{2,2}(Y)=2g \ . }}
Finally, the non-vanishing Hodge numbers of the intersection $E$, which is a $\IP^1\times\IP^1$ bundle of the curve $\Cg$, are given by
\eqn\HodgeE{\eqalign{ 
  &h^{0,0}(E)=h^{3,3}(E)=1 \ , \cr
  &h^{1,1}(E)=h^{2,2}(E)=3 \ , \cr
  &h^{1,0}(E)=h^{0,1}(E)=h^{3,2}(E)=h^{2,3}(E)=g \ , \cr
  &h^{2,1}(E)=h^{1,2}(E)=2g \ . }}
With the help of eqs.~\HodgeY, \HodgeX\ and \HodgeE, we will evaluate the maps in~eq.~(D.1)\yyy, where $\mathcal{X}^{[0]}$ is the disjoint union of $X$ and $Y$ and $\mathcal{X}^{[1]}\equiv E$.

%%%%%%%%%%%%%%%%%%%%%%%%%%%%%%%%%%%%%
\appendix{B}{Conifold flop transitions in local Calabi--Yau fourfolds}
%%%%%%%%%%%%%%%%%%%%%%%%%%%%%%%%%%%%%
%
To describe the flop transition between the two small resolutions $\Xlsharpi1$ and $\Xlsharpi2$, we describe the conifold fibers of the genus $g$ curve $\Cg$ as a symplectic quotient $V /\!\!/ U(1)$ as in refs.~\refs{\GuilleminAA,\WittenYC}. To this end we introduce gauged linear $\sigma$-model fields $s_1, s_2$ and $s_3, s_4$ with $U(1)$ charge $+1$ and $-1$, respectively. As usually, these gauged linear $\sigma$-model fields are constrained by the D-term
\eqn\Dterm{ | s_1 |^2 + | s_2 |^2 - | s_3 |^2 - | s_4 |^2 \,=\, r \ , }
where the parameter $r$ distinguishes among the singular phase $\Xlsing$ (for $r=0$) and the two small resolutions $\Xlsharpi1$ (for $r>0$) and $\Xlsharpi2$ (for $r<0$)  \WittenYC. In the following we collectively denote the geometry of these three phases by $\widetilde {X_r}$. 

In addition to their $U(1)$ charges the fields $s_1$ to $s_4$ transform as sections of line bundles $\cS_1$ to $\cS_4$ over the curve $\Cg$, and therefore the local fourfold $\widetilde{X_r}$ is realized as the non-trivial fibration
\eqn\FibSympl{\xymatrix{ 
  V /\!\!/ U(1) \ar[r] & \widetilde{X_r} \ar[d]^\pi \cr & \Cg } \ . }
The coordinates $x_1$ to $x_4$ in eq.~\ConSing\ arise as the gauge invariant combinations
\eqn\xCoord{ x_1 \,=\, s_1 s_3 \ , \quad x_2\,=\,s_2 s_4 \ , \quad x_3 \,=\, s_1 s_4 \ , \quad x_4 \,=\, s_2 s_3 \ , }
which fulfill the relation~\ConSing\ by construction. Moreover, the line bundles $\cS_\ell$ are related to the line bundles $\cL_\ell$ according to\foot{In terms of the line bundles $\cL_\ell$, these relations specify the line bundles $\cS_\ell$ up to a line bundle $\cP$, i.e., $\cS_{1/2} \sim \cS_{1/2} \otimes \cP$ and $\cS_{3/4} \sim \cS_{3/4} \otimes \cP^{-1}$. Note, however, that the line bundle $\cP$ cancels out in the symplectic quotient $V /\!\!/ U(1)$.}
\eqn\LSRel{\cL_1 \,=\, \cS_1 \otimes \cS_3 \ , \quad \cL_2\,=\,\cS_2  \otimes \cS_4 \ , \quad \cL_3 \,=\, \cS_1 \otimes  \cS_4 \ , \quad \cL_4 \,=\, \cS_2 \otimes \cS_3 \ . }

The (compact) surfaces $S^\sharp_1$ and $S^\sharp_2$ of the small resolutions $\Xlsharpi1$ and $\Xlsharpi2$ are now given by
\eqn\Sdef{ r>0 : \ \ S^\sharp_1 \,=\, \{ s_3 = s_4 = 0 \} \ , \qquad r<0 : \ \ S^\sharp_2 \,=\, \{ s_1 = s_2 = 0 \} \ . }
Furthermore, we define the divisor class $D_\ell = \{ s_\ell = 0 \}$ and $D_p = \pi^{-1}(p)$ in terms of a point $p$ on the curve $\Cg$. Note that these divisor classes do not depend on the parameter $r$ because the fourfold $\widetilde{ X_r}$ is a normal variety for all values of $r$ (as the conifold singularities arise in $\widetilde{X_r}$ for $r=0$ at codimension two). As a consequence, we can also define the cohomology elements $H^2(\widetilde{X_r})$ (because $H^{2,0}(\widetilde{X_r})=0$) in a $r$-independent way by means of Poincar\'e duality. Thus, we define the $(1,1)$-forms $\om_\ell$ and $\om_p$ as duals of the divisors $D_\ell$ and $D_p$.  

Let us first concentrate on the phase $r>0$. The surface $S^\sharp_1$ intersects $D_p$ at a generic $\IP^1$-fiber $F_1$, i.e., 
\eqn\defFone{ F_1\,=\,S^\sharp_1.D_p \ . }
The generic fiber $F_1$ yields according to ref.~\WittenYC\ the intersection numbers
\eqn\intSi{
  F_1.D_{1/2}=S^\sharp_1.D_p.D_{1/2}=1 \ , \quad  
  F_1.D_{3/4}=S^\sharp_1.D_p.D_{3/4}=-1 \ , \quad 
  F_1.D_p=S^\sharp_1.D_p.D_p=0 \ . }
The vanishing intersection is a consequence of the fact that two generic fibers $F_1$ are non-intersecting. Furthermore, $S^\sharp_1$ intersects the divisors $D_{1/2}$ in the two sections $C_1'$ and $C_1$, i.e.,
\eqn\defCone{ C_1\,=\,S^\sharp_1.D_2 \ , \quad C_1'\,=\,S^\sharp_1.D_1 \ , }
and we arrive at their intersections
\eqn\intSii{\eqalign{
   C_1.D_p&=1 \ , \quad C_1.D_1=0 \ , \quad C_1.D_2=\deg \cL_4 - \deg \cL_1=-n \ , \cr
   C_1'.D_p&=1 \ , \quad C_1'.D_2=0 \ , \quad C_1'.D_1=\deg \cL_1 - \deg \cL_4=n \ , }}
and the self-intersection
\eqn\intSelfi{S^\sharp_1.S^\sharp_1\,=\,2-2g \ . }
The vanishing intersections are a consequence of the fact that the two sections $C_1$ and $C_1'$ are disjoint in $S_1^\sharp$. The intersections $C_1.D_2=S^\sharp_1.D_2.D_2$ and $C_1'.D_1=S^\sharp_1.D_1.D_1$ and the self-intersection of $S^\sharp_1$ are calculated by using eqs.~\LSRel\ and the fact that -- on the level of the performed intersection calculus -- we have the relations $D_1 + D_3 \sim (\deg \cS_1 + \deg \cS_3) D_p$, $D_2 + D_3 \sim (\deg \cS_2 + \deg \cS_3) D_p$ and so on. By inspecting the resulting intersection numbers~\intSi\ and \intSii, we readily identify the curves $C_1$ and $F_1$ with the curves $C$ and $F$ (and $C_1'$ with $C'$) of eq.~\DivInt. The K\"ahler form is again given by $J(S^\sharp_1)=J^F_1(\om_2 + n\,\om_p) + J^C_1 \om_p$ (c.f., eq.~\VolS), and, as before, we find the K\"ahler volume
\eqn\VolSii{
 {1\over2} \int_{S^\sharp_1} J(S^\sharp_1)\wedge J(S^\sharp_1)\,=\,  {n\over 2}(J^F_1)^2 + J^F_1 J^C_1 \ , }
where $J^F_1$ and $J^C_1$ measure the volumes of the curves $F_1$ and $C_1$. 

We now turn to the phase $r<0$. In particular, we are interested how the volume integral \VolSii\ behaves as we traverse from the $r>0$ phase to the $r<0$ phase. The divisors and their dual $(1,1)$-forms remain invariant, but we need to recalculate the intersection numbers. Using similar arguments as for the intersection numbers in the phase $r>0$, we find for $r<0$
\eqn\intSiii{
  S_2^\sharp.D_p.D_{1/2}=-1 \ , \quad S_2^\sharp.D_p.D_{3/4}=1 \ , \quad 
  S_2^\sharp.D_3.D_4=0 \ , \quad S_2^\sharp.D_2.D_2=n-(2g-2) \ , }
and
\eqn\intSelfii{S^\sharp_2.S^\sharp_2\,=\,2-2g \ . }
With these intersection numbers we immediately infer the volume integral \VolSiii\ of the surface $S^\sharp_2$ expressed in terms of the K\"ahler coordinates $J_1^F$ and $J_1^C$.

%%%%%%%%%%%%%%%%%%%%%%%%%%%%%%%%%%%%%
\appendix{C}{The quarternionic Hopf fibration and the Milnor fibration}
%%%%%%%%%%%%%%%%%%%%%%%%%%%%%%%%%%%%%
In order to control 
the intersection properties of the
four-cycles $B_\ell^\flat$ on $\widetilde{X^\flat}$, we must take some
care in how they are defined.  Our main tools are
the Milnor fibration \MilnorAA\ 
and
the quaternionic Hopf fibration.

Locally near a zero of $\epsilon$, we can use $\epsilon$
as a local coordinate on the curve $\Cg$, and
regard $\epsilon$ as locally describing the map $\widetilde{X^\flat}\to\Cg$.
The fiber of the map over $0$ has a singular point, and
locally near that singular point, there are four coordinates
$x_\ell$ on $\widetilde{X^\flat}$ with respect to which the function
$\epsilon$ takes the form
$$ \epsilon \,=\, x_1x_2-x_3x_4 \ . $$
We will use these coordinates to describe the fourfold near
such a zero, which is
isomorphic to a neighborhood of the origin in $\IC^4$.

The function $\epsilon=\epsilon(x_\ell):\IC^4\to\IC$ 
defines an isolated hypersurface singularity,
and Milnor found a very elegant way to describe the topology near
such a singularity.  Let 
$S^7_r=\{\|x_1\|^2+\|x_2\|^2+\|x_3\|^2+\|x_4\|^2=r^2\}$ be the sphere
of radius $r$ in $\IC^4$.  The {\it Milnor fibration}\/ is the map 
$$ \left.\frac{\epsilon}{\|\epsilon\|}\right|_{S^7_r-(S^7_r\cap\{\epsilon=0\})} : S^7_r-(S^7_r\cap\{\epsilon=0\})\to S^1,$$
and its fibers, the {\it Milnor fibers}, are in general homotopic to
a wedge of $3$-spheres.  In our case, the function $\epsilon$ 
describes the
simplest isolated hypersurface singularity, and 
the Milnor fiber is diffeomorphic to $T^*S^3$, the cotangent bundle 
of the $3$-sphere.
A $3$-sphere in the Milnor fiber is a {\it vanishing cycle},
since there is a four-chain whose boundary is the $3$-sphere, given by
taking the cone over $S^3\subset S^7_r$ to get 
a bounding four-chain $\Sigma^4\subset \IB^8_r$,
where $\IB^8_r$ is the $8$-ball of radius $r$.

As we will show momentarily, a generating $3$-sphere in the Milnor
fiber can be deformed to a $3$-sphere contained in the fiber
of the map $\epsilon$.  Our strategy for giving an explicit description
of the four-cycles $B^\flat_\ell$ is as follows.  We have a path
joining $p_0$ to $p_\ell$ on $\Cg$.  In the middle of this path,
we follow a $3$-sphere within the fiber of the map $\widetilde{X^\flat}\to\Cg$.
Near each endpoint, though, we stop following the path, and follow instead the
deformation of the $3$-sphere in the fiber of $\epsilon$
to a $3$-sphere in the Milnor fiber of small radius, concluding by
using the bounding four-chain $\Sigma^4$ to close off the four-cycle.

In order to describe the behavior explicitly near the origin,
we use
the quaternionic Hopf fibration.
Let us introduce quaternion variables $q_1=x_1+x_4j$ and $q_2=x_3+x_2j$
which allow us to regard $\IC^4$ as $\IH^2$.  Note that the
quaternionic conjugate $\overline{q_1}=\bar{x}_1-x_4j$ satisfies
$q_1\overline{q_1}=\|q_1\|^2=\|x_1\|^2+\|x_4\|^2$.

There is a natural map $\IH^2\to\IH\IP^1\cong S^4$ defined by
$(q_1,q_2)\mapsto[q_1,q_2]$. If $q_1\ne0$ then
$$\eqalign{[q_1,q_2]\,&=\,[1,\frac{q_2}{q_1}] 
  \,=\,[1,\frac{q_2\overline{q_1}}{\|q_1\|^2}]
  \,=\,[1,\frac{(x_3+x_2j)(\bar{x}_1-x_4j)}{\|x_1\|^2+\|x_4\|^2}] \cr
  \,&=\,[1,\frac{(\bar{x}_1x_3+x_2\bar{x}_4)+(x_1x_2-x_3x_4)j}{\|x_1\|^2+\|x_4\|^2}] \ .}
$$
If we restrict this map to the sphere $S^7_r\subset \IH^2$ of radius $r$,
we get the {\it quaternionic Hopf fibration}\/ $S^7\to S^4$ whose
fibers are $3$-spheres.  More explicitly, if we fix 
$\sigma+\tau j\in\IH\subset S^4$, then the quaternionic Hopf fiber
over $\sigma+\tau j$ defined by
$$
\frac{\bar{x}_1x_3 + x_2\bar{x}_4}{\|x_1\|^2+\|x_4\|^2} \,=\, \sigma \ , \quad
\frac{x_1x_2-x_3x_4}{\|x_1\|^2+\|x_4\|^2} \,=\, \tau \ , \quad
\|x_1\|^2+\|x_2\|^2+\|x_3\|^2+\|x_4\|^2\,=\,r^2 
$$
is a three-sphere.

We have chosen our coordinates very carefully, to insure that
$\tau/\|\tau\|=\epsilon/\|\epsilon\|$.  Thus, {\it the three-spheres 
in the quaternionic Hopf fibration are contained in the Milnor fibers
for the function $\epsilon=x_1x_2-x_3x_4$.}  (Note that the
complex variable $\sigma$ labels a real $2$-parameter family
of such $3$-spheres within a fixed Milnor fiber labeled by $\tau$.  
We denote that $3$-sphere by
$S^3_\sigma$ if we need to emphasize this dependence on parameters.)
The advantage of this description
is that we can immediately see that the bounding four-chains
$\Sigma^4$ are quaternion-linear subspaces of $\IH^2$, and so
are nonsingular four-manifolds with boundary.   Moreover,
any two such bounding four-chains meet transversally in a single
point, with intersection number $+1$ (using the natural orientation
of the quaternions).  This reflects a well-known property of
the quaternionic Hopf fibration, analogous to the same property of
the ordinary Hopf fibration: any two fibers $S^3_j\subset S^7$
($j=1,2$)
of the quaternionic Hopf fibration have linking number $1$
in the $7$-sphere.

To finish our story, we must show that the Hopf fiber $S^3_\sigma$
in the Milnor fiber over $\tau/\|\tau\|$ can be naturally
deformed to a $3$-sphere in the fiber $\epsilon^{-1}(\tau)$
of the holomorphic map $\epsilon$.
Assume that $\sigma\ne0$, which imples that $x_1$ and $x_4$ do not simultaneously
vanish on $S^3_\sigma$.
We rescale, defining
$$
  \widehat{x}_\ell\,=\, \frac{x_\ell}{\sqrt{\|x_1\|^2+\|x_4\|^2}} \ , \quad \ell=1,2,3,4 \ .
$$
The defining equations for the fiber of the Hopf fibration become
$$
\bar{\widehat{x}}_1\widehat{x}_3 + \widehat{x}_2\bar{\widehat{x}}_4 \,=\, \sigma \ , \quad
\widehat{x}_1\widehat{x}_2 - \widehat{x}_3\widehat{x}_4 \,=\, \tau \ , \quad
\|\widehat{x}_1\|^2+\|\widehat{x}_4\|^2\,=\,1 \ .
$$
That is, the rescaled $3$-sphere $\widehat{S^3_\sigma}$ is contained 
in $\epsilon^{-1}(\tau)$.
(Note that $\|\widehat{x}_1\|^2+\|\widehat{x}_2\|^2+\|\widehat{x}_3\|^2+\|\widehat{x}_4\|^2=r^2(\|x_1\|^2+\|x_4\|^2)$
does not constrain the variables, but rather, allows the definition 
$$
  \|x_1\|^2+\|x_4\|^2\,=\,\frac{\|\widehat{x}_1\|^2+\|\widehat{x}_2\|^2+\|\widehat{x}_3\|^2+\|\widehat{x}_4\|^2}{r^2} \ ,
$$ 
which can be used to  construct the inverse transformation.)

Let us now consider the intersection number of two such cycles
$B^\flat_\ell.B^\flat_{\ell'}$.  Near each point $p_j$,
the bundle of $3$-spheres in fibers is an oriented bundle, so we
can fix a consistent orientation of the $3$-spheres throughout
a neighborhood of $p_j$.  We can also fix a common orientation
for all paths emanating from $p_j$, either pointing towards the
point or pointing away from the point.  Note that changing the
orientation of all of the $3$-spheres changes the orientations of
both $B^\flat_\ell$ and $B^\flat_{\ell'}$, and thus does not
change the intersection number.  Similarly, changing the orientation
of all of the paths emanating from $p_j$ does not change the intersection
number.

We computed a local contribution to the intersection number at a point
$p_j$ by using the natural orientation of the quaternions.  Given a 
choice of how to orienting paths emanating from $p_j$, the quaternion
orientation determines an orientation of all of the $3$-spheres.
Whichever orientation it is, it is the same orientation for both
four-cycles, so the local contribution of ${}+1$ to the intersection number
is correct.

Globally, if $\ell\ne\ell'$ then $B^\flat_\ell$ and     $B^\flat_{\ell'}$
are given by paths from $p_0$ to $p_\ell$ and $p_{\ell'}$, respectively.
If we choose the tangent directions of those two paths at $p_0$ to
be different, then we can use the computation above to conclude that
the intersection number is $1$.

On the other hand, if $\ell=\ell'$, we can choose two different paths
from $p_0$ to $p_\ell$, and we can take them to have differnt tangent
directions at $p_0$ and also at $p_\ell$.  Thus, at each endpoint we
get a contribution of $1$, for a total intersection number of $2$.

%%%%%%%%%%%%%%%%%%%%%%%%%%%%%%%%%%%%%
\appendix{D}{The Clemens-Schmid exact sequence}
%%%%%%%%%%%%%%%%%%%%%%%%%%%%%%%%%%%%%

%%%
\subsec{Triple-point-free Clemens-Schmid exact sequences}
%%%
The original sources for the Clemens--Schmid exact sequence are
refs.~\refs{\SchmidAA,\ClemensAA}; we follow the exposition in ref.~\MorrisonAA,
which is based in part on refs.~\refs{\GriffithsAA,\CornalbaAA,\PerssonAA}.

A {\it semistable degeneration} is 
 a K\"ahler manifold $\mathcal{X}$ of dimension $d+1$ together 
with a map $\mathcal{X}\to\Delta$ to the unit disk such that the fibers
$\mathcal{X}_t:=\pi^{-1}(t)$ for $t\ne0$ are compact complex manifolds of dimension $d$
and $\mathcal{X}_0=\bigcup X_i$ is reduced divisor
with each $X_i$ 
a compact complex manifold of dimension $d$, such that all intersections
of distinct components $X_{i_1}$, \dots, $X_{i_k}$ are
transverse.  (``Reduced'' means that the function $t$ has a simple zero
along each $X_i$.)

We will consider a special case of this, in which $X_i$ meets $X_j$
transversally for $i\ne j$, but all triple intersections
$X_i\cap X_j \cap X_k$ ($i$, $j$, $k$ distinct) are empty.
In this case, we call the degeneration {\it triple-point-free}
following refs.~\refs{\CrauderAA,\CrauderAB}.

For a triple-point-free degeneration, 
define $\mathcal{X}^{[0]}$ to be the disjoint union of the components
$X_i$, and $\mathcal{X}^{[1]}$ to be the disjoint union of the
intersections $X_{ij}:=X_i\cap X_j$.  There are restriction maps
$$
  H^m(\mathcal{X}^{[0]}) \longrightarrow H^m(\mathcal{X}^{[1]}) \ ,
$$  
and we define
\eqn\MVSeq{\eqalign{
E_2^{0,m} \,&:=\, \operatorname{Ker}(H^m(\mathcal{X}^{[0]}) \longrightarrow H^m(\mathcal{X}^{[1]}))\ ,\cr
E_2^{1,m} \,&:=\, \operatorname{Coker}(H^m(\mathcal{X}^{[0]}) \longrightarrow H^m(\mathcal{X}^{[1]})) \ . }}
As the notation suggests, these are the $E_2$ terms in a spectral sequence,
which degenerates at $E_2$ and
converges to the cohomology of $\mathcal{X}_0$.  In practice, this means
that there are short exact sequences
\eqn\TPFXzero{
  0\longrightarrow E_2^{1,m-1} \longrightarrow H^m(\mathcal{X}_0) \longrightarrow E_2^{0,m}\longrightarrow 0 \ . }
For brevity, we denote $H^m(\mathcal{X}_0)$ by $H^m$.  It turns out
that this is also isomorphic to the cohomology of the
total space  $H^m(\mathcal{X})$.
(For general semistable degenerations, the construction involves more
strata $\mathcal{X}^{[k]}$, and is much more complicated.)

These cohomology groups carry {\it mixed Hodge structures}.  This means
that there is a ``weight''
filtration on cohomology whose graded pieces carry Hodge structures
of the given weight.\foot{A {\it Hodge structure of weight $k$}
on a vector space $V$ is a decomposition $V\otimes \mathbb{C}
\cong \bigoplus_{p+q=k}V^{p,q}$ with $V^{q,p}=\overline{V^{p,q}}$.}
In the triple-point-free case, the weight filtration has
only two non-trivial terms: $W_{m-1}H^m=E_2^{1,m-1}$, and
$W_mH^m=H^m$ (with $W_{m-2}H^m=\{0\}$); 
the corresponding graded pieces $Gr_{m-1}^WH^m:=W_{m-1}H^m$ 
and $Gr_m^WH^m:=W_mH^m/W_{m-1}H^m$ carry
Hodge structures of weights $m-1$ and $m$, respectively.

There is an induced filtration, with induced Hodge structures
of negative weight, on the homology $H_m$.
The weights of the Hodge structures are $-m$ and $-m+1$.

On the other hand, the cohomology of the general fibers $H^m(\mathcal{X}_t)$
admit a monodromy transformation $T$ with a logarithm $N$, and the limit
as $t\to0$ of the Hodge structures on $H^m(\mathcal{X}_t)$ gives another
mixed Hodge structure.  Let us denote this limit by $H^m_{\text{lim}}$.
The {\it monodromy weight filtration} is in general a somewhat complicated
linear algebra construction using $N$, but in the case that $N^2=0$
(which corresponds to our triple-point-free situation)
it takes a simple form: 
\eqn\Ntpf{
W_{m-1}H^m_{\text{lim}}\,=\,\operatorname{Im}(N) \ , \quad
W_{m}H^m_{\text{lim}}\,=\,\operatorname{Ker}(N) \ , \quad
W_{m+1}H^m_{\text{lim}}\,=\,H^m_{\text{lim}}  \ . }
So there is a Hodge structure of weight $m-1$ on $\operatorname{Im}(N)$,
a Hodge structure of weight $m$ on $\operatorname{Ker}(N)/\operatorname{Im}(N)$
and a Hodge structure of weight $m+1$ on $H^m_{\text{lim}}/\operatorname{Ker}(N)$.

The Clemens--Schmid exact sequence is an exact sequence of mixed Hodge
structures:
\eqn\CSsqn{
\cdots \longrightarrow H_{2n-m+2} \overset{\alpha}{\longrightarrow} H^m \overset{i^*}{\longrightarrow} H^m_{\text{lim}}
\overset{N}{\longrightarrow} H^m_{\text{lim}} \overset{\beta}{\longrightarrow} H_{2n-m} \overset{\alpha}{\longrightarrow} H^{m+2} \longrightarrow \cdots \ . }
Let us describe the various maps in this sequence.  $N$ is the logarithm
of monodromy, as described above.  $i^*$ is just the restriction of
a cohomology class from the total space $\mathcal{X}$ to the fiber
$\mathcal{X}_t$. $\alpha$ is the composition of Poincar\'e
duality on the total space
$$
  H_{2n-m+2}(\mathcal{X})\longrightarrow H^m(\mathcal{X},\partial \mathcal{X}) \ ,
$$  
with the natural map from relative cohomology to absolute cohomology
$$
  H^m(\mathcal{X},\partial \mathcal{X})\longrightarrow H^m(\mathcal{X}) \ , 
$$  
while $\beta$ is the composition of Poincar\'e duality on the fiber
$$
  H^m(\mathcal{X}_t) \longrightarrow H_{2n-m}(\mathcal{X}_t) \ ,
$$  
with the homology push-forward
$$
  H_{2n-m}(\mathcal{X}_t)\overset{i_*}{\longrightarrow}  H_{2n-m}(\mathcal{X}) \ . 
$$  

The Clemens--Schmid exact sequence
 induces exact sequences on the graded pieces, which are morphisms
of (pure) Hodge structures.  There are four such 
exact sequences: three isomorphisms
\eqn\sqni{
 0 \longrightarrow
Gr_{m+1}^WH^m_{\text{lim}} \overset{N}{\longrightarrow}
Gr_{m-1}^WH^m_{\text{lim}} \longrightarrow
0 \ , }
\eqn\sqnii{
0 \longrightarrow 
Gr_{m-1}^WH^m \overset{i^*}{\longrightarrow}
Gr_{m-1}^WH^m_{\text{lim}} \longrightarrow
0 \ , }
\eqn\sqniii{
 0 \longrightarrow 
Gr_{m+1}^WH^m_{\text{lim}} \overset{\beta}{\longrightarrow}
Gr_{m-2n+1}^WH_{2n-m} \longrightarrow
0 \ , }
and a more interesting one
\eqn\sqniv{
 0 \longrightarrow
Gr_{m-2}^WH^{m-2}_{\text{lim}} \overset{\beta}{\longrightarrow}
Gr_{m-2n-2}^WH_{2n-m+2} \overset{\alpha}{\longrightarrow}
Gr_{m}^WH^{m} \overset{i^*}{\longrightarrow}
Gr_{m}^WH^{m}_{\text{lim}} \longrightarrow
 0 \ . }
Thus, the crucial things to calculate in any example are the kernel
and cokernel of
\eqn\sqnalpha{
 \alpha:\ Gr_{m-2n-2}^WH_{2n-m+2} \longrightarrow
Gr_{m}^WH^{m} \ , }
for each $m$.

%%%
\subsec{Conifold transition in Calabi--Yau threefolds}
%%%
The Clemens--Schmid exact sequence can be used to study
 three-dimensional conifold transitions
\refs{\CandelasUG,\GreeneHU} in the general geometric setting of 
refs.~\refs{\ClemensAB,\ClemensAC,\FriedmanAA,\MorrisonAB}.
This has previously been worked out in ref.~\DiaconescuJV; we
review it here to establish notation and familiarity with our setup.

We take a family $\widetilde{\mathcal{X}}$
of Calabi--Yau threefolds $\mathcal{X}_s$ depending on $s\in\Delta$
which acquires $\delta$ nodes at $s=0$.  Our key assumption is that
the $\delta$ vanishing cycles for these nodes only span a subspace
of dimension $\sigma:=\delta-\rho$ within $H^3(\mathcal{X}_s)$.
Since the vanishing cycles span the image of $N$ on $H^3_{\text{lim}}$,
this implies that $W_2H^3_{\text{lim}}=\operatorname{Im}(N)$ has
rank $\sigma$, and so that $Gr_{2}^WH^3\cong Gr_2^WH^3_{\text{lim}}$
also has rank $\sigma$.

To obtain a semistable degeneration, we would like to blowup the nodes,
but a simple computation shows that the resulting central fiber would
not be reduced (i.e., the function $s$ would have a double zero along
the exceptional divisor).  The way forward is pointed to by 
{\it Mumford's semistable reduction theorem} \KempfAA, 
and we make the ``basechange''
$s=t^2$ before blowing up the nodes.

It is worth making the local computation to see what is going on.
We have
$$
  x_1x_2-x_3x_4\,=\,t^2 \ , 
$$  
and we are blowing up the origin in that space. At $t=0$ we see two
(local) components: one of them, $X$, is the blowup of the original fiber
$\widetilde{\mathcal{X}}_0$ at the node.  Note that this is the {\it not}
the familiar small blowup which replaces the node by a $\mathbb{P}^1$
but rather a bigger blowup which replaces it by a quadric surface
$\mathbb{P}^1\times\mathbb{P}^1$.  The other local component $Y$ is
the exceptional divisor of the blowup, isomorphic to a nonsingular
quadric hypersurface in $\mathbb{P}^4$.
Note that $X\cap Y = E \cong \mathbb{P}^1\times \mathbb{P}^1$.

More globally, we will have $X$ and $\delta$ exceptional divisors
$Y_1$, \dots, $Y_\delta$, one for each node, with
$\mathcal{X}_0 = X \cup \bigcup Y_i$.
$X$ is the blowup of $X^\sharp$ along the rational curves
$S^\sharp_i$ with exceptional divisors $E_i\subset X$, and $\mathcal{X}_t$
coincides with $X^\flat$.

To compute the cohomology of $\mathcal{X}_0$ and its weight filtration,
we must study the maps $H^m(\mathcal{X}^{[0]})\to H^m(\mathcal{X}^{[1]})$,
which in this case can be written as
$$
  H^m(X^\sharp)\oplus Bl^m \oplus \bigoplus H^m(Y_i) \longrightarrow \bigoplus H^m(E_i)\ , 
$$  
where $Bl^m$ represents the addition to the cohomology of $X^\sharp$
caused by the blowup.  Explicitly, if $e_i\subset E_i$ is the fiber of
$E_i\to S^\sharp_i$, then $Bl^2=\bigoplus \mathbb{Z}[E_i]$ and
$Bl^4=\bigoplus \mathbb{Z}[e_i]$.

In the cases $m=1$ and $m=5$, all of the constituents of the cohomology
vanish, so we conclude that $W_1H^1$, $W_1H^2$, $W_5H^5$, and $W_5H^6$
all vanish.  In the case $m=3$, the only nonvanishing constitutent
is $H^3(X^\sharp)$, so $W_3H^3=H^3(X^\sharp)$ and $W_3H^4$ vanishes.

In the case $m=0$, the maps $H^0(Y_i)\to H^0(E_i)$ are isomorphisms,
so the kernel is just $W_0H^0=H^0(X^\sharp)$ and the cokernel $W_0H^1$
vanishes.  In the case $m=6$, the right side vanishes and we simply get
$W_6H^6 = H^6(X^\sharp)\oplus\bigoplus H^6(Y_i)=\mathbb{Z}^{\delta+1}$.
 
In the case $m=4$, since $Gr_4^WH^5$ vanishes, the map
$$
  H^4(X^\sharp)\oplus \bigoplus \mathbb{Z}[e_i] \oplus \bigoplus H^4(Y_i) \longrightarrow \bigoplus H^4(E_i)
$$
is surjective, and its kernel must be 
$W_4H^4 = H^4(X^\sharp)\oplus \mathbb{Z}^\delta$.

Finally, in the case $m=2$, since $Gr_{2}^WH^3$ has rank $\sigma$ and it is the
cokernel of
the map
$$
  H^2(X^\sharp)\oplus \bigoplus \mathbb{Z}[E_i] \oplus \bigoplus H^2(Y_i) \longrightarrow
\bigoplus H^2(E_i) \ ,
$$
the kernel of that map is
$W_2H^2 = H^2(X^\sharp) \oplus \mathbb{Z}^\sigma$.

Now we can use the Clemens--Schmid exact sequence to relate the cohomology
and Hodge structures of $X^\sharp$ and $X^\flat$.  We focus on the
``interesting'' part of the sequence.  If $m=4$, this reads
$$
  0 \longrightarrow Gr_2^WH^2_{\text{lim}} \longrightarrow Gr_{-4}^WH_4 \longrightarrow Gr_4^WH^4
  \longrightarrow Gr_4^WH^4_{\text{lim}} \longrightarrow 0 \ . 
$$
Note that there is no monodromy here, so the first and last terms coincide
with $H^2(X^\flat)$ and $H^4(X^\flat)$, respectively.  Since the middle
two groups have the same rank, the kernel and cokernel must also have
the same rank.  This simply expresses  Poincar\'e duality for $H^*(X^\flat)$,
and we learn nothing new.

If $m=3$, our sequence reads
$$
  0 \longrightarrow Gr_1^WH^1_{\text{lim}} \longrightarrow Gr_{-5}^WH_5 \longrightarrow Gr_3^WH^3
  \longrightarrow Gr_3^WH^3_{\text{lim}} \longrightarrow 0 \ .
$$
The first two terms vanish, so this gives an isomorphism between
the last two terms.
The last term is only part of $H^3(X^\flat)$, and in fact we have
$$
H^3(X^\flat)\,=\,Gr_3^WH^3_{\text{lim}} \oplus \mathbb{Z}^{2\sigma}
\,=\,H^3(X^\sharp) \oplus \mathbb{Z}^{2\sigma} \ . 
$$
Note that at the level of Hodge structures, the limit of the Hodge
structures on $H^3(\mathcal{X}_t)$ as $t\to0$ retains a piece
of smaller rank which coincides with $H^3(X^\sharp)$ and
has a Hodge structure of weight $3$, while the other part of the cohomology
goes to parts of the limiting mixed Hodge structure of weights $2$
and $4$.

Finally, if $m=2$, our sequence reads
$$
  0 \longrightarrow Gr_0^WH^0_{\text{lim}} \longrightarrow Gr_{-6}^WH_6 \longrightarrow
  Gr_2^WH^2 \longrightarrow Gr_2^WH^2_{\text{lim}} \longrightarrow 0 \ .
$$
The first two terms have ranks $1$ and $\delta+1$, respectively, 
and the third term is isomorphic to $H^2(X^\sharp) \oplus \mathbb{Z}^\sigma$.
Thus,
$$
  H^2(X^\sharp) 
=Gr_2^WH^2_{\text{lim}} \oplus \mathbb{Z}^{\delta-\sigma}
= H^2(X^\flat)\oplus \mathbb{Z}^{\delta-\sigma} \ .
$$

%%%
\subsec{Conifold transition along a genus $g$ curve in global Calabi--Yau fourfolds}
%%%
To apply the Clemens--Schmid exact sequence
to our fourfold extremal transition between the Calabi--Yau fourfolds $\Xs$ and
$\Xf$, we must first construct a {\it semistable degeneration}, which
relates one to the other.

Our deformed Calabi--Yau fourfold has a local equation of the form
$$
  x_1x_2-x_3x_4\,=\, \epsilon \ ,
$$
where $x_1$, $x_2$, $x_3$, $x_4$, and $\epsilon$ are sections of the bundles
$\mathcal{L}_1$, \dots, $\mathcal{L}_4$, and $K_\Cg$
over $\Cg$, respectively.  We make a one-parameter deformation
of this for $t\in \Delta$ (the unit disk), 
approaching the singular Calabi--Yau space at $t=0$, as follows.
For that purpose, we use the equation
\eqn\defeq{ x_1x_2-x_3x_4\,=\,t^2\epsilon \ , }
where $t$ is the coordinate on the disk $\Delta$.  We will resolve
singularities by blowing up $x_1=\ldots =x_4=t=0$, which gives a variety that
is still fibered over $\Cg$.  (We can treat $\epsilon$ as
a local coordinate on $\Cg$.)

We can in fact regard eq.~\defeq\ as defining the deformation
more globally.  When we do the blowup, on the central fiber we don't
get the usual ``small'' blowup $\Xs$, but rather, the proper
transform in that blowup is a variety $X$ which is the blowup of
$\Xs$ along $\Ssharp$.  There is also an exceptional divisor $Y$ of this blowup
map in the ambient space, and all together the blown up total space
gives a family $\mathcal{X}$ mapping to the disk via $t$, such that
the central fiber is $\mathcal{X}_0= X \cup Y$.

To see the structure more clearly, we blow up in the ambient space in terms of the local coordinates $z_0$ to $z_4$ with $z_0={t\over x_1}$, $z_1=x_1$, $z_2={x_2\over x_1}, \ldots, z_4={x_4\over x_1}$. (These local coordinates are again appropriate sections over the curve $\Cg$.) Then the equation~\defeq\ becomes
\eqn\defeqloc{z_2-z_3 z_4 - z_0^2 \eps \,=\, 0 \ ,}
while the reduced form of the resulting {\it semistable degeneration} reads
$$
  t\,=\,z_0\,z_1 \ .
$$
The components $z_0=0$ and $z_1=0$ intersect the (local) hypersurface equations \defeqloc\ transversly and gives rise to the varieties $X$ and $Y$, respectively. 

Intrinsically, we can describe $Y$ -- locally given as $z_1=0$ -- as a quadratic hypersurface inside
a $\mathbb{P}^4$ bundle over $\mathcal{C}$.  More precisely, the 
bundle is $\mathbb{P}(\mathcal{L}_1\oplus \mathcal{L}_2\oplus \mathcal{L}_3\oplus \mathcal{L}_4\oplus \mathcal{O})$ (spanned by $x_1,\ldots,x_4$, and $t$) when $\epsilon$ giving a coefficient in the equation.  As a 
consequence, the fibers of $Y\to\mathcal{C}$ are quadrics of rank $5$
for generic points of $\mathcal{C}$, dropping to rank $4$ precisely
at the $2g-2$ zeros of the section $\epsilon$ of the canonical bundle.

We let $E=X\cap Y$, which locally is given by $z_0=z_1=0$.  This is a $\mathbb{P}^1\times \mathbb{P}^1$ bundle
over $\mathcal{C}$; more precisely, it is the fibration $\{x_1x_2-x_3x_4=0\}\subset \mathbb{P}(\mathcal{L}_1\oplus \mathcal{L}_2\oplus \mathcal{L}_3\oplus \mathcal{L}_4)$.

The cohomology of the central fiber $\mathcal{X}_0$ is governed by the short exact sequence~(D.2)\yyy, which -- together with the relations~(D.1)\yyy{} for the disjoint union $\mathcal{X}^{[0]}$ of $X$ and $Y$, and for $\mathcal{X}^{[1]}\equiv E$ -- results into the (non-vanishing) graded pieces $Gr^W_{m-1}H^m(\mathcal{X}_0)$ and $Gr^W_mH^m(\mathcal{X}_0)$. The for us relevant graded cohomology groups together with their mixed Hodge structures -- carrying weight and Hodge filtrations -- are recorded here\foot{To derive the mixed Hodge structure of the central fiber $\cX_0$, we need further cohomology data of the component varieties $X$ and $Y$, which we have collected in Appendix~A\yyy.} 
\eqn\MHXzeroi{\eqalign{
  Gr_1^WH^2(\cX_0)\,&=\,\bigoplus_{p+q=1} V_1^{p,q} \ ,  \quad 
     \dim V_1^{p,q}\,=\,\left( 0 ,\, 0 \right) \ , \cr
  Gr_2^WH^2(\cX_0)\,&=\,\bigoplus_{p+q=2} V_2^{p,q} \ , \quad
     \dim V_2^{p,q}\,=\,\left( 0,\,h^{1,1}_{\Xs},\,0 \right) \ ,
}}
and
\eqn\MHXzeroii{\eqalign{
  Gr_3^WH^4(\cX_0)\,&=\,\bigoplus_{p+q=3} V_3^{p,q} \ ,  \quad 
     \dim V_3^{p,q}\,=\,\left( 0 ,\, g-\tilde h^{2,1},\,g-\tilde h^{2,1},\, 0 \right) \ , \cr
  Gr_4^WH^4(\cX_0)\,&=\,\bigoplus_{p+q=4} V_4^{p,q} \ , \quad
     \dim V_4^{p,q}\,=\,\left( 1,\, h^{3,1}_{\Xs}\,,h^{2,2}_{\Xs}+2g-1,\,h^{3,1}_{\Xs},\,1 \right) \ .
}}
These graded cohomology groups are expressed in terms of the Hodge numbers $h^{p,q}_{\Xs}$ of the Calabi--Yau fourfold $\Xs$. $\tilde h^{2,1}$ denotes the number of harmonic $(2,1)$-forms participating in the extremal transition. That is to say, $0\le \tilde h^{2,1}\le g$ refers to the number of $(2,1)$-forms that are in the image of the canonical map $H^{2,1}(\Xs)\rightarrow H^{2,1}(\Ssharp)$ and therefore disappear together with the cycle $\Ssharp$ in the extremal transition to the fourfold $\Xf$.  

First, let's see how the vanishing cycles of the family behave.  The limiting
mixed Hodge structure on $H^4(\mathcal{X}_t)$ as $t\to0$ will have
Hodge structures of three weights (c.f., eqs.~\Ntpf), determined by the behavior of $N$,
the logarithm of monodromy.  The vanishing cycles are in $\operatorname{Im}(N)$,
and give a Hodge structure of weight $3$, which, according to the Clemens--Schmid exact
sequence (c.f., eq.~\sqnii), is identified with $Gr_3^WH^4(\cX_0)$ in \MHXzeroii. Thus, there
is no $(3,0)$ part of this Hodge structure; the $(2,1)$ and $(1,2)$ parts of dimension $(g-\tilde h^{2,1})$
contribute to $H^{3,1}_{\text{lim}}$ and $H^{2,2}_{\text{lim}}$, respectively.

The cokernel of $N$ gives a Hodge structure of weight $5$. This cokernel is
isomorphic to $\operatorname{Im}(N)$ due to eq.~\sqni. The non-vanishing $(g-\tilde h^{2,1})$-dimensional
$(3,2)$ and $(2,3)$ parts contribute to $H^{2,2}_{\text{lim}}$ and to $H^{1,3}_{\text{lim}}$, respectively.

The remaining portion of $H^{4}_{\text{lim}}$ -- corresponding to a Hodge structure of weight $4$ --
goes over to $H^4(\mathcal{X}_0)$, and we shall see how much of it matches with $H^4(X^\sharp)$.
The Clemens--Schmid exact sequence tells us together with \MHXzeroii\ (by evaluating the
cokernel~\sqnalpha\ in the sequence~\sqniv) that the weight $4$ contribution of the limiting
mixed Hodge contains $(4,0)$, $(3,1)$, $(2,2)$, $(1,3)$ and $(0,4)$ pieces. The the two one-dimensional
$(4,0)$ and $(0,4)$ parts are associated to the holomorphic $(4,0)$-form and the anti-holomorphic $(0,4)$-form

of $\Xs$. The $h^{3,1}_{\Xs}$ dimensional $(3,1)$ and $(1,3)$ pieces are identified with the complex structure
deformations of the fourfold $\Xs$. Finally, the $(2,2)$ part is $(h^{2,2}_{\Xs}+2g - 4)$-dimensional. Only $(h^{2,2}_{\Xs}-1)$
of these forms are associated with $(2,2)$-forms of $\Xs$. The remaining $(2g-3)$ $(2,2)$-pieces, which go
over to $H^4(\mathcal{X}_0)$, are associated with the $(2g-3)$ four-forms $[B^\flat_\ell]$ and map to forms in $H^4(Y)$.

Thus, in summary the Hodge diamond of $\Xf$ expressed in terms of the Hodge numbers of $\Xs$ and decomposed
into the pieces of the limiting mixed Hodge structure of $H^4_{\text{lim}}$ reads
\eqn\HDXf{ H^{p,q}(\Xf)\,=\, \bigoplus_m Gr_m^WH^{p,q}_{\text{lim}} \ , }
with
\eqn\HDXfgraded{
  \dim Gr_m^WH^{p,q}_{\text{lim}}\,=\!\!\!
    \vcenter{\offinterlineskip
    \halign{\hbox to 7ex{#}&\hbox to 7ex{#}&\hbox to 7ex{#}&\hbox to 7ex{#}&\hbox to 7ex{#}&\hbox to 7ex{#}&\hbox to 7ex{#}&\hbox to 7ex{#}&\hbox to 7ex{#}&\hbox to 7ex{#}\cr
    % 0-forms
    &&&& \sentry{0} \cr
    &&&& \sentry{1} \cr
    &&&& \sentry{0} \cr
    % 1-forms
    &&& \sentry{0} && \sentry{0} \cr
    &&& \sentry{0} && \sentry{0} \cr
    &&& \sentry{0} && \sentry{0} \cr
    % 2-forms
    && \sentry{0} &&  \sentry{0} &&  \sentry{0} \cr
    && \sentry{0} &&  \sentry{h^{1,1}_{\Xs}\!-\!1} &&  \sentry{0} \cr
    && \sentry{0} &&  \sentry{0} &&  \sentry{0} \cr
    % 3-forms
    & \sentry{0} &&  \sentry{0} &&  \sentry{0} &&  \sentry{0}  \cr
    & \sentry{0} &&  \sentry{\hskip-2exh^{2,1}_{\Xs}\!-\!\tilde h^{2,1}\hskip-2ex} &&  \sentry{\hskip-2exh^{2,1}_{\Xs}\!-\!\tilde h^{2,1}\hskip-2ex} &&  \sentry{0}  \cr
    & \sentry{0} &&  \sentry{0} &&  \sentry{0} &&  \sentry{0}  \cr
    % 4-forms
    \sentry{0} && \sentry{\,\,g\!-\!\tilde h^{2,1}} &&  \sentry{g\!-\!\tilde h^{2,1}} &&  \sentry{0} &&  \sentry{0} \cr
    \sentry{1} && \sentry{h^{3,1}_{\Xs}} &&  \sentry{\hskip-3exh^{2,2}_{\Xs}\!+\!2g\!-\!4\hskip-3ex} &&  \sentry{h^{3,1}_{\Xs}} &&  \sentry{1} \cr
    \sentry{0} && \sentry{0} &&  \sentry{g\!-\!\tilde h^{2,1}} &&  \sentry{\,\,g\!-\!\tilde h^{2,1}} &&  \sentry{0} \cr
    % 5-forms
     & \sentry{0} &&  \sentry{0} &&  \sentry{0} &&  \sentry{0}  \cr
    & \sentry{0} &&  \sentry{\hskip-2exh^{2,1}_{\Xs}\!-\!\tilde h^{2,1}\hskip-2ex} &&  \sentry{\hskip-2exh^{2,1}_{\Xs}\!-\!\tilde h^{2,1}\hskip-2ex} &&  \sentry{0}  \cr
    & \sentry{0} &&  \sentry{0} &&  \sentry{0} &&  \sentry{0}  \cr
    % 6-forms
    && \sentry{0} &&  \sentry{0} &&  \sentry{0} \cr
    && \sentry{0} &&  \sentry{h^{1,1}_{\Xs}\!-\!1} &&  \sentry{0} \cr
    && \sentry{0} &&  \sentry{0} &&  \sentry{0} \cr
    % 7-forms
    &&& \sentry{0} && \sentry{0} \cr
    &&& \sentry{0} && \sentry{0} \cr
    &&& \sentry{0} && \sentry{0} \cr
    % 8-forms
    &&&& \sentry{0} \cr
    &&&& \sentry{1} \cr
    &&&& \sentry{0} \cr
    }} \!\! . }
The triplet at a position $(p,q)$ in the Hodge diamond corresponds to the three weights with respect to the limiting monodromy weight filtration, i.e., $\vcenter{\offinterlineskip\halign{\hfil$\smmath{#}$\hfil\cr \dim Gr_{p+q-1}^WH^{p,q}_{\text{lim}}\cr \dim Gr_{p+q}^WH^{p,q}_{\text{lim}}\cr \dim Gr_{p+q+1}^WH^{p,q}_{\text{lim}}\cr}}$. Note that the indicated decomposition~\HDXf\ only holds in the limit $t\to0$, whereas for finite $t$ the Hodge type of the individual graded pieces get corrected. Nevertheless, also for finite $t$ the Hodge numbers $h^{p,q}_{\Xf}$ are obtained by summing up the entries of the triplets, that is to say $h^{p,q}_{\Xf}=\dim Gr_{p+q-1}^WH^{p,q}_{\text{lim}}+\dim Gr_{p+q}^WH^{p,q}_{\text{lim}}+\dim Gr_{p+q+1}^WH^{p,q}_{\text{lim}}$.

%%%
\listrefs
\end